\documentclass[12pt]{article}
\pdfoutput=1
 
\usepackage{cite} 
\usepackage{amsmath}
\usepackage{amssymb}
\usepackage{amsfonts}
\usepackage{amsthm}
\usepackage{cancel}
\usepackage{pstricks}
\usepackage{graphicx}
\usepackage{dcolumn}
\usepackage{bm}
\usepackage{slashed}
\usepackage{bigints}
\usepackage{fullpage}
\usepackage{authblk}
\usepackage{empheq}
\usepackage{mathrsfs}
\usepackage{sectsty}
\allsectionsfont{\centering\normalfont}
\usepackage[hidelinks]{hyperref}
\hypersetup{
    colorlinks,
    linkcolor={blue!60!black},
    citecolor={blue!60!black},
    urlcolor={blue!60!black}
}

\newcommand*\wc{{\mkern 1mu\cdot\mkern 1mu}}

\renewcommand{\Re}{\operatorname{Re}}

\renewcommand{\v}[0]{\bm}

\DeclareMathOperator{\Tr}{Tr}

\def\be#1\ee{\begin{equation}#1\end{equation}}
\def\ba#1\ea{\begin{align}#1\end{align}}
\def\[#1\]{\begin{align}#1\end{align}}

\title{QED as a many-body theory of worldlines:\\ II. All-order S-matrix formalism}
\author[1]{Xabier Feal}
\author[2,3]{Andrey Tarasov}
\author[3,4]{Raju Venugopalan}
\affil[1]{\small \emph{IGFAE, Universidade de Santiago de Compostela, R\'ua de Xoaqu\'in D\' iaz de R\'abago, s/n\protect\\ 15705 Santiago de Compostela, SPAIN}
}%
\affil[2]{\small \emph{ Department of Physics, North Carolina State University, Raleigh, NC 27695, USA}}
\affil[3]{\small \emph{
Center for Frontiers in Nuclear Science (CFNS) at Stony Brook University,\protect\\
Stony Brook, NY 11794, USA}}
\affil[4]{\small \emph{Physics Department, Bldg. 510A, 
Brookhaven National Laboratory, 
 Upton, NY 11973, USA}
}%

\date{\today}

\begin{document}

\maketitle

\begin{abstract}

In \cite{Feal:2022iyn}, we developed a first-quantized worldline formalism for all-order  computations of amplitudes in QED. In particular, we demonstrated in this framework an all-order proof of the infrared safety of the Faddeev-Kulish (FK) S-matrix for virtual exchanges in the scattering of charged fermions. In this work, we extend the worldline formalism for both the Dyson and FK S-matrix to consider further the emission and absorption of arbitrary numbers of photons. We show how Low's theorem follows in this framework and derive Weinberg's theorem for the exponentiation of IR divergences. In particular, we extend our all-order proof of the IR safety of the FK S-matrix to both virtual exchanges and real photon emissions. We argue that the worldline approach 
leads to a modern Wilsonian interpretation of the IR safety of the FK S-matrix and provides a novel template for the treatment of IR divergences in real-time problems. Using Grassmannian integration methods, we derive a simple and powerful result for N-th rank vacuum polarization tensors. Applications of these methods will be discussed in follow-up work. 

\end{abstract}

\newpage
\begingroup
\hypersetup{linkcolor=black}
\endgroup

\newpage 

\section{\label{sec:introduction}Introduction}

In our previous paper \cite{Feal:2022iyn} (henceforth Paper I), we employed a powerful worldline formalism ~\cite{Edwards:2019eby,Strassler:1992zr,Schubert:2001he} to reformulate QED as a first quantized many-body Lorentz covariant theory of 0+1-dimension super-pairs of spinning charged worldlines with pair-wise worldline interactions described by Lorentz forces.  As we noted, the quest for such a first-quantized many-body framework can be traced back to the foundational papers of QED~~\cite{PhysRev.80.440,PhysRev.84.108,Schwinger:1951nm} and inspired attempts to extend such approaches to QCD~\cite{Polyakov:1980ca}.
The semi-classical description of this worldline formalism including internal degrees of freedom also motivated what is sometimes called ``old-fashioned" string theory~\cite{Berezin:1976eg,Casalbuoni:1975bj,Brink:1976uf,Barducci:1976qu,Balachandran:1976ya,Barducci:1976xq,Ohnuki:1978jv}. Other early attempts in this direction include replacing conventional Wilson loop operators by exact, gauge-invariant and renormalizable worldline path integrals in a supersymmetric loop-space formulation of QED \cite{Rajeev:1985ii} and a systematic strong-coupling $\hbar$-expansions of QED and QCD with point-like particle worldline variables \cite{PhysRevD.16.2486,PhysRevD.16.2476}.

In Paper I, we demonstrated several novel features of this reformulation of QED that are valuable for a deeper understanding of the structure of the infrared (IR) of the theory, and its extensions to other gauge theories and gravity. They also hold the promise of more efficient computations 
with phenomenological consequences. Examples of these novel results include recent work by us on the role of the chiral anomaly in the proton's spin~\cite{Tarasov:2020cwl,Tarasov:2021yll} and in a covariant wordline formulation of chiral kinetic theory~\cite{Mueller:2017arw,Mueller:2017lzw} relevant for heavy-ion collisions and in astrophysical contexts.

A significant result in Paper I suggestive of the power of this formalism was the reformulation of perturbation theory in QED to all-loop order. In particular, we showed that multi-loop vacuum-vacuum amplitudes, to all orders in perturbation theory, can be expressed compactly as nth-rank polarization tensors of worldline currents. As a limiting case of these  results~\cite{Strassler:1992zr}, one recovers the Bern-Kosower expression for the one-loop polarization tensor with arbitrary numbers of external photon legs~\cite{Bern:1991aq}. An explicit mapping of the polarization tensor to the language of Feynman diagrams is given in \cite{Tarasov:2019rfp}. For the one-loop case, these techniques can be extended to QCD to compute multi-leg one-loop amplitudes~\cite{Kosower:1987ic,Bern:1990cu,Bern:1991aq}; a recent review of the state-of-the art going beyond one-loop can be found in \cite{Kosower:2022iju}.

Another key result of Paper I was a demonstration of a formal proof of the infrared  finiteness of the Faddeev-Kulish (FK) S-matrix~\cite{Kulish:1970ut} to all orders in perturbation theory. The original argument of Faddeev and Kulish was that the IR divergences of QED present in the convential Dyson S-matrix formalism can be accounted for by dressing the {\it in} and {\it out} scattering states of the latter, to take into account the asymptotic interactions that are the underlying 
reason of these divergences. The FK S-matrix of these modified {\it in} and {\it out} states is then IR finite. In the worldline framework, the corresponding IR divergences can be removed by modifying the worldline currents to include their asymptotic interactions at very early and late times. We showed in Paper I that accounting for these asymptotic worldline currents provided a remarkably simple proof of the IR safety of the FK S-matrix for the scattering of charged fermion currents including virtual photon exchanges (dressed by virtual fermion loops) to all orders in perturbation theory. We discussed as a specific example the case of  M\"{o}ller scattering and demonstrated the equivalence of our results to those of Hannesdottir and Schwartz~\cite{Hannesdottir:2019opa}, who showed that the IR finiteness of the FK S-matrix can be obtained by absorbing 
the IR divergences of the Dyson S-matrix into asymptotic soft Wilson line factors that dress the latter. 

In this paper, we will extend the discussion of Paper I to the general case, where in addition to virtual photon exchanges one also includes the emission and absorption of arbitrary numbers of photons. A classic theorem in this regard is Low's theorem~\cite{PhysRev.140.B516}, which states that the bremsstrahlung amplitude of low energy photons in any transition can be  extracted, to all-loop orders in perturbation theory, into a soft factor multiplying the amplitude of the transition without the emission or absorption of the low energy photon\footnote{The latter is usually called the non-radiative amplitude, albeit it can include the radiation of an arbitrary numbers of \textit{hard} bosons.}. More generally, radiative amplitudes admit an expansion in powers of the emitted soft photon energy $\omega_{\v{k}}$ where the 
leading power $\mathcal{O}(\omega_{\v{k}}^{-1})$ \textit{soft theorem} reproduces classical long-distance radiation \cite{PhysRev.52.54,PhysRev.173.1505}, and depends only on the momenta of the \textit{in} and \textit{out} charged particles of the transition at infinity.

A problem of active interest is that of the systematic computation of the next-to-leading power corrections, known as \text{next-to-soft theorems}. Using Ward's identity, Low showed \cite{Low:1958sn} that for particles of spin-0 the leading $\mathcal{O}(\omega_{\v{k}}^{-1})$ {\it and}  $\mathcal{O}(\omega_{\v{k}}^{0})$ terms in this multipole expansion are  exactly determined by the non-radiative amplitude. This result was subsequently extended to spin-1/2 particles by Burnett and Kroll in \cite{PhysRevLett.20.86}, and generalized subsequently by Bell and Van Royen \cite{Bell:1969yw} to particles of arbitrary spin. Discussions of the structure of this multipole expansion have been extended to high-energy hadronic interactions \cite{Lipatov:1988ii,DelDuca:1990gz}; recent reviews of various aspects of subleading soft theorems can be found 
in \cite{McLoughlin:2022ljp,White:2022wbr,Bonocore:2021cbv}. A particularly interesting aspect of these investigations is the interpretation of the leading and subleading soft theorems in terms of the conformal dynamics of Goldstone modes of spontaneously broken BMS-like symmetries on the celestial sphere~\cite{Lysov:2014csa,Kapec:2017tkm,Arkani-Hamed:2020gyp,Lippstreu:2021avq,McLoughlin:2022ljp}. The connections of this work to the language of worldlines is under investigation and will be reported separately. 

A related fundamental result  is the generalization of Low's theorem by Weinberg who demonstrated the Abelian exponentiation and cancellation of IR divergences {\it in both QED and in gravity} to all loop orders in the Dyson S-matrix~\cite{PhysRev.140.B516}. In the modern language of gauge theories, the surviving infrared safe exponentiated soft factors 
can be expressed in terms of so-called cusp anomalous dimensions we discussed previously in Paper I. In this paper, we will first show how both Low's theorem and the Abelian exponentiation theorem are recovered in the worldline version of the Dyson S-matrix formalism. We will then construct the corresponding FK S-matrix for soft photon absorptions and emissions and demonstrate explicitly that it is infrared safe to all orders. Just as for the case of virtual exchanges, the proof follows fundamentally from the simple asymptotic classical structure of the first-quantized worldline currents. In particular, we will show how both the asymptotic real and virtual contributions of the FK S-matrix are expressed naturally in the first-quantized language of cusps in worldline currents. 
The results of this paper complete the proof initiated in Paper I of the 
IR safety of the FK S-matrix in QED to all orders in perturbation theory.

In addition, we will demonstrate explicitly how one recovers Weinberg's result for the 
emission rate for arbitrary numbers of soft photons in the $n\rightarrow m$ scattering of charged fermions. This computation is illustrative because it demonstrates that the IR safety of the Faddeev-Kulish S-matrix is due to its dependence on the finite initial and final times at which the {\it in} and {\it out} states of this S-matrix are defined; modes with energies below the characteristic scale therefore do not contribute. 
This FK strategy can be understood fundamentally as the Minkowski spacetime analog of the Wilsonian approach to the lattice regularization of ultraviolet (UV) and IR divergences in the Euclidean formulation of gauge theories. In the latter, computations of physical quantities are performed at fixed volume and lattice spacing, with the results then extrapolated to infinite volume and zero lattice spacing. The existence of the former guarantees infrared safety, while the latter limit corresponds to the UV renormalization of bare quantities into physical ones. Likewise, for real-time scattering computations, the FK S-matrix can be employed at all intermediate steps in the computation of physical observables with the 
initial and final times taken to infinity in the final step of the computation. As we will  demonstrate, worldlines provide the natural framework for a concrete implementation of this Wilsonian philosophy to real-time problems. In forthcoming work, we will demonstrate the power of this strategy in high order computations of cusp anomalous dimensions in QED.

The paper is organized as follows. In Section~\ref{sec:dyson_qed_s_matrix}, we will revisit the derivation of the Dyson S-matrix in the worldline formalism and generalize it to include the emission and absorption of arbitrary numbers of photons. This is obtained straightforwardly by including the coupling of  gauge fields to external currents and then performing the functional integral over the gauge fields; the final expressions are derived simply by taking functional derivatives of the generalized Wilson lines obtained in Paper I and then setting the external currents to zero. The low energy limit of worldline currents is examined in Section~\ref{sec:soft_theorems}; these classical currents can be factorized from the worldline path integrals in the Dyson S-matrix to all loop orders. This provides a straightforward proof of Low's theorem. We next demonstrate Abelian exponentiation of infrared divergences, and recover Weinberg's result for the cancellation of the exponentiated real and virtual divergences. Further, we clarify that these infrared divergences are unrelated to the infrared divergences in the 
field-strength renormalization factors that appear in the Lehmann-Symanzik-Zimmermann (LSZ) reduction formula that relates the S-matrix to the residues of the poles of time ordered Green functions. We show, following Weinberg, that these divergences are cancelled at the level of the S-matrix itself.

Section~\ref{sec:faddeev_kulish} spells out the proof the infrared safety of the FK S-matrix. This follows from the inclusion of asymptotic worldline currents which play the role of an infrared regulator of infrared divergences. Both real and virtual exponentiated soft contributions are shown to be infrared finite and combine to give Weinberg's result for the cross-section. We discuss in detail the implications of our results in the context of the Wilsonian interpretation of the FK S-matrix we outlined above. 

The paper contains two Appendices. The first of these discusses the explicit computation of N-th rank polarization tensors discussed previously in Paper I. We show for the first time that the Grassmannian integrals for these can be performed explicitly resulting in elementary expressions for vacuum polarization tensors of arbitrary rank. A straightforward consequence is the worldline proof of Furry's theorem \cite{PhysRev.51.125}. More non-trivially (as we will demonstrate in Paper III) our expressions can be used to efficiently compute light-by-light scattering and cusp anomalous dimensions to high orders. The second Appendix illustrates the cancellation between virtual and real divergences for the Dyson S-matrix and its comparison to the Faddeev and Kulish S-matrix result for concrete example of the cross-section for single photon emission in M\"oller scattering.

\section{\label{sec:dyson_qed_s_matrix}The Dyson S-matrix: radiation and absorption of real photons to all-loop orders}

In Paper I \cite{Feal:2022iyn}, we derived a general expression for the Dyson S-matrix involving the exchange of virtual photons to all loop orders. We will here extend this result to include the absorption and emission of real photons to all orders. 

Our notations are similar to those defined previously.  We will denote as $N_i^c$, $\bar{N}_i^c$, and $N_i^\gamma$ the number of \textit{in} charged particles, antiparticles and real photons, respectively, and $N_o^c$, $\bar{N}_o^c$, and $N_o^\gamma$ the number
of \textit{out} charged particles, antiparticles and real photons.  The total number of \textit{in} and \textit{out} charges are given by $N_{i,T}^c=N_i^c+\bar{N}_i^c$ and $N_{o,T}^c=N_o^c+\bar{N}_o^c$, respectively. 
Incoming and outgoing external charges are identified with indices $n,m=1,\ldots,r$, with the required number of real worldlines to describe the process, given then by $r=(N_{o,T}^c+N_{i,T}^c)/2=(\bar{N}_{o,T}^c+\bar{N}_{i,T}^c)/2$. A pair of virtual charges will be denoted with indices $i,j=1,\ldots,\ell$, with $\ell$ the required number of virtual worldlines to a given loop order within the loop expansion of the S-matrix.

The Dyson S-matrix element for emitting $N_o^\gamma$ photons and absorbing $N_i^\gamma$ photons in the initial to final ($i\to f$) transition amplitude of $r$ real charges of spin-1/2 is given by 
\ba 
\label{eq:s_fi_ph_nigamma_nfgamma}
&S_{fi}^{(r)}(N_o^\gamma,N_i^\gamma)= \lim_{\substack{t_f\to+\infty\\ t_i\to -\infty}}\Big\langle \,\underbrace{p_f^{N_o^c},s_f^{N_o^c},\,\ldots\,,\,p_f^1\,,\,s_f^1\,}_{\text{\textit{out}  charges}},\,\underbrace{\bar{p}_f^{\bar{N}_o^c},\bar{s}_f^{\bar{N}_o^c},\,\ldots\,,\,\bar{p}_f^1\,,\,\bar{s}_f^1\,}_{\text{\textit{out} anticharges}},\,\underbrace{k_f^{N_o^\gamma},\lambda_f^{N_o^\gamma}\,,\,\ldots\,,\,k_f^{1}\,,\,\lambda_f^1\,}_{\text{\textit{out} photons}}\,\nonumber\\&;t_f\,\Big|\,\underbrace{p_i^{N_i^c}, s_i^{N_i^c}\,,\,\ldots\,,\, p_i^1\,,\,s_i^1\,}_{\text{\textit{in} charges}},\, \underbrace{\bar{p}_i^{\bar{N}_i^c}, \bar{s}_i^{\bar{N}_i^c}\,,\,\ldots\,,\, \bar{p}_i^1\,,\, \bar{s}_i^1\,}_{\text{\textit{in} anticharges}},\, \underbrace{k_i^{N_i^\gamma},\lambda_i^{N_i^\gamma}\,,\, \ldots\,,\, k_i^1\,,\, \lambda_i^1}_{\text{\textit{in} photons}}\,;\, t_i\, \Big\rangle\,,
\ea 
where $|p,s\rangle$ denotes single fermion state of momentum $p$ and spin $s$, and $|k,\lambda\rangle$ a photon state of momentum $k$ and polarization $\lambda$. 

Our starting point in constructing this Dyson S-matrix element will be the QED Euclidean  path integral\footnote{We will not consider here the limit of masslesss QED where, as is well known, the emergence of additional  collinear divergences makes the treatment of IR divergences more problematic \cite{PhysRev.133.B1549,Kinoshita:1962ur,PhysRev.140.B516} relative to the discussion in this work.}
\ba
\label{eq:QED_euclidean_generating_functional}
\text{Z}[\mathcal{J},\bar{\eta},\eta] =& \int \mathcal{D}A \mathcal{D}\bar{\Psi}\mathcal{D}\Psi 
\exp\bigg\{-\frac{1}{4}\int d^4x F_{\mu\nu}^2-\frac{1}{2\zeta}\int d^4x
  \,(\partial_\mu A_\mu)^2+\int d^4x \mathcal{J}_\mu A_\mu
  \\
&-\int
  d^4x\bar{\Psi}(\slashed{D}+m)\Psi+ \int d^4x \bar{\eta}\Psi+ \int d^4x \bar{\Psi}\eta\bigg\}\,,\nonumber
\ea
where $g=\pm e$, $\zeta$ is the gauge-fixing parameter, $m$ represents the lepton mass and $\slashed{D}=D_\mu\gamma_\mu$ and 
$D_\mu = \partial_\mu -igA_\mu$. 
In addition to the two anticommuting sources $\eta(x)$ and $\bar{\eta}(x)$ required to describe real external fermions, we introduced here an  auxiliary photon source $\mathcal{J}(x)$ to describe real external photons. We will follow here the strategy in Paper I and work in Euclidean spacetime, where the path integral in Eq.~\eqref{eq:QED_euclidean_generating_functional} is well defined non-perturbatively and the gauge and matter fields can be completely integrated out to express the resulting amplitudes as a theory of worldlines. The corresponding on-shell Dyson S-matrix elements can be always obtained later, to any given order in perturbation theory, by Wick rotation of the Euclidean amplitude to Minkowski spacetime and performing successive LSZ reductions as discussed in Paper I. 

Integrating out the fermion field first in Eq.~\eqref{eq:QED_euclidean_generating_functional} gives
\ba 
\label{eq:QED_euclidean_generating_functional_psiintegrated}
\mathrm{Z}[\mathcal{J},\bar{\eta},\eta]=& \int \mathcal{D}A \exp\bigg\{-\frac{1}{4}\int d^4x F_{\mu\nu}^2-\frac{1}{2\zeta}\int d^4x (\partial_\mu A_\mu)^2+\int d^4x \mathcal{J}_\mu A_\mu\nonumber\\
&+\ln \det (\slashed{D}+m) + \int d^4x \int d^4y \bar{\eta}(x) D_F^A(x,y) \eta(y)\bigg\}\,,
\ea 
where the dressed Euclidean fermion Green function satisfies
\ba 
(\slashed{D}+m)D_F^A(x,y) = \delta^4(x-y)\,.
\ea 

To obtain the radiative amplitudes in  Eq.~\eqref{eq:s_fi_ph_nigamma_nfgamma}, we will require, as the main building blocks, the evaluation of gauge field expectation values of the propagation of $r$ spin-1/2 fields from $\{x_i^n\}$ to $\{x_f^n\}$. Let us consider here the emission and absorption of real photons for the scattering problem of $r$ positive energy charges, so that $N_i^c=N_o^c=r$. From  Eq.~\eqref{eq:QED_euclidean_generating_functional_psiintegrated}, we then obtain
\ba 
\label{eq:dlogZ_dJ_detas}
&\frac{1}{\mathrm{Z}[\mathcal{J},\bar{\eta},\eta]}\frac{\delta^{N_o^\gamma+N_i^\gamma+N_i^c+N_o^c} \mathrm{Z}[\mathcal{J},\bar{\eta},\eta]}{\delta\mathcal{J}(y_f^{N_o^\gamma})\cdots \delta\mathcal{J}(y_f^1)\delta\mathcal{J}(y_i^{N_i^\gamma})\cdots\delta\mathcal{J}(y_i^1)\delta \eta(x_i^{N_i^c})\cdots\delta\eta(x_i^{1})\delta\bar{\eta}(x_f^1)\cdots\delta\bar{\eta}(x_f^{N_o^c})}\Bigg|_{\substack{\mathcal{J}=0\\\eta=\bar{\eta}=0}}\nonumber\\
&=\frac{\delta}{\delta\mathcal{J}(y_f^{N_o^\gamma})\cdots \delta\mathcal{J}(y_f^1)\delta\mathcal{J}(y_i^{N_i^\gamma})\cdots\delta\mathcal{J}(y_i^1)}\sum_{\text{perm}}\epsilon_{n_f\ldots 1} \bigg\langle \prod_{n=1}^{r} D_F^A(x_f^n,x_i^n)\bigg\rangle_A[\mathcal{J}]\,\Bigg|_{\mathcal{J}=0}\,,
\ea 
where $\epsilon_{N_o^c\ldots 1}$ denotes the totally antisymmetric symbol, and the sum runs over the $r!$ permutations of their $r$ spacetime points given by the dressed Green functions. The generalization of the present procedure to negative energy plane wave solutions is straightforward. The expectation value $\big\langle\cdots\big\rangle_A$ contains the $A_\mu$-path integral and is given by 
\ba 
\label{eq:langle_prod_d_f_a_rangle}
\bigg\langle \prod_{n=1}^r D_F^A(x_f^n,x_i^n)\bigg\rangle_A [\mathcal{J}] = & \frac{1}{\mathrm{Z}[0,0,0]}\int \mathcal{D}A \exp\bigg\{-\frac{1}{4}\int d^4x F_{\mu\nu}^2 -\frac{1}{2\zeta} \int d^4x (\partial_\mu A_\mu)^2\nonumber\\
&+\int d^4x\mathcal{J}_\mu A_\mu + \log \det(\slashed{D}+m) \bigg\}\prod_{n=1}^r D_F^A(x_f^n,x_i^n)\,,
\ea 
expressed as a  functional of $\mathcal{J}$. 
It describes the many-body Green function of $r$ spinning charges propagating from $x_i^n$ to $x_f^n$ coupled to a fully dynamical gauge field $A_\mu$, that further interacts with $\mathcal{J}$, and polarizes from the vacuum an arbitrary number of virtual fermions within the fermion loop determinant.

To perform the $A_\mu$-integral above, we will express the fermion loop determinant and the $r$ fermion dressed Green functions in Eq.~\eqref{eq:langle_prod_d_f_a_rangle} as closed and open worldline propagators of virtual and real point-like spinning charges, respectively. As discussed in detail in \cite{Feal:2022iyn}, the fermion loop determinant can be rewritten exactly as
\ba 
\label{eq:det_d_m_worldlineform}
\det \big(\slashed{D}+m\big) =& \sum_{\ell=0}^\infty \frac{(-1)^\ell}{\ell!}\exp\bigg\{\frac{1}{2}\Tr \int^\infty_0 \frac{d\varepsilon_0}{\varepsilon_0}e^{-\varepsilon_0}\bigg\} \\
\times& \prod_{i=1}^\ell \int^\infty_0 \frac{d\varepsilon_i^0}{2\varepsilon_i^0}\int\mathcal{D}\varepsilon_i\frac{\mathcal{D}\pi_i}{2\pi}\int_\text{P} \mathcal{D}^4x_i \int_\text{AP} \mathcal{D}^4\psi_i \exp\bigg\{-S_{V,0}^i+i\int d^4x J_{V,\mu}^i A_\mu\bigg\}\,,\nonumber
\ea 
where the $\ell$-th term in the sum includes $\ell$ closed worldline propagators of $\ell$ virtual fermions in interaction with $A_\mu$, with the $i$-th virtual fermion free worldline action given by
\ba
\label{eq:s_v_0_i}
S_{V,0}^i= m^2\int_0^1 d\tau \varepsilon_i(\tau)+\frac{1}{4}\int^1_0 d\tau \frac{\dot{x}_i^2(\tau)}{\varepsilon_i(\tau)}+\frac{1}{4}\int^1_0 d\tau \psi_\mu^i(\tau)\dot{\psi}_\mu^i(\tau)\,,
\ea 
and its charged current by
\ba 
\label{eq:j_v_mu_i}
J_{V,\mu}^i(x)= g\int^1_0 d\tau \dot{x}_\mu^i(\tau)\delta^4\big(x-x_i(\tau)\big)-g\int^1_0 d\tau\varepsilon_i(\tau)\psi_\mu^i(\tau)\psi_\nu^i(\tau)\frac{\partial}{\partial x_\nu}\delta^4(x-x(\tau))\,.
\ea 

The various elements entering the worldline representation of the fermion determinant in Eq.~\eqref{eq:det_d_m_worldlineform} deserve a brief explanation, with the reader referred to \cite{Feal:2022iyn} for details. The $\ell$-th loop term in the infinite sum in Eq.~\eqref{eq:det_d_m_worldlineform} can be thought as the amplitude of finding $\ell$ virtual fermions polarized from the vacuum  coupled to $A_\mu$, which, in light of Eq.~\eqref{eq:langle_prod_d_f_a_rangle}, is dynamical and couples to $\mathcal{J}_\mu$. Each of the $i=1,\ldots,\ell$ virtual fermions paths are described by a commuting 0+1-dimensional closed bosonic wordline, $x_\mu^i(\tau)$, and an anticommuting Grassmanian worldline $\psi_\mu^i(\tau)$, with $\tau\in [0,1]$. The former encodes the local 4-position of the virtual point-like spinning charge $i$ in spacetime during its interaction with $A_\mu$. The later, its spin precession, with local spin tensor $\sigma_{\mu\nu}^i(\tau)=i[\psi_\mu^i(\tau),\psi_\nu^i(\tau)]/2$ couples to the magnetic and boosted electric components of the gauge field in $F_{\mu\nu}$.

Besides the worldline super-pair $\{x_\mu^i(\tau),\psi_\mu^i(\tau)\}$, a commuting \textit{einbein} worldline $\varepsilon_i(\tau)$ and its conjugate partner $\pi_i(\tau)$ are introduced
to account for the diffeomorphism invariance of the worldline parameter $\tau$. The \textit{einbeins} have trivial dynamics and can be replaced by their zero-modes $\varepsilon_i^0\equiv \varepsilon_i(0)$, that correspond
to the proper times of each virtual particle within the loop; alternately, they correspond to a Schwinger parameter in the corresponding Feynman diagram. 

The $\ell$-th loop contribution includes $\ell$ path integrals for each loop fermion present, taken over all possible closed path configurations are given by periodic (P) and anti-periodic (AP) boundary conditions for $x_\mu(\tau)$ and $\psi_\mu(\tau)$, 
respectively,
\ba 
\label{eq:p_ap_boundary_conditions}
x_\mu^i(1) = x_\mu^i(0)\,,\,\,\,\, \psi_\mu^i(1) = -\psi_\mu^i(0)\,.
\ea 
It includes also the required integrals over all possible proper-times $\varepsilon_i^0$, in which each individual fermion can complete the closed loop. The common UV divergent factor in the loop expansion in Eq.~\eqref{eq:det_d_m_worldlineform} removes the UV poles of the free loop contributions, subtracting thus the zero-point energy of the vacuum.

The $r$ dressed Green functions in Eq.~\eqref{eq:langle_prod_d_f_a_rangle} can be expressed as open worldline path integrals as well, describing in a first-quantized fashion the amplitude of the $r$ real spinning charges to go from $x_i^n$ to $x_f^n$ while interacting with the gauge field $A_\mu$. Using the Fradkin and Gitman result  \cite{Fradkin:1966zz} (see Paper I for a pedagogical derivation and further details) one obtains, 
\ba
 \bar{D}_F^A(x_f^n,x_i^n) = &\frac{1}{N_5} \exp\bigg\{\bar{\gamma}_\lambda
  \frac{\partial}{\partial \theta_\lambda^n}\bigg\}\int_0^\infty d\varepsilon^0_n
  \int d\chi^0_n \int \mathcal{D}\varepsilon_n \frac{\mathcal{D}\pi_n}{2\pi}\int \mathcal{D}\chi_n
  \mathcal{D}\nu_n\int \mathcal{D}^4x_n \int\mathcal{D}^5\psi_n\nonumber&
 \\
  &\times\exp\bigg\{-S_{R,0}^n+i\int d^4x J_{R,\mu}^n A_\mu^n \bigg\} \Bigg|_{\theta=0}\,,
\label{eq:Fradkin-Gitman}
\ea
with $\bar{D}_F^A(x_f^n,x_i^n)= D_F^A(x_f^n,x_i^n)\gamma_5$, $\bar{\gamma}_\mu=-i\gamma_\mu\gamma_5$ and $N_5$ is the free anticommuting path integral normalization. 

The various terms in the above expression also require a detailed discussion. 
Analogously to virtual fermions, a real fermion is fully described by a super-pair of {\it open} 0+1-dimensional worldlines $x_\mu^n(\tau)$ and $\psi_\lambda^n(\tau)$, but with the dimension of the fermionic degrees of freedom now labeled by $\lambda=1,\ldots,5\,$, the fifth element being necessary to implement the helicity-momentum constraint on their trajectories. Further, in distinction to virtual fermions traversing closed loops, the super-pair $\{x_\mu^n(\tau),\psi_\mu^n(\tau)\}$ of a real external spinning charge in the scattering satisfies the open boundary conditions,
\ba 
\label{eq:open_boundary_conditions}
x_\mu^n(1)=x_\mu^{f,n}\,,\,\,\, x_\mu^n(0)=x_{\mu}^{i,n}\,,\,\,\,\psi_\lambda^n(1)=-\psi_\lambda^n(0)+2\theta_\lambda^n\,.
\ea 
We have left explicit here as well the time reparametrization invariance of the real spinning charge worldline action by introducing a super-pair of \textit{einbeins}, the commuting worldine $\varepsilon_n(\tau)$ and the anticommuting worldline $\chi_n(\tau)$, with trivial dynamics, as well as their momentum conjugate partners $\pi_n(\tau)$ and $\nu_n(\tau)$. The commuting \textit{einbein} $\varepsilon_n(\tau)$ implements, as in the case of virtual fermions, the energy-momentum constraint of the real spinning charge, while $\chi_n(\tau)$ implements the required helicity-momentum constraint. The \textit{einbein} zero-modes are denoted $\varepsilon_n(0)=\varepsilon^n_0$ and $\chi_n(0)=\chi^n_0$ and correspond to a commuting and an anticommuting Schwinger parameter, respectively, in the corresponding Feynman diagrams, to any given order in perturbation theory.

The $A_\mu$-independent terms in Eq.~(\ref{eq:Fradkin-Gitman}) are collected in the super-gauge unfixed worldline action of the free real spinning charge 
\ba 
\label{eq:s_r_0}
S_{R,0}^n =& \frac{1}{4}\psi_\lambda^n(1)\psi_\lambda^n(0)+\int^1_0 d\tau\bigg\{i\pi^n(\tau)\dot{\varepsilon}^n(\tau)+ \nu^{n}(\tau)\dot{\chi}^n(\tau)+ \varepsilon^n(\tau) m^2+\frac{\dot{x}_{\mu}^{n,2}(\tau)}{4\varepsilon^n(\tau)}\nonumber\\
+&\frac{1}{4}\psi_\lambda^n(\tau)\dot{\psi}_\lambda^n(\tau)-\chi^n(\tau)\bigg(m\psi_5^n(\tau)+\frac{i}{2\varepsilon^n(\tau)}\dot{x}_{\mu}^n(\tau)\psi_\mu^n(\tau)\bigg)\bigg\}\,.
\ea 
The interaction-dependent terms in Eq.~(\ref{eq:Fradkin-Gitman}) are expressed entirely in terms of the coupling to $A_\mu$ of the real spinning charge's worldline current 
\ba
\label{eq:j_r_mu_n}
J_{\mu,R}^n ( x) =g\int^1_0 d\tau \dot{x}_\mu^n(\tau) \delta^4(x-x^n(\tau))- g\int^1_0 d\tau \epsilon(\tau) \psi_\mu^n(\tau)\psi_\nu^n(\tau)\frac{\partial}{\partial x_\nu^n} \delta^4(x-x^n(\tau))\,,
\ea 
which is then of identical form as the currents created by virtual spinning charges in  Eq.~\eqref{eq:j_v_mu_i}.

Substituting Eqs.~\eqref{eq:det_d_m_worldlineform} and \eqref{eq:Fradkin-Gitman} into Eq.~\eqref{eq:langle_prod_d_f_a_rangle}, and performing the quadratic $A_\mu$-integral yields,
\ba 
\label{eq:prod_d_f_a_xfn_xin_result}
\bigg\langle \prod_{n=1}^r \bar{D}_F^A(x_f^n,x_i^n)\bigg\rangle_A [\mathcal{J}] =&\frac{\mathrm{Z}_\text{MW}}{\mathrm{Z}[0,0,0]} \prod_{n=1}^r\bigg\{\exp\bigg\{\bar{\gamma}_\lambda\frac{\partial}{\partial \theta_\lambda^n}\bigg\}\bigg\}\nonumber\\
&\times\sum_{\ell=0}^\infty \frac{(-1)^\ell }{\ell!} \mathrm{W}^{(r,\ell)}(x_f^r,x_i^r,\theta^r,\ldots,x_f^1,x_i^1,\theta^1)\bigg|_{\theta_n=0}[\mathcal{J}]\,,
\ea 
where the $r$ and $\ell$ generalized Wilson lines and loops are encoded in the many-body propagator $\mathrm{W}^{(r,\ell)}$ for $r$ real charge trajectories from $x_i^n$ to $x_f^r$ and for $\ell$ virtual fermions to describe a closed loop, while interacting non-locally, to all orders in perturbation theory, amongst themselves, and with the external current $\mathcal{J}$:
\ba 
\label{eq:w_r_ell}
&\mathrm{W}^{(r,\ell)}(x_f^r,x_i^r,\theta^r,\ldots,x_f^1,x_i^1,\theta^1)[\mathcal{J}] \nonumber\\
&= \bigg\langle \exp\bigg\{+\frac{1}{2}\int d^4x \int d^4y \Big(iJ_\mu^{(r,\ell)}(x)+\mathcal{J}_\nu(x)\Big)D_{\mu\nu}^B(x-y)\Big(iJ_\nu^{(r,\ell)}(y)+\mathcal{J}_\nu(y)\Big)\bigg\}\bigg\rangle\,,
\ea 
where 
\ba 
\label{eq:d_munu_b_x_euclidean}
D_{\mu\nu}^B(x)=\frac{1}{4\pi^{{d}/2}}\Gamma\bigg(\frac{{d}-2}{2}\bigg)\frac{1}{(x^2)^{\frac{{d}}{2}-1}}\bigg\{\frac{1+\zeta}{2}\eta_{\mu\nu}+({d}-2)\frac{1-\zeta}{2}\frac{x_\mu x_\nu}{x^2}\bigg\}\,,
\ea
is the Euclidean free photon propagator in $d$-dimensions and the charged worldline currents of the $r$ real and $\ell$ virtual fermions in Eq.~\eqref{eq:j_v_mu_i} and \eqref{eq:j_r_mu_n} are combined into the net worldline current 
\ba 
\label{eq:j_mu_r_ell_def}
J_\mu^{(r,\ell)}(x) =  \sum_{n=1}^r J_{\mu,R}^n(x)+\sum_{i=r+1}^{r+\ell} J_{\mu,V}^i(x)\,.
\ea 
Eq.~\eqref{eq:w_r_ell} has the structure of a normalized worldline expectation value $\langle \cdots \rangle$ of a functional $\mathcal{O}$ of the $r+\ell$ real and virtual worldline currents $J^{(r,\ell)}$ in Eq.~\eqref{eq:j_mu_r_ell_def} and the external photon source $\mathcal{J}$, with the later remaining fixed. 
It is a many-body path integral
over all possible closed path configurations of the $\ell$ virtual loop particles, with the  periodic boundary conditions specified in Eq.~\eqref{eq:p_ap_boundary_conditions}, and over all possible open paths of the $r$ real particles, with the open boundary conditions given in Eq.~\eqref{eq:open_boundary_conditions}, with a final integration over each particle's proper time:
\ba 
\Big\langle &\mathcal{O}\big[J_\mu^{(r,\ell)},\mathcal{J}_\mu\big]\Big\rangle = \exp\bigg\{\frac{1}{2}\Tr \int^\infty_0 \frac{d\varepsilon_0}{\varepsilon_0^i}e^{-\varepsilon_0^i}\bigg\}\prod_{i=r+1}^{r+\ell}\bigg\{\int_0^\infty\frac{d\varepsilon_0^i}{2\varepsilon_0^i} \int \mathcal{D}\varepsilon_i \frac{\mathcal{D}\pi_i}{2\pi}\int_\text{P} \mathcal{D}^4x_i\int_\text{AP} \mathcal{D}^4\psi_i \bigg\}\nonumber\\
&\times \prod_{n=1}^r \bigg\{\frac{1}{N_5}\int_0^\infty d\varepsilon_0^n\int d\chi_0^n\int\mathcal{D}\varepsilon_n \frac{\mathcal{D}\pi_n}{2\pi}\int \mathcal{D}\chi_n \mathcal{D}\nu_n \int\mathcal{D}^4x_n \int \mathcal{D}^5\psi_n\bigg\}\nonumber\\
&\times \exp\left({-S^{(r,\ell)}_0}\right) \mathcal{O}\big[J_\mu^{(r,\ell)},\mathcal{J}_\mu\big]\,.\label{eq:langle_rangle_def}
\ea 
The free worldline action of the system of $r+\ell$ charges in the many-body path integral is simply the sum of the individual free actions in Eqs.~\eqref{eq:s_v_0_i} and \eqref{eq:s_r_0},
\ba 
\label{eq:s_r_l_j_r_l_definition}
S^{(r,\ell)}_0=\sum_{i=r+1}^{r+\ell} S_{V,0}^i +\sum_{n=1}^r S_{R,0}^n\,.
\ea 
Note that Eq.~\eqref{eq:w_r_ell} contains the full exponentiation, with spin precession and back-reaction, of all photon lines, hard or soft, connecting any of the $r$ real and $\ell$ virtual particles. It contains as well as all the photon lines connecting these with the external auxiliary current $\mathcal{J}$, and the external auxiliary current with itself. It is a straightforward extension of $\mathrm{W}^{(r,\ell)}$ in Paper I to the general case of real photon absorption and emission.  Finally, we note that the contribution of disconnected graphs in Eq.~\eqref{eq:prod_d_f_a_xfn_xin_result} will be adequately removed by the normalization $\mathrm{Z}_\text{MW}/\mathrm{Z}[0,0,0]$. 
\begin{figure}[ht] 
\centering
\includegraphics[scale=1]{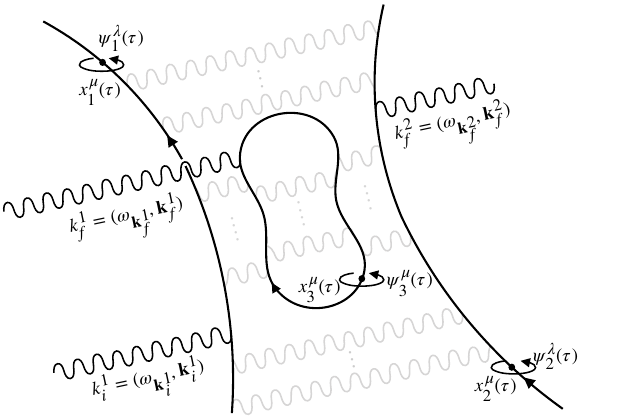}
\caption{A one-fermion loop contribution  $\mathcal{S}^{(2,1)}_{fi}(2,1)$ to the amplitude for the emission of two real photons with on-shell 4-momenta $k_f^{1,2}$ and the absorption of a real photon with four-momentum $k_i^1$, in the scattering of two real fermions. Each fermion, real or virtual, is described by a 0+1-dimensional super-pair $\{x_\mu(\tau),\psi_\mu(\tau)\}$, of open or closed worldlines respectively, exchanging arbitrary numbers of virtual photons. The worldline amplitude  implements the possibility that the three real photons are emitted or absorbed at any proper time $\tau\in[0,1]$, from any real or virtual worldline. The solid lines depict real and virtual fermion worldlines, the solid wavy line the real photons, and the faded wavy lines the virtual photon exchanges between the fermions.}
\label{fig:figure_1}
\end{figure}

Having integrated the matter and gauge fields out of $\mathrm{Z}[\mathcal{J},\bar{\eta},\eta]$, we will now proceed to construct on-shell amplitudes in physical time, as in the amplitude depicted in Fig.~\ref{fig:figure_1}. In general, the  radiative Dyson S-matrix element for the amplitude for the absorption of $N_i^\gamma$ and emission of $N_o^\gamma$ real photons during the scattering of 
$N_i^c$ to $N_o^c=N_i^c$ positive energy charges
($\bar{N}_i^c=\bar{N}_o^c=0$) in Eq.~\eqref{eq:s_fi_ph_nigamma_nfgamma} can be written as 
\ba 
&\mathcal{S}_{fi}^{(r)}(N_o^\gamma,N_i^\gamma) = \prod_{p=1}^{N_o^\gamma}\Bigg\{\frac{-i\epsilon^*_{\mu_f^p}(\v{k}_f^p,\lambda_f^p)}{\sqrt{2\omega_f^p}(2\pi)^{3/2}}\int d^4y_f^p e^{-ik_f^p\wc y_f^p}\partial_{y_f^p}^2\Bigg\}\prod_{p=1}^{N_i^\gamma}\Bigg\{\frac{-i\epsilon_{\mu_i^p}(\v{k}_i^p,\lambda_i^p)}{\sqrt{2\omega_i^p}(2\pi)^{3/2}}\int d^4y_i^p e^{+ik_i^p\wc y_i^p}\partial_{y_i^p}^2\Bigg\} \nonumber\\ &\times\prod_{n=1}^r \Bigg\{\lim_{\substack{t_f^n\to+\infty\\t_i^n\to-\infty}}\int d^3\v{x}_i^n \int d^3\v{x}_f^n \Psi^{(+),\dag}_{f_n,\beta_n}(x_f^n) \gamma^0_{\beta_n\alpha_n}\Psi_{i_n,\alpha_n}^{(+)}(x_i^n)\Bigg\}\frac{1}{\mathrm{Z}[\mathcal{J},\bar{\eta},\eta]}\bigg(\frac{1}{i}\bigg)^{N_o^\gamma+N_i^\gamma+N_o^c+N_i^c} \nonumber\\
\label{eq:QED_S_matrix_reduction}
&\times \prod_{p=1}^{N_o^\gamma}\bigg\{\frac{\delta}{\delta \mathcal{J}_{\mu_f^p}(y_f^p)} \bigg\} \prod_{p=1}^{N_i^\gamma}\bigg\{\frac{\delta}{\delta \mathcal{J}_{\mu_i^p}(y_i^p)} \bigg\}\frac{\delta^{N_i^c+N_o^c} \mathrm{Z}[\mathcal{J},\bar{\eta},\eta]}{\delta \eta_{\gamma_{N_i^c}}(x_i^{N_i^c})\ldots\delta \eta_{\gamma_1}(x_i^1)\delta \bar{\eta}_{\beta_1}(x_f^1)\cdots\delta \bar{\eta}_{\beta_{N_0^c}}(x_f^{N_o^c})}\bigg|_{\substack{\mathcal{J}=0\\\bar{\eta}=\eta=0}}
\ea 
where $\mathrm{Z}[\mathcal{J},\bar{\eta},\eta]$ is the QED generating functional in Minkowski spacetime, and 
\ba 
\Psi_{f_n}^{(+)}(x_f^n)= u(\v{p}_f^n,s_f^n)e^{-ip_f^n\wc x_f^n}\,,\,\,\,\,\,\, \Psi_{i_n}^{(+)}(x_i^n)= u(\v{p}_i^n,s_i^n)e^{-ip_i^n\wc x_i^n}\,,
\ea 
are free plane wave solutions of positive energy of momentum $p_{f,i}^n$ and spin $s_{f,i}^n$, and $\alpha_n$, $\beta_n$ and $\gamma_n$ are spin indices. The LSZ reduction formula requires that one introduce the corresponding wave-function renormalization factors for each real fermion and photon field present.  
We have left these factors implicit because, as we will discuss at length in 
section 3.3, they do not play any role in the proof of the IR safety of cross-sections. 

Consider first the case of a single outgoing photon ($N_o^\gamma=1$ and $N_i^\gamma=0$). In Euclidean time, the gauge field expectation value at point $x$ within the propagation of the $r$ charges gives, using Eqs.~\eqref{eq:dlogZ_dJ_detas} and \eqref{eq:prod_d_f_a_xfn_xin_result} (and noting $\gamma_5^2=+1$),
\ba 
\label{eq:delta_Z_delta_J_delta_etas_euclidean}
& \frac{1}{\mathrm{Z}[\bar{\eta},\eta,\mathcal{J}]}\frac{\delta^{1+N_o^c+N_i^c} \mathrm{Z}[\bar{\eta},\eta,\mathcal{J}]}{\delta\mathcal{J}_\mu(y)\delta \eta_{\gamma_r}(x_i^{N_i^c})\cdots\delta\eta_{\gamma_1}(x_i^{1})\delta\bar{\eta}_{\beta_1}(x_f^1)\cdots\delta\bar{\eta}_{\beta_r}(x_f^{N_o^c})}\Bigg|_{\substack{\mathcal{J}=0\\\eta=\bar{\eta}=0}}=\sum_\text{perm} \epsilon_{N_o^c\ldots 1}\\
&\times \prod_{n=1}^r\bigg\{\exp\bigg\{\bar{\gamma}_\lambda\frac{\partial}{\partial \theta_\lambda^n}\bigg\}\gamma_5\bigg\}_{\beta_n\gamma_n}\frac{\mathrm{Z}_\text{MW}}{\mathrm{Z}[0,0,0]} \sum_{\ell=0}^\infty \frac{(-1)^\ell }{\ell!}
\frac{\delta \mathrm{W}^{(r,\ell)}(x_f^r,x_i^r,\theta^r,\ldots,x_f^1,x_i^1,\theta^1)[\mathcal{J}]}{\delta \mathcal{J}_\mu(y)}\Bigg|_{\substack{\theta_n=0\\\mathcal{J}=0}}\,.\nonumber
\ea 
Using  further Eq.~\eqref{eq:w_r_ell}, the functional derivative of the many-body generalized Wilson lines and loops for the $r$ real and $\ell$ virtual charges can be expressed as 
\ba 
\label{eq:d_w_r_l_d_j_result}
&\frac{\delta \mathrm{W}^{(r,\ell)}(x_f^r,x_i^r,\theta^r,\ldots,x_f^1,x_i^1,\theta^1)[\mathcal{J}]}{\delta \mathcal{J}_\mu(y)}\Bigg|_{\mathcal{J}=0}\\
&= \bigg\langle \int \frac{d^4q}{(2\pi)^4}e^{-iq\wc y}\tilde{D}_{\mu\nu}^B(q) i\tilde{J}_\nu^{(r,\ell)}(q)\exp\bigg\{\frac{1}{2}\int \frac{d^4q}{(2\pi)^4}i\tilde{J}_\mu^{(r,\ell)}(-q)\tilde{D}_{\mu\nu}^B(q)i\tilde{J}_\nu^{(r,\ell)}(+q)\bigg\}\bigg\rangle\nonumber\,.
\ea 
where $\tilde{D}^B_{\mu\nu}(q)$ and $\tilde{J}_\mu^{(r,\ell)}(q)$ are respectively the Fourier transforms of the free Euclidean photon propagator in Eq.~\eqref{eq:d_munu_b_x_euclidean} and the Euclidean charged worldline currents in Eq.~\eqref{eq:j_mu_r_ell_def}.

The evaluation\footnote{See Appendix A of \cite{Feal:2022iyn} for the notations and conventions employed as well as full details of this procedure.} of the on-shell amplitude requires the Wick rotation of this quantity before plugging it into Eq.~\eqref{eq:delta_Z_delta_J_delta_etas_euclidean}. This involves replacing each free photon propagator by its Minkowski counterpart, 
\ba 
\label{eq:d_munu_b_q_minkowski}
\tilde{D}_{\mu\nu}^B(q) = -\frac{g_{\mu\nu}}{q^2+i\epsilon}+(1-\zeta)\frac{q_\mu q_\nu}{(q^2+i\epsilon)^2}\,,
\ea 
multiplied by a factor of $i$, replacing the real and virtual Euclidean worldline currents in Eq.~\eqref{eq:j_mu_r_ell_def} by currents in Minkowski spacetime, 
\ba 
\label{eq:j_mu_r_ell_q_Minkowski}
\tilde{J}_\mu^{(r,\ell)}(q) &= \sum_{n=1}^r g\int^1_0 d\tau \Big(\dot{x}_\mu^n(\tau)+\epsilon_n^0q^\nu \psi_\mu^n(\tau)\psi_\nu^n(\tau)\Big)e^{+iq\wc x_n(\tau)}\\
&+\sum_{i=r+1}^{r+\ell} g\int^1_0 d\tau \Big(\dot{x}_\mu^i(\tau)+\epsilon_i^0q^\nu \psi_\mu^i(\tau)\psi_\nu^i(\tau)\Big)e^{+iq\wc x_i(\tau)}\nonumber\,,
\ea 
and likewise,  for each dressed fermion Green function in the $2r$-point correlator.  
Finally, one replaces the Euclidean Dirac matrix $\gamma_5$ by $-i\gamma_5$. One then obtains\footnote{It is to be understood henceforth that the currents and propagators are in Minkowski spacetime unless specified otherwise.}
\ba 
\label{eq:delta_Z_delta_J_delta_etas_minkowski}
&\frac{1}{\mathrm{Z}[\bar{\eta},\eta,\mathcal{J}]}\frac{\delta^{1+N_i^c+N_o^c} \mathrm{Z}[\bar{\eta},\eta,\mathcal{J}]}{\delta\mathcal{J}_\mu(y)\delta \eta(x_i^{N_i^c})\cdots\delta\eta(x_i^{1})\delta\bar{\eta}(x_f^1)\cdots\delta\bar{\eta}(x_f^{N_o^c})}\Bigg|_{\substack{\mathcal{J}=0\\\eta=\bar{\eta}=0}}\\
&=\sum_\text{perm} \epsilon_{N_o^c\ldots 1} \prod_{n=1}^r\bigg\{\exp\bigg\{\bar{\gamma}_\lambda\frac{\partial}{\partial \theta_\lambda^n}\bigg\}\gamma_5\bigg\}_{\beta_n\gamma_n}\frac{\mathrm{Z}_\text{MW}}{\mathrm{Z}[0,0,0]}\sum_{\ell=0}^\infty \frac{(-1)^\ell }{\ell!}\nonumber\\
&\times
\bigg\langle \int \frac{d^4q}{(2\pi)^4}e^{-iq\wc y}i\tilde{D}_{\mu\nu}^B(q) i\tilde{J}_\nu^{(r,\ell)}(q)\exp\bigg\{\frac{1}{2}\int \frac{d^4q}{(2\pi)^4}i\tilde{J}_\mu^{(r,\ell)}(-q)i\tilde{D}^{\mu\nu}_B(q)i\tilde{J}_\nu^{(r,\ell)}(+q)\bigg\}\bigg\rangle\nonumber\,.
\ea 
Note that the normalized expectation value $\langle \cdots \rangle$ defined in Eq.~\eqref{eq:langle_rangle_def} has to be replaced at this point by its rotation to Minkowski time following the rules given in  Appendix A of Ref. \cite{Feal:2022iyn}.
Plugging Eq.~\eqref{eq:delta_Z_delta_J_delta_etas_minkowski} into Eq.~\eqref{eq:QED_S_matrix_reduction} and defining $\bar{\gamma}_0=\gamma_5\gamma_0$ one finally gets a loop expansion for the Dyson S-matrix element for the radiation of a single real photon off the scattering of $r$ positive energy charges of the form
\ba 
\label{eq:s_fi_r_l_1_0_loopexpansion}
\mathcal{S}_{fi}^{(r)}(1,0) =\sum_{\ell=0}^\infty \mathcal{S}_{fi}^{(r,\ell)}(1,0)\,,
\ea 
with the $\ell$-th loop contribution given by
\ba 
\label{eq:s_fi_r_ell_1_0}
&\mathcal{S}_{fi}^{(r,\ell)}(1,0) =\frac{\mathrm{Z}_\text{MW}}{\mathrm{Z}[0,0,0]}\frac{(-1)^\ell}{\ell!} \prod_{n=1}^r \Bigg\{\lim_{\substack{t_f^n\to+\infty\\t_i^n\to-\infty}}\int d^3\v{x}_i^n \int d^3\v{x}_f^n \Psi^{(+),\dag}_{f_n}(x_f^n) \exp\bigg\{\bar{\gamma}_\lambda\frac{\partial}{\partial \theta_n}\bigg\}\bar{\gamma}_0\Psi_{i_n}^{(+)}(x_i^n)\Bigg\}\nonumber\\
&\cdot \frac{i\epsilon_{\mu}^*(\v{k},\lambda)}{\sqrt{2\omega}_{\v{k}}(2\pi)^{3/2}}k^2i\tilde{D}^{\mu\nu}_B(k)\bigg\langle i\tilde{J}_\nu^{(r,\ell)}(-k)\exp\bigg\{\frac{1}{2}\int \frac{d^4q}{(2\pi)^4}i\tilde{J}_\mu^{(r,\ell)}(-q)i\tilde{D}^{\mu\nu}_B(q)i\tilde{J}_\nu^{(r,\ell)}(+q)\bigg\} \bigg\rangle +\text{perm.}
\ea 

We can further simplify Eq.~\eqref{eq:s_fi_r_ell_1_0} and simultaneously confirm that it is  gauge invariant to all-loop orders in perturbation theory. Using Eq.~\eqref{eq:d_munu_b_q_minkowski}, the real photon vertex in Eq.~\eqref{eq:s_fi_r_ell_1_0} can be rewritten as
\ba 
k^2D^{\mu\nu}_B(k)J^{(r,\ell)}_\nu(k) =  k^2\bigg\{-\frac{g^{\mu\nu}}{k^2}+(1-\zeta)\frac{k^\mu k^\nu}{k^4}\bigg\} J_{\nu}^{(r,\ell)}(k)= -J^{\mu}_{(r,\ell)}(k) +k^\mu \frac{1-\zeta}{k^2}\Big(k\wc \tilde{J}^{(r,\ell)}(k)\Big)
,\ea 
Next, using Eq.~\eqref{eq:j_mu_r_ell_q_Minkowski} the factor in front of the gauge dependent term gives
\ba 
\label{eq:k_dot_j_r_l}
&k\wc \tilde{J}^{(r,\ell)}(k)=\sum_{i=r+1}^{r+\ell} g\int^1_0 d\tau \bigg\{k\wc \dot{x}_i(\tau)+\varepsilon_i^0 \Big(k\wc \psi_i(\tau)\Big)^2\bigg\}e^{ik\wc x_i(\tau)}+\sum_{n=1}^r g\int^1_0 d\tau \bigg\{k\wc \dot{x}_n(\tau)\nonumber\\
&+\varepsilon_n^0 \Big(k\wc \psi_n(\tau)\Big)^2\bigg\}e^{ik\wc x_n(\tau)}=\frac{g}{i} \sum_{i=r+1}^{r+\ell}\bigg\{e^{ik\wc x_i(1)}-e^{ik\wc x_i(0)}\bigg\}+\frac{g}{i}\sum_{n=1}^r \bigg\{e^{ik\wc x_n(1)}-e^{ik\wc x_n(0)}\bigg\}\,.
\ea 
To obtain the final equality in the above equation, we used the fact that for either  virtual or real fermions the  quantity $k_\mu k_\nu \psi^\mu \psi^\nu$ vanishes. This is because $k^\mu k^\nu $ is a symmetric tensor whereas the Grassmanian tensor $\psi_\mu^i\psi_\nu^i$ is antisymmetric. 

The first term in brackets of the final expression on the r.h.s. of Eq.~\eqref{eq:k_dot_j_r_l} corresponds to a Ward identity for the scalar QED vertex, and has been rewritten as a total derivative of a pure phase. It vanishes since the periodic boundary conditions in Eq.~\eqref{eq:p_ap_boundary_conditions} hold for the $i=r+1,\ldots,r+\ell$ virtual fermion worldlines; this simply reflects the fact that the charge flux that goes out from $x_i(0)$ in the loop comes back in at $x_i(1)$ and is therefore a conserved quantity. 

For the second term in brackets, the $r$ real charges worldlines have open boundary conditions, $x^n(1)=x_f^n$ and $x^n(0)=x_i^n$; as noted in Eq.~\eqref{eq:open_boundary_conditions}, 
\ba
\label{eq:open_boundary_conditions_infinity}
x_f^n=(t_f^n,\v{x}_f^n)\,,\,\,\, t_f^n\to +\infty\,,\,\,\, x_i^n=(t_i^n,\v{x}_i^n)\,,\,\,\, t_i^n\to-\infty\,.
\ea 
These two remaining boundary terms $e^{ik\wc x_{n}(1)}$ and $e^{ik\wc x_{n}(0)}$ can be neglected as well\footnote{The requirement that the Ward identity be valid for external charges, and its relation to IR divergences, will be addressed at length in the next section.}; in this case however it is on account of the rapid oscillatory behavior of their phases as $t_{f,i}^n\to \pm\infty$. With this understanding, the second term in brackets in the r.h.s. of Eq.~\eqref{eq:k_dot_j_r_l} also vanishes. 

The net result is that $k\wc J^{(r,\ell)}(k)=0$ for any real or virtual worldline fermion current, and independently of its particular configuration inside the path integral. Since the previous statements are valid for off-shell photon momenta $q$ as well, the same identity holds for the gauge-dependent terms in the virtual exchange vertices in the exponential within the normalized expectation value of Eq.~\eqref{eq:s_fi_r_ell_1_0}. Therefore the only surviving terms in Eq.~\eqref{eq:s_fi_r_ell_1_0} are the gauge-independent terms that acquire the simple gauge-invariant worldline expression:
\ba 
&\mathcal{S}_{fi}^{(r,\ell)}(1,0) =\frac{\mathrm{Z}_\text{MW}}{\mathrm{Z}[0,0,0]}\frac{(-1)^\ell}{\ell!}\prod_{n=1}^r \Bigg\{\lim_{\substack{t_f^n\to+\infty\\t_i^n\to-\infty}}\int d^3\v{x}_i^n \int d^3\v{x}_f^n \Psi^{(+),\dag}_{f_n}(x_f^n) \exp\bigg\{\bar{\gamma}_\lambda\frac{\partial}{\partial \theta_n}\bigg\}\bar{\gamma}_0\Psi_{i_n}^{(+)}(x_i^n)\Bigg\}\nonumber\\
&\times \frac{i\epsilon_{\mu}^*(\v{k},\lambda)}{\sqrt{2\omega}_{\v{k}}(2\pi)^{3/2}}\bigg\langle \tilde{J}^\mu_{(r,\ell)}(-k)\exp\bigg\{\frac{i}{2}\int \frac{d^4q}{(2\pi)^4}\frac{1}{q^2+i\epsilon}\tilde{J}_\mu^{(r,\ell)}(-q)\tilde{J}^\mu_{(r,\ell)}(+q)\bigg\} \bigg\rangle+\text{perm.} 
\ea 

The procedure outlined for the emission of a single real photon can be straightforwardly extended to the emission and absorption of an arbitrary number of quanta in the scattering,
\ba 
\label{eq:s_fi_r_nfgamma_nigamma_loopexpansion}
\mathcal{S}_{fi}^{(r)}(N_o^\gamma,N_i^\gamma) = \sum_{\ell=0}^\infty \mathcal{S}^{(r,\ell)}_{fi}(N_o^\gamma,N_i^\gamma)\,,
\ea 
with the $\ell$-th loop contribution given by
\ba 
\label{eq:s_fi_rell_nfgamma_nigamma}
&\mathcal{S}_{fi}^{(r,\ell)}(N_o^\gamma,N_i^\gamma) = \frac{\mathrm{Z}_\text{MW}}{\mathrm{Z}[0,0,0]}\frac{(-1)^\ell}{\ell!}\prod_{p=1}^{N_o^\gamma} \Bigg\{\frac{i\epsilon_{\mu_f^p}^*(\v{k}_f^p,\lambda_f^p)}{\sqrt{2\omega_{\v{k}_f^p}}(2\pi)^{3/2}} \Bigg\}\prod_{p=1}^{N_i^\gamma}\Bigg\{ \frac{i\epsilon_{\mu_i^p}(\v{k}_i^p,\lambda_i^p)}{\sqrt{2\omega_{\v{k}_i^p}}(2\pi)^{3/2}}\Bigg\}\nonumber\\
\times & \prod_{n=1}^r\Bigg\{\lim_{\substack{t_f^n\to+\infty\\t_i^n\to-\infty}}\int d^3\v{x}_i^n \int d^3\v{x}_f^n \Psi^{(+),\dag}_{f_n}(x_f^n) \exp\bigg\{\bar{\gamma}_\lambda\frac{\partial}{\partial \theta_n}\bigg\}\bar{\gamma}_0\Psi_{i_n}^{(+)}(x_i^n)\Bigg\} \\
\times &\, \bigg\langle \prod_{p=1}^{N_o^\gamma} \bigg\{\tilde{J}_{(r,\ell)}^{\mu_f^p}(-k_f^p)\bigg\} \prod_{p=1}^{N_i^\gamma} \bigg\{ \tilde{J}_{(r,\ell)}^{\mu_i^p}(+k_i^p)\bigg\}\exp\bigg\{\frac{i}{2}\int \frac{d^4q}{(2\pi)^4}\frac{1}{q^2+i\epsilon}\tilde{J}_{\mu}^{(r,\ell)}(-q)\tilde{J}_{(r,\ell)}^{\mu}(+q)\bigg\}\bigg\rangle \nonumber\\
+ & \text{perm.}\nonumber\,.
\ea 

Eqs.~\eqref{eq:s_fi_r_nfgamma_nigamma_loopexpansion} and  \eqref{eq:s_fi_rell_nfgamma_nigamma} are the central result of the present work. They provide compact expressions that capture in full generality  the $i\to f$ scattering of a system of $r$ real fermions accompanied by the radiation or absorption of any number of hard or soft real photons and an arbitrary number of virtual soft or hard photons exchanged between real and virtual fermions during the scattering.
The normalized expectation value is finally taken averaging over all possible worldline paths of the $r$ real and $\ell$ virtual charges within each term in the loop expansion. An example of one of these radiative amplitudes is illustrated in Fig. \ref{fig:figure_1}. In the next section we will show how this all-order compact form of the radiative amplitudes provides an intuitive infrared safe formulation of these amplitudes in QED.

\section{\label{sec:soft_theorems} Abelian exponentiation of real and virtual IR divergences in the Dyson S-matrix}

Now that we have derived the general worldline form of the Dyson S-matrix in QED with any number of real and virtual photons attached, we will turn to addressing the problem of its IR structure. We will first show how virtual and real divergences in the S-matrix can be easily isolated in this worldline formulation from an analysis of the low energy limit of the currents, and how this behavior naturally leads to the well-known Abelian soft theorems and their exponentiation. We will then review in this context the Dyson S-matrix approach 
to IR regularization; specifically, we will show how real and virtual IR divergences cancel amongst themselves in the cross-section to all-loop orders in perturbation theory when any number of real photons of arbitrarily low energies accompany  the scattering of charged particles. In Section \ref{sec:faddeev_kulish}, we will confront this conventional solution to the IR problem in QED with the alternative formulation in terms of infrared safe Faddeev-Kulish (FK) S-matrix elements. We will further employ the worldline formalism to demonstrate how this FK S-matrix can be constructed to be manifestly free of real and virtual IR singularities to all orders in perturbation theory.

\subsection{\label{sec:abelian_exponentiation} IR structure of worldline currents in the Dyson S-matrix and Low's theorem}

We wrote the Dyson S-matrix in Eq.~\eqref{eq:s_fi_rell_nfgamma_nigamma} as a quantum theory of worldline currents, as was also the case in Paper I, with the novel feature being the absorption and emission of  real photons attached to the charged currents. Our strategy to analyze its IR structure will follow the same logic as in our earlier work. We will inspect the low energy $k\to 0$ limit\footnote{In what follows, by the limit $k\to 0$ we mean asymptotic equivalence. That is, there is always an implicit IR scale $\Lambda$ 
below which the currents can be safely replaced by their IR equivalents $\tilde{J}_\mu^{IR}(k)$, where the photon momenta $k<\Lambda$. We will have an extensive discussion of this infrared scale later on in the manuscript.} of the net current given in Eq.~\eqref{eq:j_mu_r_ell_q_Minkowski}:
\ba 
\label{eq:lim_k_0_j_mu_r_ell_k}
\tilde{J}_\mu^{IR}(k)= \lim_{k\to 0} \tilde{J}_{\mu}^{(r,\ell)}(k) = \lim_{k\to 0} \sum_{i=r+1}^{r+\ell}\tilde{J}_{\mu,V}^i(k)+\lim_{k\to 0} \sum_{n=1}^r \tilde{J}_{\mu,R}^n(k) \,.
\ea

Virtual IR divergences will be introduced in the Dyson S-matrix element when any current in Eq.~\eqref{eq:lim_k_0_j_mu_r_ell_k} contains a $\sim 1/q$ contribution that survives in the $q\to 0$ limit, corresponding to the divergence of the loop integral for a virtual photon 4-momenta $q$. 
Real IR divergences appear when a worldline current in Eq.~\eqref{eq:lim_k_0_j_mu_r_ell_k} contains a $\sim 1/k$ factor. For each real photon present, there is a corresponding $\sim 1/k$ factor  multiplying the amplitude without real photons. When the modulus squared is taken of this amplitude, and integrated over the 
phase space of the emitted photons, the contribution of the $1/k$ terms results in the well-known IR divergence of soft photon amplitudes. As we emphasized in Paper I, it is therefore vital to better understand the nature of the $1/k$ terms in the worldine currents.

Firstly, in Eq.~\eqref{eq:j_mu_r_ell_q_Minkowski}, the fermion piece of the currents can be neglected in the $k\to 0$ limit; this is because they are subleading relative to the bosonic contribution. For the latter, it is instructive to sample any virtual or real bosonic worldline $x_\mu(\tau)$ in the path integral at points $x_\mu^\kappa = x_\mu(\tau_\kappa)$, $\kappa=1,\ldots,N$, and connect pairs of points with straight line propagation in infinitesimal proper time 
$\delta\tau_\kappa = \tau_{\kappa+1}-\tau_\kappa$, with their path described by 
\ba 
\label{eq:x_mu_tau_discretized}
x^\mu(\tau) = x_{\kappa}^\mu + \frac{\delta x_{\kappa}^\mu}{\delta \tau_\kappa} (\tau-\tau_\kappa)\,,\,\,\,\, \tau\in (\tau_{\kappa+1},\tau_\kappa)\,.
\ea 
Here $\delta x^\mu_\kappa\equiv x^\mu_{\kappa+1}-x^\mu_\kappa$, $\tau_1=0$ and $\tau_N=1$. Path integration in this piece-wise representation of the worldlines corresponds to the sum over all possible intermediate positions  $x_\kappa^\mu$. 

\begin{figure}
    \centering
    \includegraphics[scale=0.8]{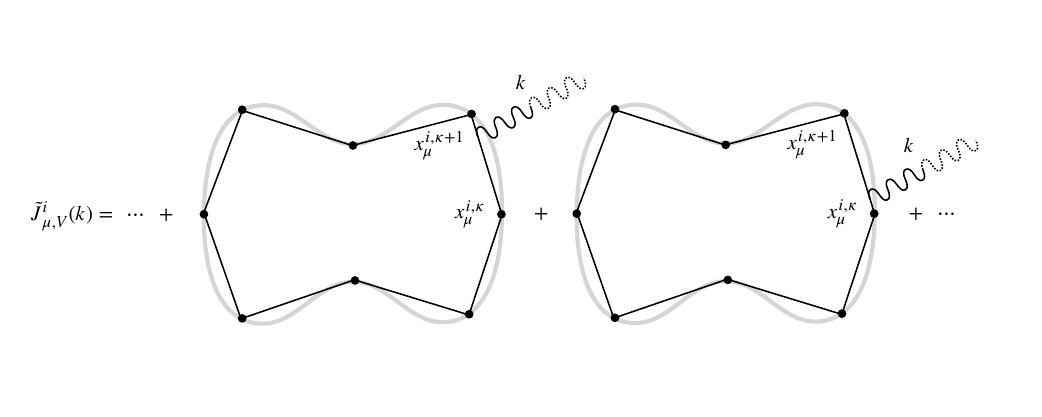}
    \caption{Diagrams generated by the $i$-th particle current $\tilde{J}^i_{\mu,V}(k)$ contribution to the Dyson S-matrix, including all possible ways of attaching a real (solid wavy line) or virtual photon (dashed wavy line) of 4-momentum $k$ to any point $\tau$ of the $i$-th particle trajectory. In the $k\to 0$ limit, the $\sim 1/k$ divergence of the soft photon graph pinched to $\kappa$ cancels the $1/k$ divergence of the graph with the soft photon pinched to $\kappa+1$, $\kappa=1,\ldots,N$. 
    }
    \label{fig:figure_2}
\end{figure}

For a virtual particle current, the discretization in Eq.~\eqref{eq:x_mu_tau_discretized} produces an incoherent sum of $2N-2$ real or virtual photon graphs, representing the $2N-2$ possible ways of attaching a photon line of 4-momentum $k$, real or virtual, to each of the two sides of the $N-1$ cusps in the virtual fermion worldline,
\ba
\label{eq:lim_k_0_j_i_mu_v}
\lim_{k\to 0} \tilde{J}^i_{\mu,V}(k)& = \lim_{k\to 0}\int^1_0 d\tau\, g\, \dot{x}^i_\mu(\tau)e^{ik\wc x^i(\tau)}= \lim_{k\to 0} \sum_{\kappa=1}^{N-1}\int^{\tau_{\kappa+1}}_{\tau_\kappa} d\tau\, g\, \dot{x}_\mu^i(\tau) e^{ik\wc x^i(\tau)}\nonumber\\
&=  \lim_{k\to 0} \frac{g}{i}\sum_{\kappa=1}^{N-1} \frac{\delta x_{\kappa,\mu}^{i}}{k\wc \delta x_{\kappa}^{i}}\Big(e^{ik\wc x_{\kappa+1}^i}-e^{ik\wc x_{\kappa}^i}\Big)\,.
\ea 
The soft photon diagrams appearing in Eq.~\eqref{eq:lim_k_0_j_i_mu_v} are illustrated in Fig.~\ref{fig:figure_2} for a given closed trajectory of the $i$-th virtual fermion contributing to the path integral. 

The phase structure of the sum of diagrams in Eq.~\eqref{eq:lim_k_0_j_i_mu_v} determines the IR finite behavior of the virtual particle currents within the Dyson S-matrix. For finite $x_\kappa^\mu$ and $x_{\kappa+1}^\mu$, the two $\sim 1/k$ IR divergent photon diagrams in the parenthesis in Eq.~\eqref{eq:lim_k_0_j_i_mu_v} coherently cancel when $k\to 0$ since their relative phases vanish. Therefore for all $x_\kappa^\mu$ finite ($\kappa=1,\ldots,N$), in other words, for virtual charges confined to a finite 4-volume, the sum in Eq.~\eqref{eq:lim_k_0_j_i_mu_v} contains no $\sim 1/k$ contributions. 

A possibility one must consider is that a virtual charge can explore a given point $x_\kappa^\mu$ located at space-like and/or  time-like infinity in the worldline path integral. The sum in Eq.~\eqref{eq:lim_k_0_j_i_mu_v} can then be organized to isolate $x_\kappa^\mu$ from finite spacetime points as follows:
\ba 
&\lim_{k\to 0}\tilde{J}^i_{\mu,V}(k) =\lim_{k\to 0} \bigg\{\frac{g}{i} \sum_{\eta=\kappa+1}^{N-1} \frac{\delta x_{\eta,\mu}^{i}}{k\wc \delta x_{\eta}^{i}}\Big(e^{ik\wc x_{\eta+1}^i}-e^{ik\wc x_{\eta}^i}\Big)+ \frac{g}{i} \frac{\delta x_{\kappa,\mu}^{i}}{k\wc \delta x_{\kappa}^{i}}\Big(e^{ik\wc x_{\kappa+1}^i}-\cancel{e^{ik\wc x_{\kappa}^i}}\Big)\nonumber\\
&+\frac{g}{i} \frac{\delta x_{\kappa-1,\mu}^{i}}{k\wc \delta x_{\kappa-1}^{i}}\Big(\cancel{e^{ik\wc x_{\kappa}^i}}-e^{ik\wc x_{\kappa-1}^i}\Big) + \frac{g}{i}\sum_{\eta=1}^{\kappa-1}\frac{\delta x_{\eta,\mu}^{i}}{k\wc \delta x_{\eta}^{i}}\Big(e^{ik\wc x_{\eta+1}^i}-e^{ik\wc x_{\eta}^i}\Big)\bigg\}\,.
\ea 
For any finite $k$, the phases with $x_\kappa^\mu$ are highly oscillatory and can therefore be dropped. Further noticing that $\delta x_{\kappa}^i/k\wc \delta x_{\kappa}^i= \delta x_{\kappa-1}^i/k\wc \delta x_{\kappa-1}^i$ as 
$x_\kappa^i\to\pm\infty$, the two $\sim 1/k$ divergent diagrams left 
unpaired, as shown above, can be paired together, with the sum then reorganized as
\ba 
&\lim_{k\to 0}\tilde{J}^i_{\mu,V}(k) =\lim_{k\to 0} \bigg\{\frac{g}{i} \sum_{\eta=\kappa+1}^{N-1} \frac{\delta x_{\eta,\mu}^{i}}{k\wc \delta x_{\eta}^{i}}\Big(e^{ik\wc x_{\eta+1}^i}-e^{ik\wc x_{\eta}^i}\Big)+ \frac{g}{i} \frac{\delta x_{\kappa,\mu}^{i}}{k\wc \delta x_{\kappa}^{i}}\Big(e^{ik\wc x_{\kappa+1}^i}-e^{ik\wc x_{\kappa-1}^i}\Big)\nonumber\\
&+ \frac{g}{i}\sum_{\eta=1}^{\kappa-1}\frac{\delta x_{\eta,\mu}^{i}}{k\wc \delta x_{\eta}^{i}}\Big(e^{ik\wc x_{\eta+1}^i}-e^{ik\wc x_{\eta}^i}\Big)\bigg\}\,.
\ea 
%
Since each of the parentheses contain only finite spacetime points by construction, the relative phases within each, as previously, cancel in 
the limit $k\to 0$. 
Hence the $\ell$ virtual fermions present in the $\ell$-th loop contribution to the Dyson S-matrix can never introduce $\sim 1/k$ IR divergences of photons, virtual or real, since
\ba
\label{eq:lim_k_0_j_i_mu_v_result}
\lim_{k\to 0}\tilde{J}^i_{\mu,V}(k) = \text{const.}\,\,\,\,\,\,\, i=r+1,\ldots,r+\ell
\ea 

\begin{figure}
    \centering
    \includegraphics[scale=0.8]{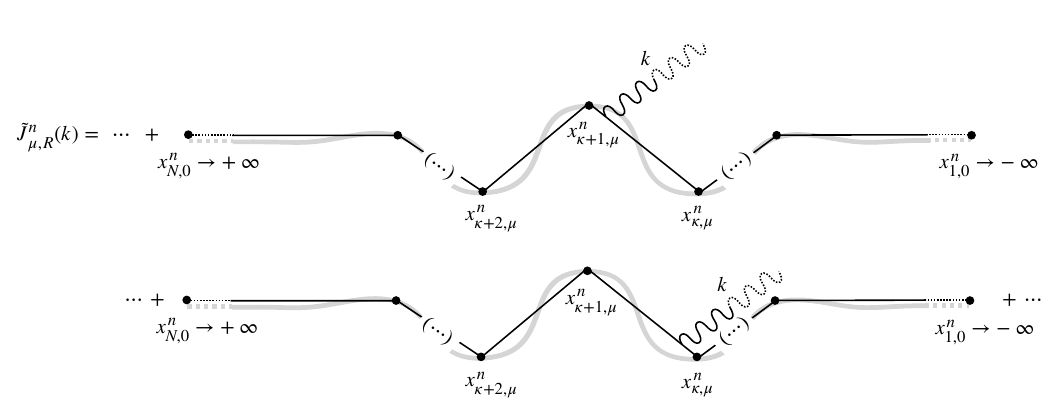}
    \caption{Diagrams generated by the $n$-th real charged fermion current $\tilde{J}^n_{\mu,R}(k)$ contribution to the Dyson S-matrix, including all possible ways of attaching a photon line of 4-momentum $k$  to any point $\tau$ of the $n$-th real particle trajectory. In the $k\to 0$ limit the $\sim 1/k$ divergences of the soft photon graphs from the internal lines cancel with each other, but the two soft photon graphs of the external lines survive, leaving two terms $\sim 1/k$  when $k\to 0$ in $\tilde{J}^n_{\mu,R}(k)$. The dots at the ends of the photon lines indicate that the photon can be either real or virtual.}
    \label{fig:figure_3}
\end{figure}

For  real fermion currents (labeled $i=1,\cdots,r$), the situation is quite different. These have open worldlines with endpoints in Eq.~\eqref{eq:open_boundary_conditions_infinity} at timelike infinity. The discretization in Eq.~\eqref{eq:x_mu_tau_discretized} then reads
\ba 
\label{eq:lim_k_0_j_n_mu_r}
&\lim_{k\to 0} \tilde{J}^n_{\mu,R}(k) = \lim_{k\to 0} g\int^{t_f^n}_{t_i^n} dt \,\dot{x}^n_\mu(t) e^{ik\wc x^n(t)-\varepsilon|t|}=\lim_{k\to 0}\sum_{\kappa=1}^{N-1}g\int^{t_{\kappa+1}^n}_{t_\kappa^n} dt\, \dot{x}^n_\mu(t) e^{ik\wc x^n(t)-\varepsilon|t|}\\
&=\lim_{k\to 0}
\frac{g}{i}\sum_{\kappa=1}^{N-1}\bigg\{\frac{\delta x_{\kappa,\mu}^n}{k\wc \delta x_\kappa^n+i\text{sign}(t_{\kappa+1}^n)\varepsilon}e^{ik\wc x_{\kappa+1}^n-\varepsilon|t_{\kappa+1}^n|}-\frac{\delta x_{\kappa,\mu}^n}{k\wc \delta x_\kappa^n+i\text{sign}(t_{\kappa}^n)\varepsilon}e^{ik\wc x_\kappa^n-\varepsilon|t_\kappa^n|} \bigg\}\nonumber\,,
\ea 
where we introduced an $\varepsilon$-regularization and performed the integral in the physical time of the charged currents. The soft photon diagrams introduced by the $n$-th real fermion current in the Dyson S-matrix in Eq.~\eqref{eq:lim_k_0_j_n_mu_r} are shown in Fig.~\ref{fig:figure_3} for a given 4-trajectory of the $n$-th virtual fermion contributing to the path integral. 

Taking the $x_{N,0}^n\equiv t_f^n\to \infty$ and $x_{1,0}^n\equiv t_i^n\to - \infty$ limits in Eq.~\eqref{eq:lim_k_0_j_n_mu_r}, the two soft photon graphs pinched to plus and minus time-like infinity drop out due to the wildly oscillatory phases, giving 
\ba 
\lim_{k\to 0}\tilde{J}^n_{\mu,R}(k) = \lim_{k\to 0}\bigg\{-\frac{g}{i} \frac{\delta x_{N-1,\mu}^n}{k\wc \delta x_{N-1}^n+i\varepsilon }e^{ik\wc x_{N-1}^n-\varepsilon |t^n_{N-1}|}+\frac{g}{i} \frac{\delta x_{1,\mu}^n}{k\wc \delta x_{1}^n-i\varepsilon }e^{ik\wc x_{2}^n-\varepsilon |t^n_{2}|}\nonumber\\
+\frac{g}{i}\sum_{\kappa=2}^{N-2}\bigg[\frac{\delta x_{\kappa,\mu}^n}{k\wc \delta x_\kappa^n+i\text{sign}(t_{\kappa+1}^n)\varepsilon}e^{ik\wc x_{\kappa+1}^n-\varepsilon|t_{\kappa+1}^n|}-\frac{\delta x_{\kappa,\mu}^n}{k\wc \delta x_\kappa^n+i\text{sign}(t_{\kappa}^n)\varepsilon}e^{ik\wc x_\kappa^n-\varepsilon|t_\kappa^n|} \bigg]\bigg\}\,,
\ea 
As shown above, this leaves the two graphs with the real or virtual soft photon attached to the two \textit{in} and \textit{out} external legs of the $n$-th real fermion unpaired, each introducing a different $\sim 1/k$ contribution. Taking then the $k\to 0$ limit, the $\sim 1/k$ divergences of the photon graphs attached to the internal lines cancel, resulting in the expression
\ba 
\label{eq:lim_k_0_j_n_mu_r_result}
&\lim_{k\to 0} \tilde{J}^n_{\mu,R}(k) =-\frac{g}{i}\lim_{k\to 0}\bigg\{\frac{\delta x^n_{N-1,\mu}}{k\wc \delta x^n_{N-1}+i\varepsilon}-\frac{\delta x^n_{1,\mu}}{k\wc \delta x^n_{1}-i\varepsilon}\bigg\} = -\lim_{k\to 0} \frac{g}{i} \bigg\{\frac{p^n_{f,\mu}}{k\wc p^n_f+i\varepsilon} -\frac{p_{i,\mu}^n}{k\wc p^n_i-i\varepsilon}\bigg\}\,.
\ea 
In the second equality we multiplied and divided numerators and denominators by the $n$-th real fermion mass $m$ and proper time $s$ respectively, which gives
\ba 
\lim_{t^n_{N}\to \infty} m \frac{\delta x_{N-1}^n}{\delta  s^n_{N-1}} = p^n_f\,, \,\,\,\,\, \lim_{t^n_{N}\to \infty} m \frac{k\wc \delta x^n_{N-1}}{\delta s^n_{N-1}} = k\wc p^n_{f}\,, 
\ea 
and
\ba 
\lim_{t^n_{1}\to -\infty} m \frac{\delta x_{1}^n}{\delta  s^n_{1}} = p^n_i\,, \,\,\,\,\, \lim_{t^n_{1}\to -\infty} m \frac{k\wc \delta x^n_{1}}{\delta s^n_{1}} = k\wc p^n_{i}\,.
\ea 
\begin{figure}
    \centering
    \includegraphics[scale=0.8]{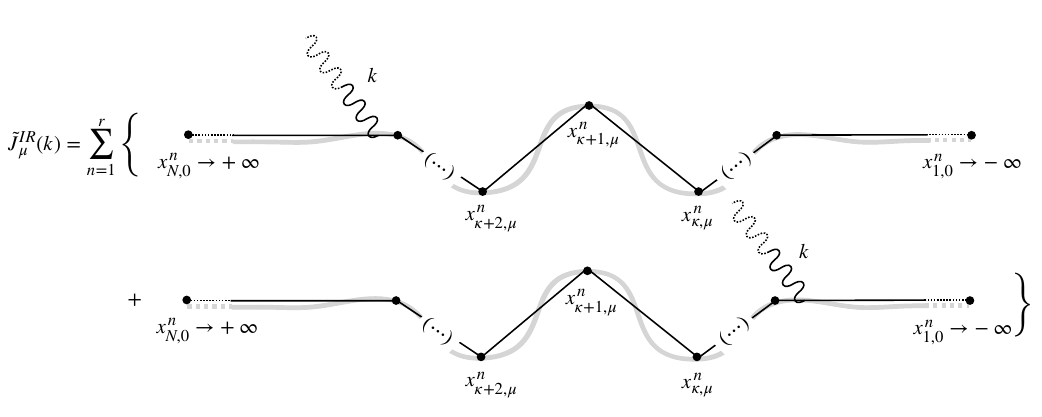}
    \caption{Graphs with soft photon lines of 4-momentum $k$ attached to the two external legs of each $n$ of the $r$ external charges that survive in the $k\to 0$ limit $\tilde{J}_\mu^{IR}(k)$ of the total current  $\tilde{J}_\mu^{(r,\ell)}(k)$ from $r$ real and $\ell$ virtual charges within the $\ell$-th term of the Dyson S-matrix. }
    \label{fig:figure_4}
\end{figure}

In summary,  using Eqs.~\eqref{eq:lim_k_0_j_i_mu_v_result} and \eqref{eq:lim_k_0_j_n_mu_r_result}, the soft photon limit of the total worldline current of the many-body system comprising the $r$ real and $\ell$ virtual particles in the $\ell$-th loop contribution to the Dyson S-matrix element in Eq.~\eqref{eq:lim_k_0_j_mu_r_ell_k} is given by the bosonic pieces  of the currents created by the external legs of the $r$ real fermions,
\ba 
\label{eq:j_mu_ir_definition}
\lim_{k\to 0} \tilde{J}_\mu^{(r,\ell)}(k) = \lim_{k\to 0} \tilde{J}_\mu^{IR}(k)\,,\,\,\,\,\, \text{with}\,\,\,\,\,
\tilde{J}_\mu^{IR}(k)= - \frac{g}{i}\sum_{n'=1}^{N_i^c+N_o^c} \frac{\eta^{n'} p^{n'}_\mu}{k\wc p^{n'}+i\eta^{n'}\varepsilon} \,,
\ea 
where the sum in $n'$ is taken over all \textit{in} and \textit{out} legs of the real worldlines, with initial or final 4-momenta $p_{n'}$, and $\eta'=+1$ or $\eta'=-1$ if the charge is outgoing or incoming, respectively. Eq.~\eqref{eq:j_mu_ir_definition} then states that real or virtual soft photons attached to internal legs  - namely, virtual charged fermion worldlines or the internal parts of real worldlines - do not introduce real or virtual IR divergences in the Dyson S-matrix. These arise solely from 
the (asymptotic) external legs of the real charges. The graphs accounted for in the IR limit of the total current of the system of $r+\ell$ charges are  illustrated  
in Fig.~\ref{fig:figure_4}.

To factor out the this IR structure of the soft photon currents in the Dyson S-matrix, we follow the strategy articulated in Paper I and separate out soft and hard, real and virtual photons in the general expression in  Eq.~\eqref{eq:s_fi_rell_nfgamma_nigamma}. The term with the virtual photon interactions in the normalized worldline expectation value in Eq.~\eqref{eq:s_fi_rell_nfgamma_nigamma} can be rewritten as
\ba 
\label{eq:soft_hard_factorization_virtual_interactions}
&\frac{i}{2}\int_0^\infty\frac{d^4q}{(2\pi)^4}\frac{1}{q^2+i\epsilon}\tilde{J}_\mu^{(r,\ell)}(-q) \tilde{J}^\mu_{(r,\ell)}(+q)\nonumber\\
&= \frac{i}{2}\int_\lambda^\Lambda \frac{d^4q}{(2\pi)^4}\frac{1}{q^2+i\epsilon}\tilde{J}_\mu^{(r,\ell)}(-q) \tilde{J}^\mu_{(r,\ell)}(+q)+\frac{i}{2}\int_\Lambda^\infty\frac{d^4q}{(2\pi)^4}\frac{1}{q^2+i\epsilon}\tilde{J}_\mu^{(r,\ell)}(-q) \tilde{J}^\mu_{(r,\ell)}(+q)
\ea 
where the parameter $\lambda\to 0$ is the IR cut-off and the scale $\Lambda$ delimits the upper limit of the IR region of low momentum virtual photons for whom the exact form of the worldline currents $\tilde{J}^{(r,\ell)}_\mu$ can be replaced by their $q\to 0$ limits in Eq.~\eqref{eq:j_mu_ir_definition}, giving
\ba
\label{eq:soft_virtual_interactions}
 &\frac{i}{2}\int_\lambda^\Lambda \frac{d^4q}{(2\pi)^4}\frac{1}{q^2+i\epsilon}\tilde{J}_\mu^{(r,\ell)}(-q) \tilde{J}^\mu_{(r,\ell)}(+q)\simeq 
\frac{i}{2}\int_\lambda^\Lambda\frac{d^4q}{(2\pi)^4}\frac{1}{q^2+i\epsilon}\tilde{J}_\mu^{IR}(-q) \tilde{J}^\mu_{IR}(+q)
\nonumber\\
&=\frac{ig^2}{2} \sum_{n',m'=1}^{N_i^c+N_o^c} \int_\lambda^\Lambda\frac{d^4q}{(2\pi)^4}\frac{1}{q^2+i\epsilon} \frac{\eta^{m'}p^{m'}}{q\wc p^{m'}-i\eta^{m'}\epsilon}\wc \frac{\eta^{n'}p^{n'}}{q\wc p^{n'}+i\eta^{n'}\epsilon} \,.
\ea 
This term is IR divergent when $\lambda\to 0$ and encodes very early and late soft interactions amongst all the real charges. Since it depends only on their asymptotic 4-momenta at infinity, the term can be factored out of the normalized worldline expectation value in Eq.~\eqref{eq:s_fi_rell_nfgamma_nigamma} because it is unaffected by the worldline path integrals. The second term in Eq.~\eqref{eq:soft_hard_factorization_virtual_interactions} is IR finite by construction, and should be retained as part of the normalized worldline expectation value encoding the hard virtual exchanges amongst all real and/or virtual worldlines.

The real photon attachments to the real and virtual charged currents that multiply the exponential factor in the normalized expectation value in Eq.~\eqref{eq:s_fi_rell_nfgamma_nigamma} can likewise be separated into soft and hard contributions. We will now discuss how this is achieved in practice.

Let $N_{o,s}^\gamma$ and $N_{i,s}^\gamma$ denote the number of \textit{in} and \textit{out} soft real photons of energy less than $\Lambda$, $N_{o,h}^\gamma$ and $N_{i,h}^\gamma$ the number of hard photons of energy greater than $\Lambda$, for a given total number of hard and soft outgoing and incoming photons, $N_o^\gamma=N_{o,s}^\gamma+N_{o,h}^\gamma$ and $N_i^\gamma=N_{i,s}^\gamma+N_{i,h}^\gamma$, respectively. For each of the  $N_{o,s}^\gamma+N_{i,s}^\gamma$ soft real photons, the corresponding worldline current $\tilde{J}_\mu^{(r,\ell)}$ can be replaced by  Eq.~\eqref{eq:j_mu_ir_definition},  and subsequently factored out of the normalized worldline expectation value in Eq.~\eqref{eq:s_fi_rell_nfgamma_nigamma}. In contrast, the $N_{o,h}^\gamma+N_{i,h}^\gamma$ hard real photons are kept as part of the normalized worldline expectation value, with the exact form for  $\tilde{J}_\mu^{(r,\ell)}(k)$; they have to be evaluated as part of the path integral. 

In light of Eqs.~\eqref{eq:lim_k_0_j_mu_r_ell_k}, \eqref{eq:soft_hard_factorization_virtual_interactions}, and \eqref{eq:soft_virtual_interactions}, the $i\to f$ amplitude in Eq.~\eqref{eq:s_fi_rell_nfgamma_nigamma} can then be expressed as the product
\ba 
\label{eq:s_fi_r_Nos_Nis}
\mathcal{S}_{fi}^{(r)}(N_{o}^\gamma,N_{i}^\gamma) = \mathcal{S}_{fi,s}^{(r)} (N_{o,s}^\gamma,N_{i,s}^\gamma)\,\,\times\,\, \mathcal{S}_{fi,h}^{(r)}(N_{o,h}^\gamma,N_{i,h}^\gamma)\,,
\ea 
where the interactions of the charged particles with soft real and virtual photons is absorbed into the IR Dyson S-matrix 
\ba 
\label{eq:s_fi_r_s_Nos_Nis}
\mathcal{S}_{fi,s}^{(r)}(N_{o,s}^\gamma,N_{i,s}^\gamma) &= \prod_{p=1}^{N^\gamma_{o,s}} \Bigg\{\frac{i\epsilon^*_{\mu_f^p}(\v{k}_f^p,\lambda_f^p)}{\sqrt{2\omega_{\v{k}_f^p}}(2\pi)^{3/2}}\tilde{J}^{\mu_f^p}_{IR}(-k_f^p) \Bigg\}
\prod_{p=1}^{N^\gamma_{i,s}}\Bigg\{ \frac{i\epsilon_{\mu_i^p}(\v{k}_i^p,\lambda_i^p)}{\sqrt{2\omega_{\v{k}_i^p}}(2\pi)^{3/2}} \tilde{J}_{IR}^{\mu_i^p}(+k_i^p)\Bigg\}\\
&\times\exp\bigg\{\frac{i}{2}\int_\lambda^\Lambda\frac{d^4q}{(2\pi)^4}\frac{1}{q^2+i\epsilon}\tilde{J}_\mu^{IR}(-q) \tilde{J}^\mu_{IR}(+q)\bigg\}\,.\nonumber
\ea 
In obtaining Eq.~\eqref{eq:s_fi_r_Nos_Nis}, we used the fact that the low energy limit of the currents in Eq.~\eqref{eq:j_mu_ir_definition} does not depend on the $\ell$ loop fermions; hence the loop expansion in $\ell$ affects only the hard part of the Dyson S-matrix, 
\ba 
\label{eq:hardS-matrix}
\mathcal{S}_{fi,h}^{(r)}(N_{o,h}^\gamma,N_{i,h}^\gamma)= \sum_{\ell=0}^\infty \mathcal{S}_{fi,h}^{(r,\ell)}(N_{o,h}^\gamma,N_{i,h}^\gamma)\,,
\ea 
with the $\ell$-th loop contribution 
\ba 
\label{eq:hardS-matrix_lth-loop}
&\mathcal{S}_{fi,h}^{(r)}(N_{o,h}^\gamma,N_{i,h}^\gamma) = \frac{(-1)^\ell}{\ell!} \frac{\mathrm{Z}_\text{MW}}{\mathrm{Z}[0,0,0]}\prod_{p=1}^{N_{o,h}^\gamma} \Bigg\{\frac{i\epsilon_{\mu_f^p}^*(\v{k}_f^p,\lambda_f^p)}{\sqrt{2\omega_{\v{k}_f^p}}(2\pi)^{3/2}} \Bigg\}\prod_{p=1}^{N_{i,h}^\gamma}\Bigg\{ \frac{i\epsilon_{\mu_i^p}(\v{k}_i^p,\lambda_i^p)}{\sqrt{2\omega_{\v{k}_i^p}}(2\pi)^{3/2}}\Bigg\} \\
&\times \prod_{n=1}^r \Bigg\{\lim_{\substack{t_f^n\to+\infty\\t_i^n\to-\infty}}\int d^3\v{x}_i^n \int d^3\v{x}_f^n \Psi^{(+),\dag}_{f_n}(x_f^n) \exp\bigg\{\bar{\gamma}_\lambda\frac{\partial}{\partial \theta_n}\bigg\}\bar{\gamma}_0\Psi_{i_n}^{(+)}(x_i^n)\Bigg\}\bigg\langle \prod_{p=1}^{N_{o,h}^\gamma} \bigg\{\tilde{J}_{(r,\ell)}^{\mu_f^p}(-k_f^p)\bigg\} \nonumber\\
&\times\prod_{p=1}^{N_{i,h}^\gamma} \bigg\{ \tilde{J}_{(r,\ell)}^{\mu_i^p}(+k_i^p)\bigg\}\exp\bigg\{\frac{i}{2}\int_\Lambda^\infty \frac{d^4q}{(2\pi)^4}\frac{1}{q^2+i\epsilon}\tilde{J}_{\mu}^{(r,\ell)}(-q)\tilde{J}_{(r,\ell)}^{\mu}(+q)\bigg\}\bigg\rangle\,+\text{perm.}\nonumber
\ea 

Eq.~\eqref{eq:s_fi_r_Nos_Nis} is a statement of the Abelian factorization of the Dyson S-matrix, and can be understood as the generalization by Weinberg~\cite{PhysRev.140.B516} of the well-known Low theorem in QED~\cite{Low:1958sn}. It demonstrates to all-loop orders in perturbation theory the factorization of real and virtual IR divergences due to soft photons from an all-loop IR finite Dyson S-matrix element containing arbitrary numbers of real and virtual hard photons. In the worldline formalism, it manifestly represents the factorization of very early and late time interactions of real spinning charges amongst themselves, and of the real photons radiated off or absorbed by these charges at very early or late times. These IR divergences are encoded in universal soft factors dressing the infrared finite hard Dyson S-matrix element. We will turn now to a discussion of the cancellation of these real and virtual IR divergences in Eq.~\eqref{eq:s_fi_r_s_Nos_Nis} in QED cross-sections.

\subsection{\label{sec:bloch_nordsieck_cancellation} Weinberg's proof of the cancellation of real and virtual IR divergences}

In the conventional understanding of the IR problem,
the amplitude of the radiative $i\to f$ transition in Eq.~\eqref{eq:s_fi_r_Nos_Nis} for the Dyson S-matrix element is zero in QED
on account of the IR singularities in the exponentiated virtual IR divergences. In this Bloch-Nordsieck picture~\cite{PhysRev.52.54}, real and virtual IR divergences only cancel amongst themselves in the total cross-section, when one considers within the transition the emission of 
an arbitrary number of low energy photons escaping undetected due to a finite detector resolution. 

To observe this cancellation between real and virtual IR divergences in our worldline framework, we will show in detail how one recovers the 
classic result of Weinberg~\cite{PhysRev.140.B516} for the photon emission rate $\Gamma^{(r)}_{fi}$ for the emission of hard photons accompanied by an arbitrary number of real soft photons whose energies add up to a total emitted energy $\omega_T<\Lambda$, in the $i\to f$ scattering the $r$ charged particles. 

From Eq.~\eqref{eq:s_fi_r_Nos_Nis}, it is clear that this rate can be factorized as 
\ba
\label{eq:gamma_r_fi_s_times_gamma_r_fi_h}
\Gamma^{(r)}_{fi} = \Gamma^{(r)}_{fi,s}\,\,\times \,\,  \Gamma^{(r)}_{fi,h}
\,.
\ea
The differential soft emission rate is given by the infinite sum over final states\footnote{
As noted by Kinoshita~\cite{Kinoshita:1962ur}, and by Lee and Nauenberg~\cite{PhysRev.133.B1549} (KLN), in general the problem of infrared singularities requires a summation over both initial and final final states that are degenerate in energy. As noted in \cite{Akhoury:1997pb}, from the KLN perspective, the Bloch-Nordsieck summation over final states discussed in this section is only valid because the emission and absorption of soft photons cannot be distinguished in the limit of vanishing photon energies. It is further observed in \cite{Akhoury:1997pb} that the link between summations of initial and final states is provided by Low's theorem. In our worldline approach, as will be discussed further in section~\ref{sec:faddeev_kulish}, both initial and final state emission and absorption are treated on the same footing in 
deriving the Faddeev-Kulish S-matrix.
}
\ba 
\frac{d\Gamma^{(r)}_{fi,s}}{d\omega_T} =& \delta(\omega_T)\Big|\mathcal{S}^{(r)}_{fi,s}(0,0)\Big|^2+\frac{1}{1!}\sum_{\lambda_f^1}\int_\lambda^\Lambda d^3\v{k}_f^1
\delta\big(\omega_T-\omega_{\v{k}_f^1}\big) \Big|\mathcal{S}^{(r)}_{fi,s}(1,0)\Big|^2\nonumber\\
&+\frac{1}{2!}\sum_{\lambda_f^1}\int_\lambda^\Lambda d^3\v{k}_f^1 \sum_{\lambda_f^2} \int_\lambda^\Lambda d^3\v{k}_f^2 \delta\big(\omega_T-\omega_{\v{k}_f^1}-\omega_{\v{k}_f^2}\big)\Big|\mathcal{S}_{fi,s}^{(r)}(2,0)\Big|^2+\cdots\,,
\label{eq:soft-differential-rate}
\ea
with $d^3\v{k}=\omega_{\v{k}}^2d\omega_{\v{k}}d\Omega_{\v{k}}$, where  $\omega_{\v{k}}$ and $\Omega_{\v{k}}$ are a real photon's energy and solid angle respectively, and the $1/N_{o,f}^\gamma!$ factor in each term accounts for the Bose statistics of photons. 
The differential rate for the hard $i\to f$ transition only includes hard  photons (with momenta $>\Lambda$), and is given by 
\ba 
\label{eq:d_Noh_Noc_Gamma}
d\Gamma^{(r)}_{fi,h} \equiv \Big|\mathcal{S}_{fi,h}^{(r)}(N_{o,h}^\gamma,N_{i,h}^\gamma) \Big|^2 \bigg\{\prod_{p=N_{o,s}^\gamma+1}^{N_{o,s}^\gamma+N_{o,h}^\gamma} d^3\v{k}_f^p \bigg\}\bigg\{\prod_{n=1}^{N_o^c} d^3\v{p}_f^n\bigg\}\,,
\ea 
where the hard $S$-matrix element on the r.h.s is given by Eqs.~\eqref{eq:hardS-matrix} and \eqref{eq:hardS-matrix_lth-loop}. We will not discuss the hard emission rate any further except to note that it can be computed order-by-order in perturbation theory employing the techniques outlined in Paper I and in the previous section. Our focus in the rest of this section will be on the soft differential emission rate, in particular the explicit cancellation of real and virtual infrared divergences. 

Using the well-known integral representation of the Dirac delta function, the sum in Eq.~\eqref{eq:soft-differential-rate} can be compactly and conveniently rewritten as
\ba 
\label{eq:dGamma_fi_r_s_domegaT_def2}
\frac{d\Gamma_{fi,s}^{(r)}}{d\omega_T} = \frac{1}{2\pi}\int^{+\infty}_{-\infty}d\sigma e^{-i\sigma\omega_T} \sum_{N_{o,s}^\gamma=0}^\infty \frac{1}{N_{o,s}^\gamma!} \prod_{p=1}^{N_{o,s}^\gamma} \Bigg\{\sum_{\lambda_f^p} \int_\lambda^\Lambda d^3\v{k}_f^p e^{+i\sigma\omega_{\v{k}_f^p}}\Bigg\} \Big|S_{fi,s}^{(r)}(N_{o,s}^\gamma,0)\Big|^2\,.
\ea 
Substituting Eq.~\eqref{eq:s_fi_r_s_Nos_Nis} into the r.h.s above, and summing in $N_{o,s}^\gamma$, Eq.~\eqref{eq:dGamma_fi_r_s_domegaT_def2} can be expressed as 
\ba 
\frac{d\Gamma^{(r)}_{fi,s}}{d\omega_T} &= \frac{1}{2\pi} \exp\Bigg\{2\Re\int^\Lambda_\lambda \frac{d^4q}{(2\pi)^4}A_V^{IR}(q)\Bigg\}\nonumber\\
&\times\int^{+\infty}_{-\infty}d\sigma e^{-i\sigma\omega_T} \exp\Bigg\{ \int_\lambda^\Lambda \frac{d^3\v{k}}{(2\pi)^3} \frac{e^{+i\omega_{\v{k}}\sigma}}{2\omega_{\v{k}}}  A_R^{IR}(\v{k})\Bigg\} \,.
\label{eq:Differential-rate}
\ea 
Here the contribution to the absolute square of the amplitude  from the exponentiation of virtual IR photon exchanges in Eq.~\eqref{eq:s_fi_r_s_Nos_Nis} is expressed in terms of the function
\ba 
\label{eq:a_v_ir_q_def}
A_V^{IR}(q) \equiv  \frac{1}{2}\frac{ig^{\mu\nu}}{q^2+i\varepsilon} \tilde{J}_\mu^{IR}(-q)\tilde{J}^{IR}_\nu(+q)= \frac{g^2}{2} \sum_{n',m'=1}^{N_i^c+N_o^c} \frac{i}{q^2+i\varepsilon} \frac{\eta^{m'}p^{m'}}{q\wc p^{m'}-i\eta^{m'}\varepsilon}\wc \frac{\eta^{n'}p^{n'}}{q\wc p^{n'}+i\eta^{n'}\varepsilon}\,.
\ea 
The contribution coming from the real IR photons in Eq.~\eqref{eq:s_fi_r_s_Nos_Nis} can also be exponentiated in the cross-section as\footnote{
In obtaining $A_R^{IR}(\v{k})$, we used the completeness relation
\ba 
\label{eq:sum_lambda}
\sum_{\lambda=\pm 1} \epsilon_\mu(\v{k},\lambda)\epsilon^*_\nu(\v{k},\lambda) = -g_{\mu\nu} -\frac{k_\mu k_\nu}{(n\wc k)^2}+\frac{k_\mu n_\nu+k_\nu n_\mu}{(n\wc k)}\,,
\ea 
where $n$ is the normalized time-like 4-vector; we used further the fact that the dependence on $n$ drops out of the amplitude. This is because the $k_\mu$ and $k_\nu$ terms, once contracted with the currents vanish, giving $k\wc J^{(r,l)}=0$, as we discussed previously.
} 
\ba 
\label{eq:a_r_ir_k_def}
A_R^{IR}(\v{k}) \equiv g^{\mu\nu} \Big(\tilde{J}^{IR}_\mu(-k)\Big)^*\tilde{J}^{IR}_\nu(-k) = -g^2\sum_{n',m'=1}^{N_o^c+N_i^c}\frac{\eta^{n'}p^{n'}}{k\wc p^{n'}+i\eta^{n'}\varepsilon}\wc \frac{  \eta^{m'} p^{m'}}{k\wc p^{m'}-i\eta^{m'}\varepsilon}\,,
\ea 
with $k_0=\omega_{\v{k}}$. 

We can now use Eq.~\eqref{eq:Differential-rate} to extract the emission rate 
for any number of real soft photons adding up to a total emitted energy less than $E$, with $E<\Lambda$. To achieve this,  we integrate the differential rate above between $\omega_T=0$ and $\omega_T=+E$; we further replace the upper limit  of soft photon energies $\Lambda$ 
by $E$. Since from Eq.~\eqref{eq:soft-differential-rate} we see that the rate  vanishes for $\omega_T<0$, in order to symmetrize the expression, 
we can formally integrate it between $\omega_T=-E$ and $\omega_T=+E$, to obtain
\ba 
\Gamma_{fi,s}^{(r)}=&
\int^{+E}_{-E} d\omega_T \frac{d\Gamma_{fi,s}^{(r)}}{d\omega_T}=\frac{1}{\pi} \exp\bigg\{2\Re \int^\Lambda_\lambda \frac{d^4q}{(2\pi)^4} A_V^{IR}(q)\bigg\}  \nonumber\\
&\times \int^{+\infty}_{-\infty} d\sigma \frac{\sin \sigma E}{\sigma } \exp\Bigg\{ \int_\lambda^E \frac{d^3\v{k}}{(2\pi)^3} \frac{1}{2\omega_{\v{k}}}  e^{+i\omega_{\v{k}}\sigma}A_R^{IR}(\v{k})
\Bigg\} \,.
\ea 

As discussed at length in section 5.2 of \cite{Feal:2022iyn}, the real part of the integral over $A_V^{IR}(q)$ comes wholly from the on-shell modes of the virtual IR photon exchanges, $q^0=\pm\omega_{\v{q}}$. It contains the interaction of the \textit{in} and \textit{out} charges at asymptotic times with their own radiative Lienard-Wiechert or ``acceleration" fields, created due to the overall shift in momenta in the scattering. The imaginary part comes wholly instead from the off-shell Coulomb modes, and encodes the interaction of the same charges at asymptotic times with the Lienard-Wiechert ``velocity" fields created when the charges come in from minus infinity and go out to plus infinity with constant momenta. As our result indicates, the  imaginary part does not contribute to the cross section. 

We can reexpress our expression for the virtual part as 
\ba 
\label{eq:2_re_int_a_v}
2\text{Re}\int^\Lambda_\lambda \frac{d^4q}{(2\pi)^4}A_V^{IR}(q)= g^2\sum_{n',m'=1}^{N_o^c+N_i^c} \eta^{m'}\eta^{n'} \int_\lambda^\Lambda  \frac{d^3\v{q}}{(2\pi)^3} \frac{1}{2\omega_{\v{q}}}
\frac{p^{m'}}{q\wc p^{m'}}\wc \frac{p^{n'}}{q\wc p^{n'}} = -\int_\lambda^\Lambda \frac{d^3\v{q}}{(2\pi)^3}\frac{1}{2\omega_{\v{q}}} A_R^{IR}(\v{q})\,.
\ea 
%
We observe thus that the radiative virtual gauge fields created by the charges and exponentiated in Eq.~\eqref{eq:2_re_int_a_v} are of precisely the same form  - but opposite in sign - as the real photon fields radiated in the scattering, and exponentiated later in the cross-section in $A_R^{IR}(\v{k})$. Hence
\ba 
\Gamma^{(r)}_{fi,s}= \frac{1}{\pi}\int^{+\infty}_{-\infty} d\sigma \frac{\sin \sigma E}{\sigma } \exp\Bigg\{ \int_\lambda^E \frac{d^3\v{k}}{(2\pi)^3} \frac{1}{2\omega_{\v{k}}}  e^{+i\omega_{\v{k}}\sigma}A_R^{IR}(\v{k})-\int^\Lambda_\lambda\frac{d^3\v{q}}{(2\pi)^3} \frac{1}{2\omega_{\v{q}}}A_R ^{IR}(\v{q}) \Bigg\} \,,
\ea 
which is IR finite as $\omega_{\v{k}}\to 0$.
The expression above can be rewritten as 
\ba 
\label{eq:Gamma_fi_r_s_result1}
\Gamma^{(r)}_{fi,s}&= \exp\bigg\{-\int^\Lambda_E \frac{d^3\v{k}}{(2\pi)^3}\frac{1}{2\omega_{\v{k}}}A_R^{IR}(\v{k})\bigg\}\nonumber\\
&\times \frac{1}{\pi}  \int^{+\infty}_{-\infty} d\sigma \frac{\sin \sigma E}{\sigma } \exp\Bigg\{ \int_\lambda^E \frac{d^3\v{k}}{(2\pi)^3} \frac{1}{2\omega_{\v{k}}}  \big(e^{+i\omega_{\v{k}}\sigma}-1\big)A_R^{IR}(\v{k}) \Bigg\} \,.
\ea 
It is clear that this expression is independent of the IR cut-off $\lambda$. The integral over the  soft photon angle can be performed and yields
\ba 
\label{eq:gamma_cusp_r_def}
\int \frac{d^3\v{k}}{(2\pi)^3} \frac{1}{2\omega_{\v{k}}} A_R^{IR}(\v{k})= \Gamma_{cusp}^{(r)} \int \frac{d\omega_{\v{k}}}{\omega_{\v{k}}}\,,\,\,\,\,\,\,\,\,\, \Gamma_{cusp}^{(r)}=-\frac{g^2}{4\pi^2} \sum_{n',m'=1}^{N_o^c+N_i^c} \eta_{n'}\eta_{m'}\gamma_{n'm'} \coth \gamma_{n'm'}\,.
\ea 
The many-body cusp anomalous dimension $\Gamma^{(r)}_{cusp}$, encodes the IR behavior of the scattering of the $r$ charges, in terms of the asymptotic 4-angles $\gamma^{n'm'}$ between \textit{in} and \textit{out} charges defined as 
\ba
\label{eq:cosh_gamma_nm}
\cosh \gamma_{n'm'} = \frac{p_{n'}\wc p_{m'}}{\sqrt{p_{n'}^2p_{m'}^2}}\,.
\ea 
With this, and reintroducing in Eq.~\eqref{eq:Gamma_fi_r_s_result1} the differential rate of the hard scattering Eq.\eqref{eq:d_Noh_Noc_Gamma}, the total differential rate of the $i\to f$ hard transition (including the undetected emission of low energy photons adding up to a total emitted energy less than $E$) acquires the simple compact form obtained by Weinberg: 
\ba 
\label{eq:Gamma_r_fi_s_result}
d\Gamma^{(r)}_{fi} =d\Gamma^{(r)}_{fi,h}\exp\Bigg\{-\Gamma_{cusp}^{(r)}\int^\Lambda_E\frac{d\omega_{\v{k}}}{\omega_{\v{k}}}\Bigg\}\mathrm{F}\big(\Gamma_{cusp}^{(r)}\big) 
\equiv d\Gamma^{(r)}_{fi,h}\bigg(\frac{E}{\Lambda}\bigg)^{\Gamma_{cusp}^{(r)}} \mathrm{F}\big(\Gamma_{cusp}^{(r)}\big)  \,,
\ea 
where $\mathrm{F}$ is the function \cite{Jauch1955}
\ba
\mathrm{F}(x)= \frac{1}{\pi}\int^{+\infty}_{-\infty} d\sigma \frac{\sin \sigma }{\sigma}\exp\Bigg\{x\int ^1_0\frac{d\omega_{\v{k}}}{\omega_{\v{k}}}\big(e^{+i\omega_{\v{k}}\sigma}-1\big)\Bigg\} = 1-\frac{1}{12}\pi^2x^2+\cdots\,.
\ea 
Eq.~\eqref{eq:Gamma_r_fi_s_result} crystalizes Weinberg's insight that the well-known cancellation between real and virtual IR divergences can be expressed in terms of the RG scaling of IR sensitive cross-sections determined solely in terms of the relative angles, encoded in universal cusp anomalous dimensions,  amongst all the \textit{in} and \textit{out} real charges at infinity.

In Section \ref{sec:faddeev_kulish}, we will adopt a different perspective on the IR problem and consider the renormalization of the IR divergences-with arbitrary numbers of real and virtual photons- within the amplitude itself, by introducing the infrared safe Faddeev-Kulish S-matrix in worldline form. 
\subsection{\label{sec:z_f_factor} IR divergences in field-strength renormalization factors} 
The previous discussion showed that IR divergences exactly cancel to all-loop orders in perturbation theory in the conventional construction of physical observables through the Dyson S-matrix. However there is an important technical point that must be resolved before this proof of the IR safety of cross sections is considered complete. As noted earlier, virtual soft photons produce IR divergences not only directly in the Dyson S-matrix but in the charged particle field-strength renormalization factors $\mathcal{Z}_2$ as well. This constant $\mathcal{Z}_2$ can be read off from the fermion wave function normalization condition which, in the worldline framework, is the Dyson S-matrix matrix element for the evolution of a single fermion ($r=1$) \cite{Feal:2022iyn}:
\ba 
\label{eq:z_2_definition}
&\mathcal{S}_{fi}^{(1)} = \frac{1}{\mathcal{Z}_2} \lim_{\substack{x_f^0\to+\infty\\x_i^0\to-\infty}} \big\langle p_f,s_f;x_f^0|p_i,s_i;x_i^0\big\rangle =\frac{1}{\mathcal{Z}_2}   \frac{\mathrm{Z}}{\mathrm{Z}_\text{MW}}\sum_{\ell=0}^\infty \frac{(-1)^\ell}{\ell!} \nonumber\\
&\times\lim_{\substack{x_f^0\to+\infty\\x_i^0\to-\infty}} \int d^3 \v{x}_f d^3\v{x}_i \bar{\Psi}_{f}(x_f) \exp\bigg\{\bar{\gamma}_\lambda\frac{\partial}{\partial \theta_\lambda} \bigg\} \bar{\gamma}_0 \Psi_i(x_i)\mathrm{W}^{(1,\ell)}(x_f,x_i,\theta)\bigg|_{\theta=0}\,,
\ea 
where $\Psi_{f,i}(x)$ represent free plane waves of an on-shell fermion of momentum $p_{f,i}$ and spin $s_{f,i}$.  Eq.~\eqref{eq:z_2_definition} describes the creation of a free fermion at past infinity and its subsequent evolution up to plus infinity whilst interacting  with the dynamical gauge field created by it. As we have shown, these self-interactions (encoded in the worldline path integral $\mathrm{W}^{(1,\ell)}(x_f,x_i,\theta)$ in Eq.~\eqref{eq:z_2_definition}) produce an IR singularity when $x_f^0\to \infty$ and $x_i^0\to -\infty$. Hence $\mathcal{Z}_2$ contains an IR divergent contribution that can eventually show up in the renormalization of the unrenormalized amplitudes we computed previously. Our focus here is to assess the sensitivity of 
the renormalization of the Dyson S-matrix in Eq.~\eqref{eq:s_fi_ph_nigamma_nfgamma} to the IR divergent contributions in $\mathcal{Z}_2$ as outlined by Weinberg \cite{Weinberg:1995mt} using standard power counting in perturbation theory. (See also the related discussion in Bjorken and Drell \cite{Bjorken:1965zz}.)

In the worldline framework, UV renormalization can be performed by introducing a UV cut-off in the discretization of the worldline paths. This method makes direct contact with the framework for the UV renormalization of Wilson loops discussed at length by several authors \cite{Dotsenko:1979wb,Polyakov:1980ca,Brandt:1982gz,Korchemsky:1987wg}. We should emphasize that the IR and UV divergences in the worldline framework can be identified as the usual ones extracted from  equivalent Feynman diagram calculations, and hence the renormalization of these can be mapped to the standard methods of quantum field theory where the field-strength, electron mass and charge multiplicative renormalization factors are defined by
\ba 
\label{eq:renormalization_constants}
\Psi=\mathcal{Z}_2^{1/2} \Psi_R\,,\,\,\, A_\mu=\mathcal{Z}_3^{1/2} A_\mu^R\,, \,\,\, m= \mathcal{Z}_m m_R\,,\,\,\, g= \mathcal{Z}_g g_R\,.
\ea 
The re-scaled QED Lagrangian then becomes
\ba 
\label{eq:renormalized_lagrangian}
\mathcal{L}= -\frac{1}{4}\mathcal{Z}_3 \big(F_{\mu\nu}^R\big)^2 + \mathcal{Z}_2\bar{\Psi}_R\slashed{\partial} \Psi_R -\mathcal{Z}_2\mathcal{Z}_m m_R\bar{\Psi}_R\Psi_R -g_R\mathcal{Z}_1\Bar{\Psi}_R \slashed{A}_\mu^R \Psi_R\,,
\ea 
where $\mathcal{Z}_1\equiv \mathcal{Z}_g\mathcal{Z}_2\mathcal{Z}_3^{1/2}$. It follows from the Ward identity that $\mathcal{Z}_1=\mathcal{Z}_2$ and hence $\mathcal{Z}_g=\mathcal{Z}_3^{-1/2}$. Unrenormalized and renormalized propagators and vertex functions are related by
\ba 
\label{eq:renormalized_propagators}
\tilde{D}_F(p) = \mathcal{Z}_2 \tilde{D}_{F,R}(p)\,,\,\, \tilde{D}_B^{\mu\nu}(q) = \mathcal{Z}_3 \tilde{D}_{B,R}^{\mu\nu}(q)\,,\,\, \tilde{\Gamma}^{\mu}(p',p)=\mathcal{Z}_1^{-1}\tilde{\Gamma}^\mu_R(p',p)\,.
\ea 
\begin{figure}
    \centering
    \includegraphics[scale=0.4]{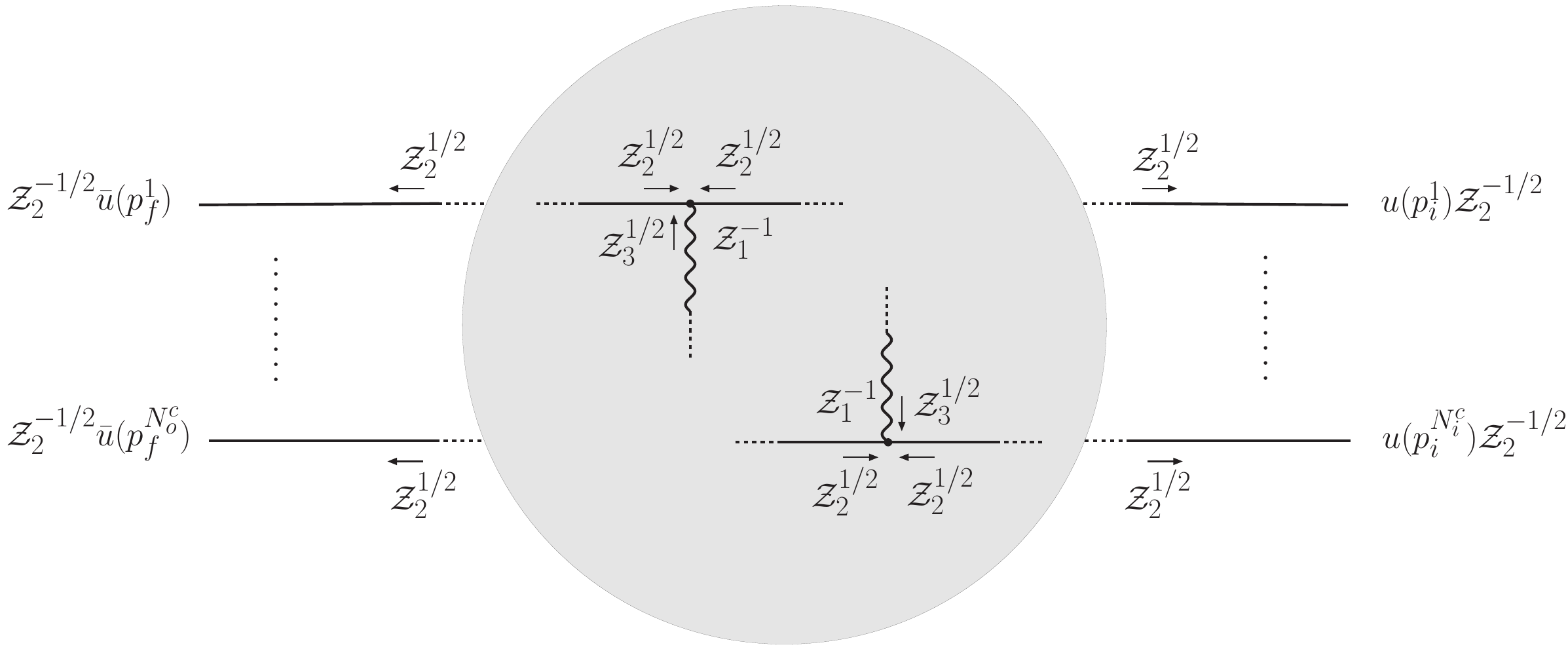}
    \caption{Schematic representation of a general QED transition with $N_i^c$ initial and $N_o^c$ final charges.  The bubble represents the skeleton of all the internal lines and vertices in the Feynman diagram. Unrenormalized quantities have been expressed in terms of their renormalized counterparts.}
    \label{fig:figure_5}
\end{figure}

Let us now consider the UV renormalization of the Dyson S-matrix element for the scattering of $r$ positive charged fermions\footnote{Since photons are neutral, $\mathcal{Z}_3$ does not contain IR divergent contributions and we can restrict our analysis to the power counting of $\mathcal{Z}_2$ factors in the above non-radiative process. }, depicted schematically in Fig.~\ref{fig:figure_5}, and given by setting $N_o^\gamma=0$ and $N_i^\gamma=0$ in Eq.~\eqref{eq:s_fi_ph_nigamma_nfgamma}. Following Eq.~\eqref{eq:renormalization_constants}, the LSZ reduction of the Dyson S-matrix in Eq.~\eqref{eq:s_fi_ph_nigamma_nfgamma} with renormalized fields requires introducing a $\mathcal{Z}_2^{-1/2}$ factor for each truncated \textit{in} and \textit{out} external fermion line. The key point here, as noted by Weinberg but frequently misunderstood in the literature, is that these $\mathcal{Z}_2$ factors will be exactly cancelled by the $\mathcal{Z}_2$ factors appearing within the internal lines and vertices. To see this,  following Eq.~\eqref{eq:renormalized_propagators}, one first observes that each vertex gets renormalized by a factor $\mathcal{Z}_1^{-1}$. Besides this factor, each photon propagator line flowing into this vertex contributes with a factor $\mathcal{Z}_3^{1/2}$. Similarly, the two fermion lines joining the vertex each contribute  with a factor $\mathcal{Z}_2^{1/2}$. (The other $\mathcal{Z}_2^{1/2}$ and $\mathcal{Z}_3^{1/2}$ factors flow either to a different vertex or to an external fermion line - or a photon line for the case where external photons are considered in the scattering process.) 

At each vertex, one finds an overall factor
\ba 
g \mathcal{Z}_1^{-1} \mathcal{Z}_2 \mathcal{Z}_3^{1/2} \equiv g_R\,.
\ea 
Therefore each fermion external line  is left with a $\mathcal{Z}_2^{1/2}$ factor. In the final LSZ truncation, these factors are exactly cancelled by  the corresponding $\mathcal{Z}_2^{-1/2}$ wavefunction renormalization constants appearing within the LSZ formula - as shown in  Fig.~\ref{fig:figure_5}. This cancellation happens at the level of the S-matrix itself. 

In contrast, as emphasized by Weinberg, the infrared divergences that appear in the soft factor in Eq.~\eqref{eq:s_fi_r_s_Nos_Nis} are unrelated to those associated with 
the  aforementioned renormalization constants of the field-strengths, the coupling and the mass. Indeed, the divergences in the soft factor would set the Dyson S-matrix to zero if the infrared cut-off $\lambda\rightarrow 0$ in section~\ref{sec:bloch_nordsieck_cancellation}. Further,  as we discussed there, this infrared divergence is cancelled by the infrared divergence in the emission of soft photons at the level of the cross-section. We will discuss in the next section the origins of the divergences in the soft factor in the worldline formalism and how they can be regulated to construct infrared safe S-matrices along the lines first proposed by Faddeev and Kulish.

\section{\label{sec:faddeev_kulish} Worldline IR renormalization of amplitudes to all-loop orders in perturbation theory: the Faddeev-Kulish S-matrix}

In Paper I, we derived in the worldline formalism for QED the 
infrared safe Faddeev-Kulish (KF) S-matrix for virtual photon exchanges to all orders in perturbation theory. We will extend this program here 
to the case of real photon absorptions and emissions thereby completing our proof of infrared safety of this S-matrix. 

As was already understood in the original Faddeev-Kulish paper~\cite{Kulish:1970ut}, the difference between the infrared divergent Dyson S-matrix and the infrared safe FK S-matrix is primarily one of the ordering of limits. In the former, one first takes the limits of infinite past and infinite future $t_{f,i}\rightarrow \pm \infty$, and then takes the infrared momentum cutoff $\lambda\rightarrow 0$, revealing a logarithmic infrared divergence in the ratio of this scale to a hard scale $\Lambda$. In the FK S-matrix, one 
instead keeps $t_{f,i} = 1/{\Lambda^\prime}$ finite. The contribution of asymptotic currents at earlier/later times than those characterized by the scale $\Lambda^\prime$ cancels the dependence of the result on $\lambda$, resulting in an infrared safe result for the FK S-matrix when $\lambda\rightarrow 0$. 

We will discuss below in detail how this works and demonstrate that one recovers Weinberg's result for the photon emission rate. We will then discuss how our results can be understood straightforwardly in the modern language of the lattice regularization of UV and IR divergences in gauge theories. The takeaway message remains the same as in Weinberg's paper: in QED, as in QCD, there is ultimately one fundamental scale and all results can in principle be expressed in terms of this scale.

%
The FK S-matrix is defined analogously to the Dyson S-matrix in Eqs.~\eqref{eq:s_fi_r_nfgamma_nigamma_loopexpansion} and \eqref{eq:s_fi_rell_nfgamma_nigamma} as 
\ba
\label{eq:s_fi_r_nfgamma_nigamma_loopexpansion_fk}
\bar{\mathcal{S}}_{fi}^{(r)}(N_o^\gamma,N_i^\gamma)=\sum_{\ell=0}^\infty \bar{\mathcal{S}}_{fi}^{(r,\ell)}(N_o^\gamma,N_i^\gamma)\,,
\ea 
where the $\ell$-th loop contribution is given by
\ba 
\label{eq:s_fi_rell_nfgamma_nigamma_fk}
&\bar{\mathcal{S}}_{fi}^{(r,\ell)}(N_o^\gamma,N_i^\gamma) = \frac{\mathrm{Z}_\text{MW}}{\mathrm{Z}[0,0,0]}\frac{(-1)^\ell}{\ell!}\prod_{p=1}^{N_o^\gamma} \Bigg\{\frac{i\epsilon_{\mu_f^p}^*(\v{k}_f^p,\lambda_f^p)}{\sqrt{2\omega_{\v{k}_f^p}}(2\pi)^{3/2}} \Bigg\}\prod_{p=1}^{N_i^\gamma}\Bigg\{ \frac{i\epsilon_{\mu_i^p}(\v{k}_i^p,\lambda_i^p)}{\sqrt{2\omega_{\v{k}_i^p}}(2\pi)^{3/2}}\Bigg\}\nonumber\\
&\times \prod_{n=1}^r\Bigg\{\int d^3\v{x}_i^n \int d^3\v{x}_f^n \Psi^{(+),\dag}_{f_n}(x_f^n) \exp\bigg\{\bar{\gamma}_\lambda\frac{\partial}{\partial \theta_n}\bigg\}\bar{\gamma}_0\Psi_{i_n}^{(+)}(x_i^n)\Bigg\} \bigg\langle \prod_{p=1}^{N_o^\gamma} \bigg\{\bar{J}_{(r,\ell)}^{\mu_f^p}(-k_f^p)\bigg\}\\
&\times  \prod_{p=1}^{N_i^\gamma} \bigg\{ \bar{J}_{(r,\ell)}^{\mu_i^p}(+k_i^p)\bigg\}\exp\bigg\{\frac{i}{2}\int \frac{d^4q}{(2\pi)^4}\frac{1}{q^2+i\epsilon}\bar{J}_{\mu}^{(r,\ell)}(-q)\bar{J}_{(r,\ell)}^{\mu}(+q)\bigg\}\bigg\rangle\,+\text{permutations}\,.\nonumber
\ea 
As stated earlier, it is important to note that the limits $t_{f,i}^n\to\pm\infty$ specified in Eq.~\eqref{eq:s_fi_rell_nfgamma_nigamma} must be taken only after all virtual and real IR divergences are cancelled in the diagrams generated by the FK S-matrix, to any particular order in perturbation theory.

Specifically, this means that the low momentum limits of the currents in Eq.~\eqref{eq:lim_k_0_j_mu_r_ell_k}, as per the FK S-matrix prescription, now contain in addition boundary terms (for $t_{i} < -1/\Lambda^\prime$ and $t_f > 1/\Lambda^\prime$) coming wholly from the $r$ real charged particle asymptotic currents that were neglected when computing the current in the conventional Dyson S-matrix in Eq.~\eqref{eq:lim_k_0_j_n_mu_r}:
\ba 
&\lim_{k\to 0} \tilde{J}_{\mu,R}^n(k) \rightarrow \lim_{k\to 0} \bar{J}_{\mu,R}^n(k) \nonumber\\
&= \lim_{k\to 0}\bigg\{
\underbrace{\frac{g}{i}\frac{p_{f,\mu}^{n}}{k\wc p_f^n+i\varepsilon}e^{ik\wc x^n_f-\varepsilon t_f^n}}_\text{Final asymptotic current}-\underbrace{\frac{g}{i}\bigg\{\frac{p_{f,\mu}^{n}}{k\wc p_f^n+i\varepsilon}-\frac{p_{i,\mu}^n}{k\wc p_i^n-i\varepsilon}\bigg\}}_{\text{IR Current in the Dyson S-matrix}}-\frac{g}{i}\underbrace{\frac{p_{i,\mu}^n}{k\wc p_i^n-i\varepsilon}e^{ik\wc x_i^n+\varepsilon t_i^n}}_\text{Initial asymptotic current}\bigg\}\,.
\ea 
In the IR limit, the charged current in Eq.~\eqref{eq:lim_k_0_j_mu_r_ell_k} should be replaced by the corresponding current for the FK S-matrix, with 
\ba 
\label{eq:lim_k_0_j_r_ell_FK}
\lim_{k\to 0}\bar{J}_{\mu}^{(r,\ell)}(k) = \lim_{k\to 0}\Big(  \tilde{J}_{\mu}^{IR}(k) + \tilde{J}_\mu^{AS}(k)\Big)\,,\,\,\text{where}\,\,\, \tilde{J}_{\mu}^{AS}(k)=+\frac{g}{i}\sum_{n'=1}^{N_o^c+N_i^c}  \frac{\eta^{n'}p_\mu^{n'}}{k\wc p^{n'}+i\eta^{n'}\varepsilon}\,\,e^{ik\wc x^{n'}}\,,
\ea 
and $\tilde{J}_\mu^{IR}$ is given, as per the Dyson S-matrix prescription, in Eq.~\eqref{eq:j_mu_ir_definition}. 

\begin{figure}[ht]
    \centering
    \includegraphics[scale=0.8]{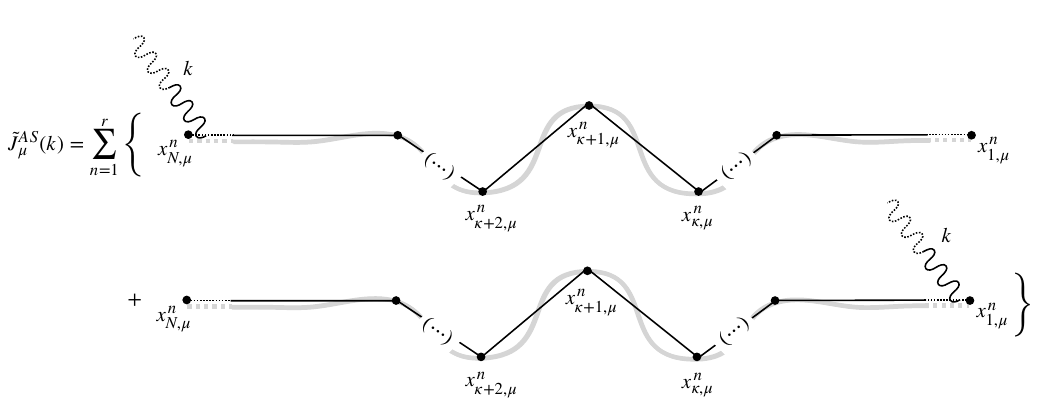}
    \caption{Asymptotic worldline (AS) graphs of a soft (real or virtual) photon line pinched at infinity 
    for each $n$ of the $r$ real charges in the scattering amplitude corresponding to the FK S-matrix prescription.}
    \label{fig:figure_6}
\end{figure}

A pair of asymptotic photon graphs at negative and positive asymptotic times generated by $J_\mu^{AS}(k)$ for each real charge $n$ of $r$ participating charges in the scattering is shown in Fig.~\ref{fig:figure_6}.
Clearly, the current above now has no $1/k$ contributions since those in $\tilde{J}^{IR}_\mu$ exactly cancel with the ones in $\tilde{J}^{AS}_\mu$ for real or virtual photons of sufficiently low 4-momentum. 
The cancellation of $1/k$ contributions in the currents occurs for momentum $k$ below the  asymptotic scale $\Lambda'$ that we introduced that correspond to the boundaries $t_{f,i}^{n}$ in the problem under consideration. This scale, by construction, is smaller than the scale $\Lambda$ we defined for the separation of the long-distance soft interactions from the hard interactions in the conventional Dyson S-matrix. We will show that physics does not depend on $\Lambda^\prime$ and shall discuss further the interpretation of this scale. 

The virtual photon exchanges within 
the normalized worldline expectation value in the FK S-matrix in Eq.~\eqref{eq:s_fi_rell_nfgamma_nigamma_fk} can then be expressed as 
\ba 
\label{eq:soft_hard_factorization_virtual_interactions_FK}
&\int_\lambda^{\Lambda_\mathrm{QED}}\frac{d^4q}{(2\pi)^4}\frac{g^{\mu\nu}}{q^2+i\varepsilon}\bar{J}^{(r,\ell)}_\mu(-q) \bar{J}^{(r,\ell)}_\nu(q) \nonumber\\
&=\bigg\{\int^{\Lambda'}_{\lambda}\frac{d^4q}{(2\pi)^4}+\int^{\Lambda}_{\Lambda'}\frac{d^4q}{(2\pi)^4}+\int^{\Lambda_\mathrm{QED}}_\Lambda \frac{d^4q}{(2\pi)^4}\bigg\}\frac{g^{\mu\nu}}{q^2+i\varepsilon}\bar{J}^{(r,\ell)}_\mu(-q) \bar{J}^{(r,\ell)}_\nu(q)\,. 
\ea 
where we replaced the upper limit of phase space of virtual photons by $\Lambda_{\text{QED}}$, the fundamental UV scale of the theory. 

The first integral on the r.h.s. contains the contribution to the virtual exchanges coming from the exchanges of photons of energy less than $\Lambda'$. Using Eqs.~\eqref{eq:lim_k_0_j_r_ell_FK} and \eqref{eq:j_mu_ir_definition} it produces
\ba 
\label{eq:soft_hard_factorization_virtual_interactions_FK_2}
&\int^{\Lambda'}_{\lambda}\frac{d^4q}{(2\pi)^4}\frac{g^{\mu\nu}}{q^2+i\varepsilon}\bar{J}^{(r,\ell)}_\mu(-q) \bar{J}^{(r,\ell)}_\nu(q) \nonumber\\
&=g^2\sum_{n',m'=1}^{N_i^c+N_o^c} \int^{\Lambda'}_\lambda \frac{d^4q}{(2\pi)^4}\frac{g^{\mu\nu}}{q^2+i\varepsilon}\frac{\eta^{n'}p_\mu^{n'}}{q\wc p^{n'}-i\eta^{n'}\varepsilon}\big(e^{-iq\wc x^{n'}}-1\big)\frac{\eta^{m'}p_\nu^{m'}}{q\wc p^{m'}+i\eta^{m'}\varepsilon}\big(e^{+iq\wc x^{m'}}-1\big)\,.
\ea 
The integrand above clearly vanishes when $q\rightarrow 0$ and is therefore manifestly independent of $\lambda$. As long as $t_{f,i}$ are finite, one can choose the corresponding $\Lambda^\prime$ to be small. The above integral appearing in the exponent
of Eq.~\eqref{eq:s_fi_rell_nfgamma_nigamma_fk} therefore gives a vanishingly small contribution to the FK S-matrix. 


The second integral on the r.h.s. of Eq.~\eqref{eq:soft_hard_factorization_virtual_interactions_FK} contains the contribution of long-distance virtual exchanges (excluding those in the asymptotic region)  amongst all the charged particles: 
\ba 
\label{eq:virtual_soft_factor_FK}
\int^{\Lambda}_{\Lambda'} \frac{d^4q}{(2\pi)^4}\frac{g^{\mu\nu}}{q^2+i\varepsilon}\bar{J}^{(r,\ell)}_\mu(-q) \bar{J}^{(r,\ell)}_\nu(q) = \int^{\Lambda}_{\Lambda'} \frac{d^4q}{(2\pi)^4}\frac{g^{\mu\nu}}{q^2+i\varepsilon}\tilde{J}^{IR}_\mu(-q) \tilde{J}^{IR}_\nu(q)\,.
\ea 
Since this expression only depends on soft exchanges, the worldline currents satisfy classical trajectories and can be factored out of the normalized worldline expectation value in Eq.~\eqref{eq:s_fi_rell_nfgamma_nigamma_fk}, just as in the Dyson S-matrix. 

A further key point is that as long as the IR regulator $\Lambda^\prime$ is kept finite (which is the case for the finite initial and final times for which the FK S-matrix is defined), the above expression is infrared finite. This can be understood in an analogous way to procedure by which the infinite volume limit is taken in the Euclidean formulation of gauge theories. In the latter, modes of order $\lambda <\Lambda^\prime$ (where the latter is the scale of the finite lattice) are automatically excluded. Computations are performed for this fixed volume, and only then is the infinite volume limit taken. The statement of infrared safety is then understood as the fact that all physical quantities are robust in the infinite volume limit. 

The worldline asymptotic currents in this language  therefore represent a particular choice of an  infrared regulator in Minkowski spacetime which ensures that modes $\lambda <\Lambda^\prime$ do not contribute. Further, the corresponding statement of infrared safety is that all physical quantities must be independent of $\Lambda^\prime$ when $t_{f,i}\rightarrow \pm\infty$ (or $\Lambda^\prime \rightarrow 0$)  {\it taken as the final step in the computation}. While this might seem suspiciously close to the usual Dyson result of the $\lambda$-dependence of the S-matrix canceling in cross-sections, there is nevertheless an important difference in the two approaches which we will return to.

The final integral in Eq.~\eqref{eq:soft_hard_factorization_virtual_interactions_FK} is IR finite by construction and identical to the hard virtual exchanges as per the Dyson S-matrix prescription in Eq.~\eqref{eq:soft_hard_factorization_virtual_interactions}. It should therefore be kept as part of the normalized worldline expectation value in Eq.~\eqref{eq:s_fi_rell_nfgamma_nigamma_fk}. It too contains divergences; these are the UV divergences of the theory, which are treated in the usual way by renormalization of bare charges and masses in the theory. 

The factorization of asymptotic and soft $N_{i,s}^\gamma$ and $N_{o,s}^\gamma$ \textit{in} and \textit{out} real photons of energies $\omega<\Lambda$ follows similarly, with the corresponding worldline current $\bar{J}_{\mu}^{(r,\ell)}$ can be replaced by Eq.~\eqref{eq:lim_k_0_j_r_ell_FK} and factored out of the normalized worldline expectation value. The $N_{i,h}^{\gamma}$ and $N_{o,h}^{\gamma}$ hard \textit{in} and \textit{out} real photons are likewise  kept as part of the normalized worldline expectation value, with the full current given by Eq.~\eqref{eq:j_mu_r_ell_q_Minkowski}. 

From Eq.~\eqref{eq:s_fi_rell_nfgamma_nigamma_fk} and Eq.~\eqref{eq:soft_hard_factorization_virtual_interactions_FK}, one then gets
\ba 
\label{eq:s_fi_r_Nos_Nis_fk}
\bar{\mathcal{S}}_{fi}^{(r)}(N_{o}^\gamma,N_{i}^\gamma) = \bar{\mathcal{S}}_{fi,s}^{(r)}(N_{o,s}^\gamma,N_{i,s}^\gamma)\,\,\times\,\, {\mathcal{S}}_{fi,h}^{(r)}(N_{o,h}^\gamma,N_{i,h}^\gamma)\,,
\ea 
where the soft factor in the FK S-matrix reads 
\ba 
\label{eq:s_fi_r_s_Nos_Nis_fk}
&\bar{\mathcal{S}}_{fi,s}^{(r)}(N_{o,s}^\gamma,N_{i,s}^\gamma) = \prod_{p=1}^{N_{o,s}^\gamma} \Bigg\{\frac{i\epsilon_{\mu_f^p}^*(\v{k}_f^p,\lambda_f^p) \Big(\tilde{J}^{\mu_f^p}_{IR}(-k_f^p)+\tilde{J}^{\mu_f^p}_{AS}(-k_f^p)\Big)}{\sqrt{2\omega_{\v{k}_f^p}}(2\pi)^{3/2}} \Bigg\} \nonumber\\
&\times\prod_{p=1}^{N_{i,s}^\gamma} \Bigg\{\frac{i\epsilon_{\mu_i^p}(\v{k}_i^p,\lambda_i^p) \Big(\tilde{J}^{\mu_i^p}_{IR}(+k_i^p)+\tilde{J}^{\mu_i^p}_{AS}(+k_i^p)\Big)}{\sqrt{2\omega_{\v{k}_i^p}}(2\pi)^{3/2}}\Bigg\} \exp\bigg\{\int^\Lambda_{\Lambda'} \frac{d^4q}{(2\pi)^4}A_V^{IR}(q)\bigg\} \,,
\ea 
with $A_V^{IR}(q)$ corresponding to the integrand of Eq.~\eqref{eq:virtual_soft_factor_FK}, defined previously in Eq.~\eqref{eq:a_v_ir_q_def}.  Since this virtual soft photon factor 
is IR finite as noted previously, and further, as also noted,  the real soft photon factors in Eq.~\eqref{eq:s_fi_r_Nos_Nis_fk} do not possess $\sim 1/k$ contributions, the 
entire soft factor is manifestly infrared finite.

The hard Dyson S-matrix element   $\mathcal{S}_{fi,h}^{(r)}(N_{o,h}^\gamma,N_{i,h}^\gamma)$ defined in Eqs.~\eqref{eq:hardS-matrix} and \eqref{eq:hardS-matrix_lth-loop} (where  $N_{i,h}^\gamma$ and $N_{o,h}^\gamma$ \textit{in} and \textit{out} are real hard photons) is likewise infrared finite by construction. 
Thus the FK S-matrix $\bar{\mathcal{S}}_{fi}^{(r)}$, with arbitrary numbers of real and virtual photons, is infrared finite to all orders in perturbation theory\footnote{Note that the previous analysis of IR divergences in the FK S-matrix did not include  the field-strength renormalization factors that 
appear in  LSZ reduction formulae.  This is because the infrared divergences in fermion wave-function normalization factors $\bar{\mathcal{Z}}_2$,  which are unrelated to those in the soft factors,  are exactly cancelled by the $\bar{\mathcal{Z}}_2$ factors introduced by the internal lines and vertices, in a completely analogous way to the discussion we had in section 3.3 for the Dyson S-matrix case. In this regard, there is no 
difference between the Dyson and F-K S-matrices; our conclusions mirror those of Weinberg and differ from those of Zwanziger~\cite{Zwanziger:1973if} who modifies the field-strength renormalization factors in the LSZ reduction formula to accommodate the FK asymptotic currents.}.
%
%

The final step in the proof of infrared safety is to show  that the emission rate corresponding to the  IR safe FK S-matrix is IR safe as well, and agrees with Weinberg's result  discussed in Section \ref{sec:bloch_nordsieck_cancellation}. Towards this end, we  observe that the photon emission rate for the FK S-matrix can be factorized as
\ba 
\label{eq:gamma_fi_r_s_times_gamma_fi_r_h_fk}
\bar{\Gamma}_{fi}^{(r)}= \bar{\Gamma}^{(r)}_{fi,s}\times 
\Gamma_{fi,h}^{(r)}\,.
\ea 
The hard emission rate $\Gamma_{fi,h}^{(r)}$ is the squared modulus of 
the hard Dyson S-matrix element in Eqs.~\eqref{eq:hardS-matrix} and \eqref{eq:hardS-matrix_lth-loop}. From our previous discussion, since 
the role of  $\tilde{J}_{AS}^{\mu}$ terms in Eq.~\eqref{eq:s_fi_r_s_Nos_Nis_fk} to to eliminate modes $\lambda < \Lambda^\prime$, it is sufficient to keep the $J_{IR}^\mu$ terms for modes $\Lambda'<\omega<\Lambda$.
Then following the same procedure as in the previous section for the Dyson S-matrix, the differential FK soft emission rate is  given by
\ba 
\frac{d\bar{\Gamma}^{(r)}_{fi,s}}{d\omega_T} =& \delta(\omega_T)\Big|\bar{\mathcal{S}}^{(r)}_{fi,s}(0,0)\Big|^2+\frac{1}{1!}\sum_{\lambda_f^1}\int_{\Lambda'}^\Lambda d^3\v{k}_f^1
\delta\big(\omega_T-\omega_{\v{k}_f^1}\big) \Big|\bar{\mathcal{S}}^{(r)}_{fi,s}(1,0)\Big|^2\nonumber\\
&+\frac{1}{2!}\sum_{\lambda_f^1}\int_{\Lambda'}^\Lambda d^3\v{k}_f^1 \sum_{\lambda_f^2} \int_{\Lambda'}^\Lambda d^3\v{k}_f^2 \delta\big(\omega_T-\omega_{\v{k}_f^1}-\omega_{\v{k}_f^2}\big)\Big|\bar{\mathcal{S}}_{fi,s}^{(r)}(2,0)\Big|^2+\cdots\,,
\label{eq:soft-differential-rate_fk}
\ea 
where we  replaced the lower limits of  integration by $\lambda \rightarrow\Lambda'$, with the latter denoting that the FK matrix element is computed for $t_{f,i}=$ finite. 

Using Eq.~\eqref{eq:s_fi_r_s_Nos_Nis_fk}, the infinite sum over final states in Eq.~\eqref{eq:soft-differential-rate_fk} can be reexponentiated to obtain
\ba 
\frac{d\bar{\Gamma}^{(r)}_{fi,s}}{d\omega_T} &= \frac{1}{2\pi} \exp\Bigg\{2\Re\int^\Lambda_{\Lambda'} \frac{d^4q}{(2\pi)^4}A_V^{IR}(q)\Bigg\}\nonumber\\
&\times\int^{+\infty}_{-\infty}d\sigma e^{-i\sigma\omega_T} \exp\Bigg\{ \int_{\Lambda'}^\Lambda \frac{d^3\v{k}}{(2\pi)^3} \frac{e^{+i\omega_{\v{k}}\sigma}}{2\omega_{\v{k}}}  A_R^{IR}(\v{k})\Bigg\} \,.
\label{eq:soft-differential-rate_fk2}
\ea
The emission rate of any number of real soft photons adding up to a total emitted energy less than $E$, with $\Lambda'<E<\Lambda$, is given by replacing the upper limit $\Lambda$ of the real photon phase space integrals above by $E$, and by integrating the resulting differential rate from $\omega_T=0$ to $\omega_T=+E$, or alternatively between $\omega_T=-E$ and $\omega_T=+E$ (as pointed out in Section \ref{sec:bloch_nordsieck_cancellation}), giving
\ba 
\bar{\Gamma}_{fi,s}^{(r)}=&
\int^{+E}_{-E} d\omega_T \frac{d\bar{\Gamma}_{fi,s}^{(r)}}{d\omega_T}=\frac{1}{\pi} \exp\bigg\{2\Re \int^\Lambda_{\Lambda'}\frac{d^4q}{(2\pi)^4} A_V^{IR}(q)\bigg\}  \nonumber\\
&\times \int^{+\infty}_{-\infty} d\sigma \frac{\sin \sigma E}{\sigma } \exp\Bigg\{ \int_{\Lambda'}^E \frac{d^3\v{k}}{(2\pi)^3} \frac{1}{2\omega_{\v{k}}}  e^{+i\omega_{\v{k}}\sigma}A_R^{IR}(\v{k})
\Bigg\} \,.
\ea 
Using the identity in Eq.~\eqref{eq:2_re_int_a_v} the expression above can be rewritten as
\ba 
\label{eq:Gamma_fi_r_s_fk_result1}
\bar{\Gamma}^{(r)}_{fi,s}&= \exp\bigg\{-\int^\Lambda_E \frac{d^3\v{k}}{(2\pi)^3}\frac{1}{2\omega_{\v{k}}}A_R^{IR}(\v{k})\bigg\}\\
&\times \frac{1}{\pi}  \int^{+\infty}_{-\infty} d\sigma \frac{\sin \sigma E}{\sigma } \exp\Bigg\{ \int_{\Lambda'}^E \frac{d^3\v{k}}{(2\pi)^3} \frac{1}{2\omega_{\v{k}}}  \big(e^{+i\omega_{\v{k}}\sigma}-1\big)A_R^{IR}(\v{k}) \Bigg\} \,,\nonumber
\ea 
and after performing the required angular integrals, gives the result 
\ba 
\label{eq:Gamma_fi_r_s_fk_result2}
\bar{\Gamma}^{(r)}_{fi,s}= \frac{1}{\pi}\exp\bigg\{-\Gamma_{cusp}^{(r)} \int^\Lambda_E  \frac{d{\omega}_{\v{k}}}{\omega_{\v{k}}}\bigg\}\int^{+\infty}_{-\infty} d\sigma \frac{\sin \sigma E}{\sigma } \exp\Bigg\{\Gamma_{cusp}^{(r)}\int_{\Lambda'}^E \frac{d\omega_{\v{k}}}{\omega_{\v{k}}}\big(e^{+i\omega_{\v{k}}\sigma}-1\big) \Bigg\}\,,
\ea 
where the cusp anomalous dimension $\Gamma^{(r)}_{cusp}$ was defined earlier in Eq.~\eqref{eq:gamma_cusp_r_def}. Finally, rescaling $\sigma E\to\sigma $ in  Eq.~\eqref{eq:Gamma_fi_r_s_fk_result2}, gives the total differential emission rate
\ba
\label{eq:Gamma_fi_r_s_fk_result3}
d\bar{\Gamma}_{fi}^{(r)}
 = \bigg[\frac{E}{\Lambda}\bigg]^{\Gamma_{cusp}^{(r)}} \bar{F}\Big(\Gamma_{cusp}^{(r)},\frac{\Lambda'}{E}\Big)\times d\Gamma_{fi,h}^{(r)}
 \,.
\ea 
The differential hard emission rate was defined in  Eq.~\eqref{eq:gamma_r_fi_s_times_gamma_r_fi_h} and the function $\bar{F}$ is 
\ba 
\bar{F}(x,\Lambda'/E) \equiv \frac{1}{\pi}\int^{+\infty}_{-\infty} d\sigma \frac{\sin \sigma}{\sigma } \exp\Bigg\{\Gamma_{cusp}^{(r)}\int_{\Lambda'/E}^1 \frac{d\omega_{\v{k}}}{\omega_{\v{k}}}\big(e^{+i\omega_{\v{k}}\sigma}-1\big) \Bigg\}\,.
\ea 
Eq.~\eqref{eq:Gamma_fi_r_s_fk_result3} approaches Weinberg's result in Eq.~\eqref{eq:Gamma_r_fi_s_result}, when at this final stage one takes the infinite time limit $t_{f,i}\to\pm \infty$; this corresponds to $\Lambda'\to 0$, resulting in  $\bar{F}(\Gamma^{(r)}_{cusp},0)=F(\Gamma^{(r)}_{cusp})$. As noted earlier, this is analogous to taking the infinite volume limit in the Euclidean ``lattice" formulation of gauge theories.

As emphasized\footnote{See the discussion after Eqs. 2.52 in \cite{PhysRev.140.B516}} by Weinberg, the scale $\Lambda$ which was introduced as the ``factorization" scale, can equivalently be thought of as a UV scale, since such a replacement of the former merely renormalizes the hard cross-section. To minimize the amount of this renormalization it is however advisable to choose a scale appropriate for the relevant energy of the process of interest. 

A skeptical reader may question the value of the FK S-matrix; one can argue that the IR divergence of the Dyson S-matrix is now  ``hidden" in the dependence of the result on $t_{f,i}$ with the divergence reappearing when these times are taken to infinity\footnote{For an interesting discussion along these lines, see \cite{Hannesdottir:2019opa}.}. Further, if the asymptotic current is seen primarily as a regulator, its exact form  does not matter as long as it cancels the $1/k$ terms that cause IR divergences. What then is the value of the interpretation in terms of asymptotic BMS-like symmetries~\cite{McLoughlin:2022ljp} of the FK S-matrix?

This hypothetical critic can be answered as follows. Firstly, by changing the order of limits, and by introducing the early and late time cut-offs, one is adapting the modern Wilsonian approach to UV and IR divergences in quantum field theory to the real-time problem of the S-matrix. The elaborate cancellations between real and virtual IR divergences contributions to cross-sections in the Dyson S-matrix approach, which becomes increasingly cumbersome at high orders, can be avoided. Further, the worldline formalism is ideally suited to an implementation of this strategy\footnote{This remark may only be made in full confidence for QED.} since closed and open worldlines can be treated on the same footing. With regard to the second point concerning the IR regulator, while it is indeed true that physical observables do not depend on its details (the form of the asymptotic currents), such a regulator is natural in the worldline formulation of QED. In high order computations, the choice of appropriate regulators of spurious divergences can greatly simplify computations, especially if one can appeal to the symmetries satisfied by such a regulator\footnote{An interesting counterpoint is provided by the observation by Faddeev and Kulish that their IR safe S-matrix has asymptotic states that are coherent states. This inspired an interesting program towards a coherent state description of the QCD analog of the FK S-matrix. However our discussion clarifies that the coherent state description of FK is not its fundamental feature; another choice of a finite time infrared regulator would in principle do as well. Thus the search for infrared safe descriptions of amplitudes in QCD~\cite{Catani:1984dp,Catani:1985xt} should be distinguished from their possible description as coherent states.}. Ultimately, the proof of this pudding will be in explicit demonstrations of the efficacy of the approach to real-time problems in gauge theories.


\section{Summary and Outlook}
In Paper I, we developed the worldline formulation of QED to express vacuum-to-vacuum amplitudes to all orders in perturbation theory as a first quantized theory of super-pairs of pointlike bosonic and fermionic worldlines interacting via pair-wise exchange of Lorentz forces. We further considered the general formulation of the Dyson S-matrix for scattering in this framework and demonstrated that the underlying origin of  infrared divergences arises from the asymptotic structure of worldline currents at very early and very late times ($t_{f,i}\rightarrow \pm \infty$) in the scattering. We further showed that keeping asymptotic currents that are discarded in the Dyson S-matrix due to their rapid oscillations  when $t_{f,i}\rightarrow \pm \infty$ is equivalent to the Faddeev-Kulish (FK) prescription for deriving an infrared safe S-matrix. Specifically, this corresponds to reversing the usual order of limits in the Dyson S-matrix  by keeping  $t_{f,i}={\rm finite}$, taking infrared cutoff $\lambda\rightarrow 0$, and only subsequently taking $t_{f,i}\rightarrow \pm \infty$ in the computation of physical observables. In Paper I, we provided a proof of the infrared safety of the Faddeev-Kulish S-matrix for virtual photon exchanges, to all orders in perturbation theory, for the scattering of arbitrary numbers of charged fermions. 

In this work, we first developed the worldline formalism to obtain general expressions for the usual Dyson S-matrix, to all loop orders, for {\it both} virtual exchanges and real radiation. For the latter, we demonstrated an explicit proof of Low's theorem; we further demonstrated Weinberg's soft photon theorem for the Abelian exponentiation of infrared divergences. In particular, we derived Weinberg's result demonstrating that the emission rate for charged lepton scattering with the emission of arbitrary numbers of photons with a fixed net energy satisfies renormalization group evolution of the energy with respect 
to the hard characteristic scale of the theory; the corresponding anomalous dimension is nothing but the cusp anomalous dimension, which has a simple interpretation in terms of the relative 4-angles between incoming and outgoing classical currents into the hard vertex for the scattering. 

We next extended the proof of the infrared safety of the FK S-matrix to the general case of both virtual exchanges and real radiation. The proof follows similarly to that of virtual exchanges considered previously. A key point is the introduction of an infrared scale $\Lambda^\prime$ corresponding to the finite asymptotic times at which the FK S-matrix is defined. With the introduction of this scale, the FK S-matrix is shown to be independent of  the soft photon momentum cutoff  as $\lambda\rightarrow 0$. We then employed this setup to recover Weinberg's result for the emission rate of soft photons; this result is recovered exactly in the final step when 
$t_{f,i}\rightarrow \pm \infty$, or equivalently, $\Lambda^\prime \rightarrow 0$. 

This derivation, and the manner in which Weinberg's result is recovered, provides a modern understanding of the FK S-matrix as a realization of 
the Wilsonian program for addressing the IR and UV divergences of gauge theories for real-time observables. The Wilsonian approach is of course standard in the Euclidean ``lattice" formulation of gauge theories where physical quantities are defined with UV and IR lattice regulators, with continuum and infinite volume extrapolations of these performed subsequently. This Wilsonian program is harder to realize in the Feynman diagram approach to real-time problems which deals with asymptotic {\it in} and {\it out} states at the very outset. 

The virtue of the  worldline formulation of the S-matrix we have developed here is that it provides a concrete path towards implementing the Wilsonian approach to regulate IR divergences in real-time problems. (UV divergences can be handled in the usual manner.)
One sees explicitly that the FK S-matrix must be understood as a finite time regularization of IR divergences. Within this logic, the detailed nature of the asymptotic currents, and any symmetries that they may satisfy, appear to be irrelevant as long as they fulfill the task of smoothly cancelling the $1/k$ divergences. Though one should therefore hesitate in ascribing physical meaning to the asymptotic symmetries of the currents, they are however not irrelevant in practice. 
Such symmetries may help facilitate high order computations just as the proper choice of IR regulators in Feynman diagram computations can transform a cumbersome computation to a simple one. 

We note finally that in Appendix A we derive, employing Grassmannian integration of worldline currents, a novel simple expression for N-th rank vacuum polarization tensors. An immediate outcome is an all-order proof of Furry's theorem. However the power of this result will be manifest in high order computations, a specific example of which is the high order computation of cusp anomalous dimensions in QED. This computation is in progress and will be reported on in follow-up work.

\section*{Acknowledgements}
We would like to thank Fiorenzo Bastianelli, Stefan Fl\"orchinger, Hofie Hannesdottir, Christian Schubert and Fanyi Zhao for discussions. We would also like to thank the referee of this manuscript for emphasizing
Weinberg's analysis of the cancellation of the infrared divergences that are absorbed in the field-strength renormalization factors in the LSZ 
formalism.  R. V. is supported by the U.S. Department of Energy, Office of Science, Office of Nuclear Physics, under contract No. DE- SC0012704. His research at Stony Brook University is supported by a grant from the Simons Foundation. 
X. F. is supported by Grant No. ED481B-2019-040 of Conseller\'ia de Cultura, Educaci\'on e Universidade of Xunta de Galicia (Spain). A. T. is supported by the U.S. Department of Energy, Office of Science, Office of Nuclear Physics through the Contract No. DE-SC0020081.

\appendix

\section{\label{sec:furry}Explicit computation of N-th rank polarization tensors and Furry's theorem}
In Paper I, we showed that multi-loop vacuum-to-vacuum amplitudes in QED (see for instance $\mathrm{Z}_{(n)}^{(\ell)}$ in Eq.~(95) of \cite{Feal:2022iyn}), encoding the contributions of all Feynman diagrams with $\ell$ fermion and $n$ photon loops, can be systematically simplified in the worldline framework to the evaluation of a product of $\ell$ one-loop $N$th rank vacuum polarization tensors, with $N=\sum_{j}(n_{ij}+n_{ji})$ the corresponding number of photons attached to the particular $i$th fermion subgraph in the amplitude. We also showed on general grounds that the relevant $N$-th rank vacuum polarization tensor in the worldline formalism in $d$ dimensions can be expressed as the normalized worldline expectation value of a product of $N$ charged particle currents:
\ba 
\label{eq:Nth-rank_polarization_tensor}
\Pi_{\mu_1\ldots\mu_N} (k_1,\ldots,k_N) = -\big\langle i\tilde{J}_{\mu_1}(k_1)\cdots i\tilde{J}_{\mu_N}(k_N)\big\rangle \,.
\ea 
Further, we showed that in order to perform the path integrals of the normalized worldline expectation value, the $i$-th current insertion at the time $\tau_i$ can be rewritten, with the aid of two independent dummy  Grassmann variables $\theta_i$ and $\bar{\theta}_i$, as
\ba
\label{eq:j_mu_i_exponentiationtrick}
&i \tilde{J}_{\mu_i}(k_i) = i\int^1_0 d\tau_i e^{ik_i\wc x(\tau_i)} \Big(\dot{x}_{\mu_i}(\tau_i) +i\varepsilon_0\psi_{\mu_i}(\tau_i)\psi_{\nu}(\tau_i)k_\nu^i\Big) \nonumber\\
&=\int^1_0d\tau_i\int d\bar{\theta}_id\theta_i \exp\bigg\{i\int^1_0 d\tau_k J_{\mu_i\rho}^{B,i}(\tau_k) x_\rho(\tau_k) -\int^1_0 d\tau_k J_{\mu_i\rho}^{F,i}(\tau_k)\psi_\rho(\tau_k)\bigg\}
\ea
where the auxiliary bosonic and fermionic virtual worldline currents are given by
\ba
\label{eq:j_b_j_f_def}
J^{B,i}_{\mu_i\rho}(\tau_k) = \Big(\eta_{\mu_i\rho}\theta_i\bar{\theta}_i\frac{d}{d\tau_i}+k_\rho^i\Big)\delta(\tau_k-\tau_i)\,,\,\,\,\, J^{F,i}_{\mu_i\rho}(\tau_k) =\Big( \eta_{\mu_i\rho}\theta_i+\varepsilon_0k_\rho^i\bar{\theta}_i\Big)\delta(\tau_k-\tau_i)\,.
\ea 
Note that a summation is implicit in the repeated index $\rho$ in Eq.~\eqref{eq:j_mu_i_exponentiationtrick}, while the index $\mu_i$ is kept fixed, corresponding to the particular $\mu_i$-component of the polarization tensor. Since the above discussion refers to vacuum-vacuum amplitudes in Euclidean spacetime, $\eta_{\mu\nu}$ is the Euclidean metric tensor. The final expressions for the amplitudes can be straightforwardly rotated to Minkowski times using the rules specified in Appendix A of \cite{Feal:2022iyn}.

After replacing each current appearing in Eq.~\eqref{eq:Nth-rank_polarization_tensor} by Eq.~\eqref{eq:j_mu_i_exponentiationtrick}, we shown that the path integrals over all possible $x_\mu(\tau)$ and $\psi_\mu(\tau)$ virtual worldlines of the normalized worldline expectation value can be performed leading to the following compact form valid for arbitrary $N$:
\ba 
\label{eq:langle_ijN_rangle}
&\big\langle i\tilde{J}_{\mu_1}(k_1)\cdots i\tilde{J}_{\mu_N}(k_N)\big\rangle = 2 \frac{g^n}{n!} (2\pi)^d\delta^d(k_1+\cdots+k_N) \int \frac{d^dp}{(2\pi)^d}\int^\infty_0 \frac{d\varepsilon_0}{\varepsilon_0}e^{-\varepsilon_0(p^2+m^2)} \int_0^1 d\tau_1\cdots \nonumber\\
&\times \int^1_0 d\tau_N \int d\bar{\theta}_1d\theta_1 \cdots \int d\bar{\theta}_Nd\theta_N \exp\bigg\{\frac{\varepsilon_0}{2} \sum_{i,j=1}^N  \int^1_0 d\tau_k \int^1_0 d\tau_l J^{B,i}_{\mu_i\rho}(\tau_k)G_B(\tau_k-\tau_l)J^{B,j}_{\mu_j\rho}(\tau_l)\nonumber\\&
-\frac{1}{2}\sum_{i,j=1}^N\int^1_0d\tau_k \int^1_0 d\tau_l J^{F,i}_{\mu_i\rho}(\tau_k)G_F(\tau_k-\tau_l)J^{F,j}_{\mu_j\rho}(\tau_l)\bigg\}\,,
\ea 
where $\theta_i$ and $\bar{\theta}_i$ for $i=1,\ldots,N$ are independent Grassmann variables, and 
\ba 
\label{eq:g_b_g_f_def}
 G_B(\tau_k-\tau_l)= |\tau_k-\tau_l|-(\tau_k-\tau_l)^2\,,\,\,\, G_F(\tau_k-\tau_l)=\mathrm{sgn}(\tau_k-\tau_l)\,,
\ea 
are the bosonic $(B)$ and fermionic $(F)$ worldline Green functions. 

The primary purpose of this appendix is to demonstrate that the Grassmann integrals in Eq.~\eqref{eq:langle_ijN_rangle} can be performed exactly for arbitrary $N$, see also \cite{Ahmadiniaz:2020wlm} for a recent discussion. This results in a novel universal expression for the $N$th rank vacuum polarization tensor in QED that can be evaluated employing only elementary integrals of polynomial functions of the $\tau_1,\ldots,\tau_N$ proper time points at which each photon is attached within the loop accounting for all possible $N!$ permutations in conventional perturbation theory. As an illustration, we will show that this expression can be employed to obtain the proof of Furry's theorem \cite{PhysRev.51.125} in the worldline formalism for general $N$. 

We start by plugging Eqs.~\eqref{eq:j_b_j_f_def}
into the exponential factor of Eq.~\eqref{eq:langle_ijN_rangle}, which gives
\ba 
&\frac{\varepsilon_0}{2} \sum_{i,j=1}^N  \int^1_0 d\tau_k \int^1_0 d\tau_l J^{B,i}_{\mu_i\rho}(\tau_k)G_B(\tau_k-\tau_l)J^{B,j}_{\mu_j\rho}(\tau_l)
 \nonumber\\
&-\frac{1}{2}\sum_{i,j=1}^N\int^1_0d\tau_k \int^1_0 d\tau_l J^{F,i}_{\mu_i\rho}(\tau_k)G_F(\tau_k-\tau_l)J^{F,j}_{\mu_j\rho}(\tau_l) \nonumber\\
&=  \frac{\varepsilon_0}{2} \sum_{i,j=1}^N \bigg\{\eta_{\mu_i\mu_j}\theta_i\bar{\theta}_i\theta_j\bar{\theta}_j \frac{d^2G_B^{ij}}{d\tau_id\tau_j}+k_{\mu_j}^i\theta_j\bar{\theta}_j \frac{dG_B^{ij}}{d\tau_j}+k_{\mu_i}^j \theta_i\bar{\theta}_i\frac{dG_{B}^{ij}}{d\tau_i}+k^i\wc k^jG_B^{ij}\bigg\}\nonumber\\
&-\frac{1}{2}\sum_{i,j=1}^N \bigg\{\eta_{\mu_i\mu_j} \theta_i\theta_j+\varepsilon_0 k^i_{\mu_j}\bar{\theta}_i\theta_j+\varepsilon_0k_{\mu_i}^j\theta_i\bar{\theta}_j+\varepsilon_0^2k^i\wc k^j \bar{\theta}_i\bar{\theta}_j\bigg\}G_F^{ij}\,,
\ea 
where we used the shorthand notation $G_B^{ij}=G_B(\tau_i-\tau_j)$ and $G_F^{ij}=G_F(\tau_i-\tau_j)$. Here as previously, the indices $\mu_i$ and $\mu_j$ correspond to the particular $\mu_i$ and $\mu_j$ components of the polarization tensor and should not be summed over.
Eq.~\eqref{eq:langle_ijN_rangle} then becomes
\ba 
\label{eq:langle_ijN_rangle_2}
&\big\langle i\tilde{J}_{\mu_1}(k_1)\cdots i\tilde{J}_{\mu_N}(k_N)\big\rangle = 2 \frac{g^N}{N!} (2\pi)^d\delta^d(k_1+\cdots+k_N) \int \frac{d^dp}{(2\pi)^d}\int^\infty_0 \frac{d\varepsilon_0}{\varepsilon_0}e^{-\varepsilon_0(p^2+m^2)} \nonumber\\
&\times\int_0^1 d\tau_1\cdots \int^1_0 d\tau_N \exp\bigg\{\frac{\varepsilon_0}{2}\sum_{i,j=1}^Nk^i\wc k^j G_B^{\tau_i\tau_j} \bigg\}I\,,
\end{align}
where the integral $I$ is always of the general form
\begin{align}
I\equiv \int d\bar{\theta}_1d\theta_1 \cdots d\bar{\theta}_Nd\theta_N\exp\bigg\{A_{ij}\theta_i\bar{\theta}_i\theta_j\bar{\theta}_j+B_{ij}\theta_i\bar{\theta}_j-C_{ij}\theta_i\theta_j-D_{ij}\bar{\theta}_i\bar{\theta}_j\bigg\}\,,\nonumber
\ea 
with $i,j=1,\ldots,N$, we adopt the repeated index summation convention, and the matrices have elements 
\ba 
\label{eq:A_B_C_D_matrixelements}
&A_{ij}=  \frac{\varepsilon_0}{2} \eta_{\mu_i\mu_j}\frac{d^2G_B^{ij}}{d\tau_id\tau_j}\,,\,\,\,\,\,\,
B_{ij}= \varepsilon_0\sum_{\alpha=1}^N k_{\mu_i}^\alpha\bigg(\delta_{ij} \frac{dG_B^{i\alpha}}{d\tau_i}+ \delta_{\alpha j} G_F^{\alpha i}\bigg)\,,\\
&C_{ij}=\frac{1}{2}\eta_{\mu_i\mu_j}G_F^{ij}\,,\,\,\,\, D_{ij}=\frac{\varepsilon_0^2}{2}k^i\wc k^j G_F^{ij}\,.\nonumber
\ea
We used here the symmetries of the worldline Green functions $G_{B}^{ij}=G_B^{ji}$ and $G_F^{ij}=-G_F^{ji}$. 

The integral $I$ is not of  standard form, since it contains terms quartic and quadratic, and in different combinations, of the Grassmann variables. However it can still be evaluated for general $N$ by standard Grassmannian integration. We refer the reader to Appendix A of \cite{Tarasov:2019rfp} for further details of the Grassmann integration procedure. We expand first the exponential to get
\ba 
\label{eq:I_def}
I = \sum_{n=0}^\infty\frac{1}{n!}\int d\bar{\theta}_1d\theta_1 \cdots d\bar{\theta}_Nd\theta_N  \bigg\{A_{ij}\theta_i\bar{\theta}_i\theta_j\bar{\theta}_j+B_{ij}\theta_i\bar{\theta}_j-C_{ij}\theta_i\theta_j-D_{ij}\bar{\theta}_i\bar{\theta}_j\bigg\}^n\,.
\ea 
It is clear that one should make a distinction at this point between even or odd $N$. This observation will be crucial later on with implications impinging directly on the proof of  Furry's theorem. If $N$ is even, only terms ranging from $n=N/2$ to $n=N$ in the sum in Eq.~\eqref{eq:I_def} do not vanish within the integral. Hence,
\ba 
\label{eq:I_step_0}
&I = \sum_{n=0}^{N/2}\frac{1}{(N/2+n)!} \int d\bar{\theta}_1d\theta_1 \cdots  d\bar{\theta}_Nd\theta_N  \bigg\{A_{ij}\theta_i\bar{\theta}_i\theta_j\bar{\theta}_j+B_{ij}\theta_i\bar{\theta}_j-C_{ij}\theta_i\theta_j-D_{ij}\bar{\theta}_i\bar{\theta}_j\bigg\}^{\frac{N}{2}+n}\,.
\ea
Expanding in powers the terms in within brackets in  Eq.~\eqref{eq:I_step_0}, one gets
\ba
\label{eq:I_step_1}
I&=\sum_{n=0}^{N/2} \frac{1}{(N/2+n)!}  \sum_{N_A=0}^{N/2+n} \frac{(N/2+n)!}{N_A!(N/2+n-N_A)!}\nonumber\\
&\times\int d\bar{\theta}_1d\theta_1 \cdots  d\bar{\theta}_Nd\theta_N \Big(A_{ij}\theta_i\bar{\theta}_i\theta_j\bar{\theta}_j\Big)^{N_A} \Big(B_{kl}\theta_k\bar{\theta}_l -C_{kl}\theta_k\theta_l-D_{kl}\bar{\theta}_k\bar{\theta}_l\Big)^{N/2+n-N_A}\,.
\ea 
For each $n$, the only non-vanishing contributions are those with $N_A=N/2-n$. This is because they carry $(N-2n)$ different $\theta$ and $\bar{\theta}$ variables, coming from the $A_{ij}\theta_i\bar{\theta}_i\theta_j\bar{\theta}_j$ factors, plus $N/2+n-N_A=2n$ other different $\theta$ and $\bar{\theta}$ coming from the required combination of the $B_{kl}\theta_k\bar{\theta}_l$, $C_{kl}\theta_k\theta_l$ and $D_{kl}\bar{\theta}_k\bar{\theta}_l$ factors. Hence picking in Eq.~\eqref{eq:I_step_1} only the $N_A=N/2-n$ terms, and redefining later on the dummy variable $n$ to $n=N/2-N_A$ gives
\ba 
\label{eq:I_step_2}
I=&\sum_{N_A=0}^{N/2} \frac{1}{N_A!(N-2N_A)!}\nonumber\\
&\times\int d\bar{\theta}_1d\theta_1 \cdots  d\bar{\theta}_Nd\theta_N \Big(A_{ij}\theta_i\bar{\theta}_i\theta_j\bar{\theta}_j\Big)^{N_A} \Big(B_{kl}\theta_k\bar{\theta}_l -C_{kl}\theta_k\theta_l-D_{kl}\bar{\theta}_k\bar{\theta}_l\Big)^{N-2N_A}\,.
\ea 
Next, expanding the second term in the parenthesis on the r.h.s. of Eq.~\eqref{eq:I_step_2} gives
\ba 
\label{eq:I_step_3}
I=&\sum_{N_A=0}^{N/2} \sum_{N_B=0}^{N-2N_A}\frac{1}{N_A!}\frac{1}{N_B!(N-2N_A-N_B)!}\nonumber\\
&\times\int d\bar{\theta}_1d\theta_1 \cdots  d\bar{\theta}_Nd\theta_N \Big(A_{ij}\theta_i\bar{\theta}_i\theta_j\bar{\theta}_j\Big)^{N_A} \Big(B_{kl}\theta_k\bar{\theta}_l\Big)^{N_B}\Big(-C_{pq}\theta_p\theta_q-D_{pq}\bar{\theta}_p\bar{\theta}_q\Big)^{N-2N_A-N_B}\,.
\ea 
Observe that if $N-2N_A-N_B$ is odd, there will be always an odd number of either $\theta_p\theta_q$ or $\bar{\theta}_p\bar{\theta}_q$ in the integral causing $I$ to 
vanish. Hence only even $N-2N_A-N_B$ contributions should be kept. Setting then $N-2N_A-N_B\equiv 2N_{CD}$, Eq.~\eqref{eq:I_step_3} can be rewritten as
\ba
\label{eq:I_step_4}
I=&\sum_{N_A=0}^{N/2}\sum_{N_B+2N_{CD}=N-2N_A}\frac{1}{N_A!}\frac{1}{N_B!(2N_{CD})!}\nonumber\\
&\times\int d\bar{\theta}_1d\theta_1 \cdots  d\bar{\theta}_Nd\theta_N \Big(A_{ij}\theta_i\bar{\theta}_i\theta_j\bar{\theta}_j\Big)^{N_A} \Big(B_{kl}\theta_k\bar{\theta}_l\Big)^{N_B}\Big(-C_{pq}\theta_p\theta_q-D_{pq}\bar{\theta}_p\bar{\theta}_q\Big)^{2N_{CD}}\,.
\ea 
Finally expanding the third and last term in parenthesis on the r.h.s. of Eq.~\eqref{eq:I_step_4} in powers gives
\ba
\label{eq:I_step_5}
I=&\sum_{N_A=0}^{N/2}\sum_{N_B+2N_{CD}=N-2N_A}\frac{1}{N_A!}\frac{1}{N_B!(2N_{CD})!}\sum_{N_C=0}^{2N_{CD}}\frac{(2N_{CD})!}{N_C!(2N_{CD}-N_C)!}(-1)^{2N_{CD}}\nonumber\\
&\times\int d\bar{\theta}_1d\theta_1 \cdots  d\bar{\theta}_Nd\theta_N \Big(A_{ij}\theta_i\bar{\theta}_i\theta_j\bar{\theta}_j\Big)^{N_A} \Big(B_{kl}\theta_k\bar{\theta}_l\Big)^{N_B}\Big(C_{pq}\theta_p\theta_q\Big)^{N_C}\Big(D_{rs}\bar{\theta}_r\bar{\theta}_s\Big)^{2N_{CD}-N_C}\,.
\ea 
Note finally that only the terms in Eq.~\eqref{eq:I_step_5} with exactly the same number of $C_{pq}\theta_p\theta_q$ and $D_{rs}\bar{\theta}_r\bar{\theta}_s$ factors survive in $I$. Namely, $N_C=2N_{CD}-N_C$, whereby $N_C=N_{CD}$. One then gets
\ba 
\label{eq:I_step_6}
I=&\sum_{2N_A+N_B+2N_{C}=N}\frac{1}{N_A!}\frac{1}{N_B!}\frac{1}{N_{C}!}\frac{1}{N_{C}!}\nonumber\\
&\times\int d\bar{\theta}_1d\theta_1 \cdots  d\bar{\theta}_Nd\theta_N \Big(A_{ij}\theta_i\bar{\theta}_i\theta_j\bar{\theta}_j\Big)^{N_A} \Big(B_{kl}\theta_k\bar{\theta}_l\Big)^{N_B}\Big(C_{pq}\theta_p\theta_q\Big)^{N_{C}}\Big(D_{rs}\bar{\theta}_r\bar{\theta}_s\Big)^{N_{C}}\,.
\ea 
To perform now the $\theta$ and $\bar{\theta}$ integrals, we conveniently rewrite Eq.~\eqref{eq:I_step_6} as
\ba 
\label{eq:I_step_7}
&I= \sum_{2N_A+N_B+2N_{C}=N}\frac{1}{N_A!N_B!N_{C}!N_{C}!} \bigg\{\prod_{\alpha=1}^{N_A} A_{i_\alpha j_\alpha  }\bigg\}\bigg\{\prod_{\alpha=1}^{N_B} B_{k_\alpha l_\alpha}\bigg\} \bigg\{\prod_{\alpha=1}^{N_{C}}C_{p_\alpha q_\alpha}\bigg\}\bigg\{\prod_{\alpha=1}^{N_{C}}D_{r_\alpha s_\alpha}\bigg\}\nonumber\\
&\times\bigg\{ \prod_{\alpha=1}^N \int d\bar{\theta}_\alpha d\theta_\alpha \bigg\} \bigg\{\prod_{\alpha=1}^{N_A} \theta_{i_\alpha}\bar{\theta}_{i_\alpha}\theta_{j_\alpha}\bar{\theta}_{j_\alpha}\bigg\}\bigg\{\prod_{\alpha=1}^{N_B}\theta_{k_\alpha}\bar{\theta}_{l_\alpha}\bigg\}\bigg\{\prod_{\alpha=1}^{N_C}\theta_{p_\alpha}\theta_{q_\alpha}\bigg\}\bigg\{\prod_{\alpha=1}^{N_C}\bar{\theta}_{r_\alpha}\bar{\theta}_{s_\alpha}\bigg\}\,.
\ea 
In Eq.~\eqref{eq:I_step_7} we will now move all $\bar{\theta}'s$ to the left and all $\theta's$ to the right. To do this systematically, note that from the anticommutation properties of any two independent Grassmann variables, one can obtain the following identities:
\ba 
\label{eq:grassmann_identities}
&\prod_{\alpha=1}^n \chi_\alpha \psi_\alpha = (-1)^{n(n-1)/2}\bigg\{\prod_{\alpha=1}^n \chi_\alpha\bigg\}\bigg\{\prod_{\alpha=1}^n\psi_{\alpha}\bigg\}\,, \\
&\bigg\{\prod_{\alpha=1}^n \chi_\alpha\bigg\}\bigg\{ \prod_{\alpha=1}^m \psi_{\alpha}\bigg\} = (-1)^{n\wc m} \bigg\{\prod_{\alpha=1}^m \psi_{\alpha} \bigg\}\bigg\{\prod_{\alpha=1}^n \chi_{\alpha}\bigg\}\nonumber\,,
\ea 
valid for any set $\{\chi_\alpha\}$ and $\{\psi_\alpha\}$ of $n$ and/or $m$ independent Grassmann variables. Applying Eqs.~\eqref{eq:grassmann_identities} to Eq.~\eqref{eq:I_step_7}, one easily obtains
\ba 
&I= \sum_{2N_A+N_B+2N_{C}=N}\frac{(-1)^{N_C}}{N_A!N_B!N_{C}!N_{C}!} \bigg\{\prod_{\alpha=1}^{N_A} A_{i_\alpha j_\alpha }\bigg\}\bigg\{\prod_{\alpha=1}^{N_B} B_{ k_\alpha l_\alpha}\bigg\} \bigg\{\prod_{\alpha=1}^{N_{C}}C_{p_\alpha q_\alpha}\bigg\}\bigg\{\prod_{\alpha=1}^{N_{C}}D_{r_\alpha s_\alpha}\bigg\}\nonumber\\
&\times\bigg\{ \prod_{\alpha=1}^{N} \int d\bar{\theta}_\alpha \bigg\} \bigg\{\prod_{\alpha=1}^{N_A} \bar{\theta}_{i_\alpha}\bar{\theta}_{j_\alpha} \bigg\}\bigg\{\prod_{\alpha=1}^{N_B}\bar{\theta}_{l_\alpha}\bigg\}\bigg\{\prod_{\alpha=1}^{N_C}\bar{\theta}_{r_\alpha}\bar{\theta}_{s_\alpha}\bigg\}\nonumber\\
&\times\bigg\{ \prod_{\alpha=1}^{N} \int d\theta_\alpha \bigg\}\bigg\{\prod_{\alpha=1}^{N_A} \theta_{i_\alpha}\theta_{j_\alpha} \bigg\}\bigg\{\prod_{\alpha=1}^{N_B}\theta_{k_\alpha}\bigg\}\bigg\{\prod_{\alpha=1}^{N_C}\theta_{p_\alpha}\theta_{s_\alpha}\bigg\}\,,
\ea 
where we used the fact that $N=2N_A+N_B+2N_C$ is even (and hence in this case $N_B$ is even as well) to obtain the overall parity factor $(-1)^{N_C}$ of the permutations. Permuting finally as
\ba 
\bigg\{\prod_{\alpha=1}^{N_A} \bar{\theta}_{i_\alpha}\bar{\theta}_{j_\alpha} \bigg\}\bigg\{\prod_{\alpha=1}^{N_B}\bar{\theta}_{l_\alpha}\bigg\}\bigg\{\prod_{\alpha=1}^{N_C}\bar{\theta}_{r_\alpha}\bar{\theta}_{s_\alpha}\bigg\}= \epsilon_{i_1j_1\ldots i_{N_A}j_{N_A}l_1\ldots l_{N_B}r_1s_1\ldots r_{N_C}s_{N_C}}  \bar{\theta}_1\bar{\theta}_2\cdots\bar{\theta}_N\,,
\ea 
and
\ba 
\bigg\{\prod_{\alpha=1}^{N_A} \theta_{i_\alpha}\theta_{j_\alpha} \bigg\}\bigg\{\prod_{\alpha=1}^{N_B}\theta_{k_\alpha}\bigg\}\bigg\{\prod_{\alpha=1}^{N_C}\theta_{p_\alpha}\theta_{s_\alpha}\bigg\} = \epsilon_{i_1j_1\ldots i_{N_A}j_{N_A}k_1\ldots k_{N_B}p_1q_1\ldots p_{N_C}q_{N_C}}  \theta_1\theta_2\cdots\theta_N\,,
\ea 
where $\epsilon_{i_1\cdots i_N}$ is the Levy-Civita symbol, and using Eq.~\eqref{eq:grassmann_identities} to reshuffle the $d\theta$ and $\theta$ variables, and the $d\bar{\theta}$ and $\bar{\theta}$ variables, respectively, one finally finds
\ba
\label{eq:I_step_8}
&I = \sum_{2N_A+N_B+2N_{C}=N}\frac{(-1)^{N_C}}{N_A!N_B!N_{C}!N_{C}!} \epsilon_{i_1j_1\ldots i_{N_A}k_{N_A}k_1\ldots k_{N_B}p_1q_1\ldots p_{N_C}q_{N_C}}   \nonumber\\
&\times \epsilon_{i_1j_1\ldots i_{N_A}j_{N_A}l_1\ldots l_{N_B}r_1s_1\ldots r_{N_C}s_{N_C}} \bigg\{\prod_{\alpha=1}^{N_A} A_{i_\alpha j_\alpha }\bigg\}\bigg\{\prod_{\alpha=1}^{N_B} B_{k_\alpha l_\alpha}\bigg\} \bigg\{\prod_{\alpha=1}^{N_{C}}C_{p_\alpha q_\alpha}\bigg\}\bigg\{\prod_{\alpha=1}^{N_{C}}D_{r_\alpha s_\alpha}\bigg\}\,.\nonumber\\
\ea 

Let us now consider the case of an odd number $N$ of photons attached to the fermion loop. In that case, one can show that the only non-vanishing contributions to the $I$ integral in Eq.~\eqref{eq:I_step_0} are those ranging from $n=(N+1)/2$ to $n=N$. The strategy to extract the different powers of $A_{ij}$, $B_{ij}$, $C_{ij}$ and $D_{ij}$ factors follows the same procedure, and one can in fact show that Eq.~\eqref{eq:I_step_8} is the final result.  The only difference is that $N$ is odd and hence $N_B$ is also odd. Therefore Eq.~\eqref{eq:I_step_8} is valid for general $N$.

Plugging this result for $I$ in Eq.~\eqref{eq:langle_ijN_rangle_2}, one gets for the normalized worldline expectation value of the product of $N$ currents a very general result valid for any $N$:
\ba
\label{eq:langle_ijN_rangle_result1}
&\big\langle i\tilde{J}_{\mu_1}(k_1)\cdots i\tilde{J}_{\mu_N}(k_N)\big\rangle = 2 \frac{g^N}{N!} (2\pi)^d\delta^d(k_1+\cdots+k_N) \int \frac{d^dp}{(2\pi)^d}\int^\infty_0 \frac{d\varepsilon_0}{\varepsilon_0}e^{-\varepsilon_0(p^2+m^2)} \\
&\sum_{2N_A+N_B+2N_{C}=N}\frac{(-1)^{N_C}}{N_A!N_B!N_{C}!N_{C}!} \epsilon_{i_1j_1\ldots i_{N_A}j_{N_A}k_1\ldots k_{N_B}p_1q_1\ldots p_{N_C}q_{N_C}}  \epsilon_{i_1j_1\ldots i_{N_A}j_{N_A}l_1\ldots l_{N_B}r_1s_1\ldots r_{N_C}s_{N_C}}\nonumber\\
&\int_0^1 d\tau_1\cdots \int^1_0 d\tau_N \exp\bigg\{\frac{\varepsilon_0}{2}\sum_{i,j=1}^Nk^i\wc k^j G_B^{ij} \bigg\}\bigg\{\prod_{\alpha=1}^{N_A} A_{i_\alpha j_\alpha }\bigg\}\bigg\{\prod_{\alpha=1}^{N_B} B_{k_\alpha l_\alpha}\bigg\} \bigg\{\prod_{\alpha=1}^{N_{C}}C_{p_\alpha q_\alpha}\bigg\}\bigg\{\prod_{\alpha=1}^{N_{C}}D_{r_\alpha s_\alpha}\bigg\}\nonumber\,.
\ea 
We have obtained thus a compact form of expression for the calculation of the $N$th rank vacuum polarization tensor in QED involving only elementary and unordered integrations of polynomials of proper time in Eq.~\eqref{eq:g_b_g_f_def} and accounting for at once all $N!$ Feynman diagrams in conventional perturbation theory, corresponding to a particular ordering of the $\tau_1,\ldots,\tau_N$ variables. 

To illustrate the versatility, power and usage of the present procedure, let us consider the simpler case of $N=2$. 
Since the possible combinations $(N_A,N_B,N_C)$ such that $2N_A+N_B+2N_C=N$ with $N=2$ are $(1,0,0)$, $(0,2,0)$ and $(0,0,1)$,  Eq.~\eqref{eq:I_step_8} gives simply
\ba 
I=\big(A_{12}+A_{21}\big)+\big(B_{11}B_{22}-B_{12}B_{21}\big)-\big(C_{12}-C_{21}\big)\big(D_{12}-D_{21}\big)\,.
\ea
Using now Eqs.~\eqref{eq:A_B_C_D_matrixelements}, this can be rewritten as
\ba
I=\varepsilon_0 \eta_{\mu_1\mu_2}\frac{d^2G_B^{12}}{d\tau_1d\tau_2}+\varepsilon_0^2k_{\mu_1}^2k_{\mu_2}^1\frac{dG_B^{12}}{d\tau_1}\frac{dG_B^{21}}{d\tau_2}-\varepsilon_0^2\Big(\eta_{\mu_1\mu_2}k_1\wc k_2 -k^2_{\mu_1}k^1_{\mu_2}\Big)\Big(G_F^{12}\Big)^2\,,
\ea 
which as shown in Paper I (c.f. Eqs.~(C22) and (C23) therein), after substitution in Eq.~\eqref{eq:langle_ijN_rangle_2} and integration in $\tau_1$ and $\tau_2$, leads immediately to the conventional form of the 2nd rank vacuum-polarization tensor obtained from conventional perturbation theory,
\begin{align}
\big\langle i\tilde{J}_\mu(k)i\tilde{J}_\nu(q)\big\rangle = (2\pi)^d\delta^d(k+q)\frac{8g^2(\eta_{\mu\nu}k^2-k_\mu k_\nu)}{(4\pi)^{d/2}}\Gamma\bigg(\frac{4-d}{2}\bigg)\int^1_0 d\tau  \frac{\tau(1-\tau)}{\big[m^2+k^2\tau(1-\tau)\big]^{2-d/2}}\,.
\end{align}

An important property of the general result we obtained in Eq.~\eqref{eq:langle_ijN_rangle_result1} for any $N$ is its behavior, as noted in \cite{Schubert:2001he,Strassler:1993km,Schubert:1997ph}, under the worldline time reparametrizations $\tau_i'=1-\tau_i$ for $i=1,\ldots,N$. The worldline Green functions $G_B$ and $G_F$ in Eq.~\eqref{eq:g_b_g_f_def} are symmetric and antisymmetric, respectively, under the previous transformations, since $G_B^{ij}=G_B^{ji}$ and $G_F^{ij}=-G_F^{ji}$. The first and second order time derivatives of the bosonic contributions 
\ba 
\frac{dG_B^{ij}}{d\tau_i}=\mathrm{sgn}(\tau_i-\tau_j)-2(\tau_i-\tau_j)\,, \,\,\, \frac{dG_B^{ij}}{d\tau_id\tau_j}=2-2\delta(\tau_i-\tau_j) = \frac{dG_B^{ji}}{d\tau_id\tau_j}
\ea 
are hence antisymmetric and symmetric, respectively. Accordingly, under the transformations $\tau_i'=1-\tau_i$ for $i=1,\ldots,N$, the term $k^i\wc k^j G_B^{ij}$ in the exponential factor in the third line of Eq.~\eqref{eq:langle_ijN_rangle_result1} remains invariant, each $A_{ij}$ factor remains invariant, the $N_B$ factors $B_{ij}$ introduce an overall parity factor of $(-1)^{N_B}$ in Eq.~\eqref{eq:I_step_8}, and the factors $C_{ij}$ and $D_{ij}$ while antisymmetric, introduce an overall parity of $(-1)^{2N_c}=+1$ since these always appear in pairs. Correspondingly, noticing that the integration measures remain invariant, one finds 
\ba 
\big\langle i\tilde{J}_{\mu_1}(k_1)\cdots i\tilde{J}_{\mu_N}(k_N)\big\rangle  = (-1)^{N_B}\big\langle i\tilde{J}_{\mu_1}(k_1)\cdots i\tilde{J}_{\mu_N}(k_N)\big\rangle \,.
\ea 
Hence in light of Eq.~\eqref{eq:langle_ijN_rangle_result1}, if $N$ is odd, then $N_B$ is also odd, and hence due to reparametrization invariance of the amplitude, the only consistent result is that it must vanish. This result is nothing but Furry's theorem \cite{PhysRev.51.125} in the worldline formalism! 

To illustrate the previous point and demonstrate further the usage of Eq.~\eqref{eq:langle_ijN_rangle_result1}, let us next consider the case of $N=3$. The triads $(N_A,N_B,N_c)$ such that $2N_A+N_B+2N_C=3$ are $(1,1,0)$, $(0,3,0)$ and $(0,1,1)$, and hence Eq.~\eqref{eq:I_step_8} becomes
\ba 
I= \epsilon_{ijk}\epsilon_{ijl}A_{ij}B_{kl}+\frac{1}{3!}\epsilon_{k_1k_2k_3} \epsilon_{l_1l_2l_3}B_{k_1l_1}B_{k_2l_2}B_{k_3l_3}-\epsilon_{mpq}\epsilon_{nrs}B_{mn}C_{pq}D_{rs}\,,
\ea 
whose contributions involve always an odd number of 
$B_{kl}$ terms making the amplitude in Eq.~\eqref{eq:I_step_8} vanish after the corresponding time integration. One can in fact continue exploring higher powers $N$. If $N=5$, for instance, the integral $I$ would involve terms of the form
\ba
\sim A^2B\,,\,\,\, \sim AB^3\,,\,\,\, \sim ABCD,\,\,\,\sim B^5,\,\,\,\sim B^3CD,\,\,\, \text{and}\,\,\,\sim BC^2D^2\,.
\ea 
All of them vanish as well for the same reason of reparametrization invariance of the  amplitude 
in Eq.~\eqref{eq:langle_ijN_rangle_result1}, satisfying Furry's theorem as anticipated. 
In follow-up work, we will apply the powerful technique demonstrated here to high order computations. 

%
%

\section{Photon radiation in M\"oller scattering.}

Just as in Paper I, it is useful to consider the concrete example of photon emission in the final state in M\"oller scattering ($e^-e^-\to e^-e^-$, illustrated in 
Fig.~\ref{fig:figure_7}) in the worldline framework.
We will first discuss the conventional approach using the Dyson S-matrix and then the infrared safe Faddeev-Kulish S-matrix.
\begin{figure}[ht]
    \centering
    \includegraphics[scale=0.6]{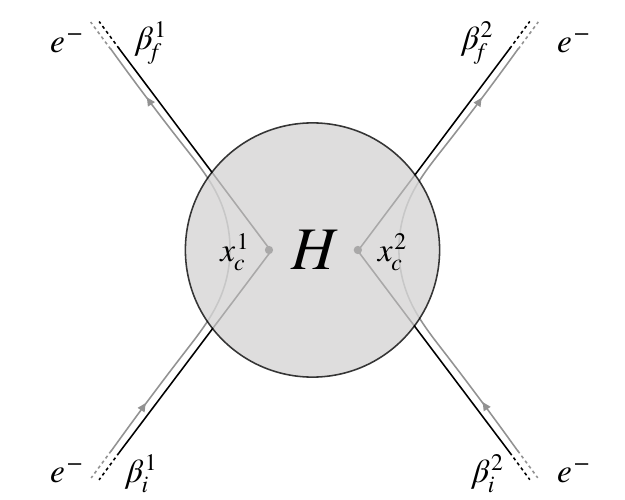}
    \caption{Worldlines for a pair of electrons with initial and final 4-velocities $\beta_{i,f}^{1,2}$ (gray), and pair of cusped worldlines with a single cusp matching the same initial and final velocities (black). In the  (real or virtual) soft photon momentum limit (out of the hard region $H$ denoted with blob) they both are described by the same charged currents.}
    \label{fig:figure_7}
\end{figure}

For  simplicity, we shall only consider the amplitude for the emission of one single IR photon. In the standard approach, the soft factor of the Dyson S-matrix in Eq.~\eqref{eq:s_fi_r_Nos_Nis} then reads (with $N_{o,s}^\gamma=1$ and $N_{i,s}^\gamma=0$) 
\ba 
\label{eq:s_fi_s_r_1_0_moller}
\mathcal{S}_{fi,s}^{(2)}(1,0) = \frac{i\epsilon_\mu^*(\v{k},\lambda)}{\sqrt{2\omega_{\v{k}}}(2\pi)^{3/2}}\tilde{J}^\mu_{IR}(-k)\exp\bigg\{\frac{1}{2}\int^\Lambda_\lambda \frac{d^4q}{(2\pi)^4}\frac{ig^{\mu\nu}}{q^2+i\varepsilon}J_\mu^{IR}(-q)J_\nu^{IR}(+q)\bigg\}\,.
\ea 
The net current at low energies, $\tilde{J}^\mu_{IR}$, comes wholly from the $k,q\to 0$ limits of the charged currents of the two external electrons. In these limits, the currents are only sensitive to the $t_{f,i}\to\pm\infty$ regions of the electron worldlines; their worldlines can be replaced by the cusped 4-trajectories of Fig.~\ref{fig:figure_7} (with initial and final velocities $\beta_{f,i}^n=p_{f,i}^n/E_{f,i}^n$):
\ba 
\label{eq:cusp_worldlines}
x_\mu^n(t) &= x_{i,\mu}^n + \beta_{i,\mu}^n(t-t_i^n)\,,\,\,\, t\in (t_i^n,t_c^n)\,,\nonumber\\
x_\mu^n(t) &= x_{c,\mu}^n+\beta_{f,\mu}^n(t-t_c^n)\,,\,\,\, t\in (t_c^n,t_f^n)\,.
\ea 
The particular form of the worldlines at short distances, replaced here by the cusps $x^n_c$ within the blob in Fig.~\ref{fig:figure_7}, are accounted for in the path integral as part of the normalized worldline expectation value in the hard S-matrix element. Using Eq.~\eqref{eq:cusp_worldlines}, one obtains
\ba 
\label{eq:j_ir_mu_k_moller}
&\tilde{J}^{IR}_\mu(k) =\lim_{k\to 0} \sum_{n=1}^2 g\int^{+\infty}_{-\infty} dt \dot{x}^n_\mu(t) e^{ik\wc x^n(t) -\varepsilon |t|} \nonumber\\
&=  - \frac{g}{i} \sum_{n=1}^2 \bigg\{\frac{\beta^n_{f,\mu}}{k\wc \beta^n_f+i\varepsilon} -  \frac{\beta^n_{i,\mu}}{k\wc \beta^n_i-i\varepsilon}\bigg\} = -\frac{g}{i}\sum_{n',m'=1}^4 \frac{\eta_{n'}\beta_{n'}}{k\wc \beta_{n'}+i\eta_{n'}\varepsilon}\,,
\ea
where the sum in $n'$ in the r.h.s. of Eq.~\eqref{eq:j_ir_mu_k_moller}, as defined previously, runs over the 4 possible \textit{in} and \textit{out} legs of the two external electron worldlines. Plugging Eq.~\eqref{eq:j_ir_mu_k_moller} into Eq.~\eqref{eq:s_fi_s_r_1_0_moller} gives 
\ba 
\label{eq:s_fi_s_r_1_0_moller2}
\mathcal{S}_{fi,s}^{(2)}(1,0) = & \frac{g\epsilon^*_\mu(\v{k},\lambda)}{\sqrt{2\omega_{\v{k}}}(2\pi)^{3/2}} \sum_{n'=1}^{4} \frac{ \eta_{n'}\beta_{n'}^{\mu}}{k\wc \beta_{n'}-i\eta_{n'}\varepsilon} \\
&\times\exp\bigg\{ \sum_{n',m'=1}^{4} \frac{g^2}{2}\int_\lambda^\Lambda\frac{d^4q}{(2\pi)^4}\frac{ig_{\mu\nu}}{q^2+i\varepsilon} \frac{\eta_{n'}\beta_{n'}^\mu}{q\wc \beta_{n'}-i\eta_{n'}\varepsilon} \frac{\eta_{m'}\beta_{m'}^\nu}{q\wc \beta_{m'}+i\eta_{m'}\varepsilon}\bigg\}\,.\nonumber
\ea 
The soft factor above contains the emitted soft photon and arbitrary numbers of virtual soft photons. The real soft photon will introduce an IR divergence in the cross-section, when integrating the modulus squared amplitude over the full phase space, including that of the soft photon. The  virtual soft photons introduce IR divergences as well due to the $q\to 0$ singularity of the integral in the loop momentum $q$ in Eq.~\eqref{eq:s_fi_s_r_1_0_moller2}. As shown previously in \cite{Feal:2022iyn}, one obtains the result 
\ba 
\label{eq:r_nm_phi_nm_moller}
&\exp\bigg\{ \sum_{n',m'=1}^{4} \frac{g^2}{2}\int_\lambda^\Lambda\frac{d^4q}{(2\pi)^4}\frac{ig_{\mu\nu}}{q^2+i\varepsilon} \frac{\eta_{n'}\beta_{n'}^\mu}{q\wc \beta_{n'}-i\eta_{n'}\varepsilon} \frac{\eta_{m'}\beta_{m'}^\nu}{q\wc \beta_{m'}+i\eta_{m'}\varepsilon}\bigg\}\\
&=\exp\bigg\{\frac{g^2}{8\pi^2}\sum_{n',m'=1}^{4} \eta_{n'}\eta_{m'}\gamma_{n'm'}\coth \gamma_{n'm'} \log \frac{\Lambda}{\lambda}-i\frac{g^2}{8\pi}\sum_{n',m'=1}^{4'}\eta_{n'}\eta_{m'}\coth\gamma_{n'm'}\log\frac{\Lambda}{\lambda} \bigg\}\,.\nonumber
\ea 
The sum in $n'$ and $m'$ in the real part of the exponential factor in the r.h.s. of  Eq.~\eqref{eq:r_nm_phi_nm_moller} runs over all pairs of \textit{in} and \textit{out} legs of the two electron worldlines, with the pairs ($n',m'$) and ($m',n'$) counted separately. The sum in the imaginary part runs only over \textit{in}-\textit{in} or \textit{out}-\textit{out} pairings of the legs. The relative cusp 4-angle between worldline legs $n'$ and $m'$ was previously defined in Eq.~\eqref{eq:cosh_gamma_nm}.

The real part of the exponent in the r.h.s. of Eq.~\eqref{eq:r_nm_phi_nm_moller} is always negative for on-shell electrons \cite{PhysRev.140.B516}. Hence the soft factor in Eq.~\eqref{eq:s_fi_s_r_1_0_moller2}, and therefore the amplitude of the transition, vanishes with $\lambda\to 0$ due to the exchange of arbitrary numbers of very soft photons during the scattering. The emission rate is still well defined in the IR limit; as discussed thoroughly in Section \ref{sec:bloch_nordsieck_cancellation}, this is the case if with the scattering one also considers the emission of any number of very soft real photons. Indeed, the rate of emission of for a photon with energy less than $E$, with $E<\Lambda$, is given by Eq.~\eqref{eq:s_fi_s_r_1_0_moller2}. Since the imaginary part drops out of the squared modulus, one has 
\ba 
&\big|\mathcal{S}_{fi,s}^{(2)}(0,0)\big|^2+\sum_{\lambda=\pm 1} \int^E_\lambda d^3\v{k} \big|\mathcal{S}^{(2)}_{fi,s}(1,0)\big|^2+\cdots =  \exp\bigg\{\frac{g^2}{4\pi^2}\sum_{n',m'=1}^{4} \eta_{n'}\eta_{m'}\gamma_{n'm'}\coth \gamma_{n'm'} \log \frac{\Lambda}{\lambda}\bigg\}\nonumber\\
&\times\bigg\{1+ g^2\int^E_\lambda \frac{d^3\v{k}}{(2\pi)^3}\frac{1}{2\omega_{\v{k}}}\bigg[\sum_{\lambda=\pm 1}\epsilon^*_\mu(\v{k},\lambda)\epsilon_\nu(\v{k},\lambda)\bigg] \sum_{n',m'=1}^4\frac{\beta_{n'}^\mu}{k\wc \beta_{n'} }\frac{ \beta_{m'}^\nu}{k\wc \beta_{m'}}+\cdots\bigg\}\,,
\ea 
The first term on the l.h.s. corresponds to the probability of not emitting a photon. The second term corresponds to the probability that a single photon is emitted and integrates over its phase space up to a maximal energy $E$ of the photon. The ellipses represent higher order terms containing the emissions of two or more soft photons. Summing over polarizations 
using Eq.~\eqref{eq:sum_lambda}, and performing the remaining $\mathbf{k}$ integral, the expression above can be expressed as 
\ba 
\label{eq:IR_cancellation_Dyson}
&\big|\mathcal{S}_{fi,s}^{(2)}(0,0)\big|^2+\sum_{\lambda=\pm 1} \int^E_\lambda d^3\v{k} \big|\mathcal{S}^{(2)}_{fi,s}(1,0)\big|^2 +\cdots = \exp\bigg\{\frac{g^2}{4\pi^2}\sum_{n',m'=1}^{4} \eta_{n'}\eta_{m'}\gamma_{n'm'}\coth \gamma_{n'm'} \log \frac{\Lambda}{\lambda}\bigg\}\nonumber\\
&\times\bigg\{1- \frac{g^2}{4\pi^2}\sum_{n',m'=1}^{4} \eta_{n'}\eta_{m'}\gamma_{n'm'}\coth \gamma_{n'm'} \log \frac{E}{\lambda}+\cdots\bigg\}\,.
\ea 

Expanding the exponential in the r.h.s. of Eq.~\eqref{eq:IR_cancellation_Dyson}, the leading $\mathcal{O}(g^2)$ term  accounts for all the diagrams wherein a single virtual soft photon is exchanged in the scattering and contains the logarithmic IR divergence $\log \Lambda/\lambda$ as $\lambda\to 0$. As anticipated, this virtual IR divergence is canceled exactly by the logarithmic IR divergence $\log E/\lambda$ resulting from the phase space integral of the real emitted photon. This well-known cancellation between virtual and real IR divergences in the scattering cross-section is schematically shown in Fig.~\ref{fig:figure_8} to $\mathcal{O}(\alpha)$ in perturbation theory.

\begin{figure}[ht]
    \centering
    \includegraphics[scale=0.35]{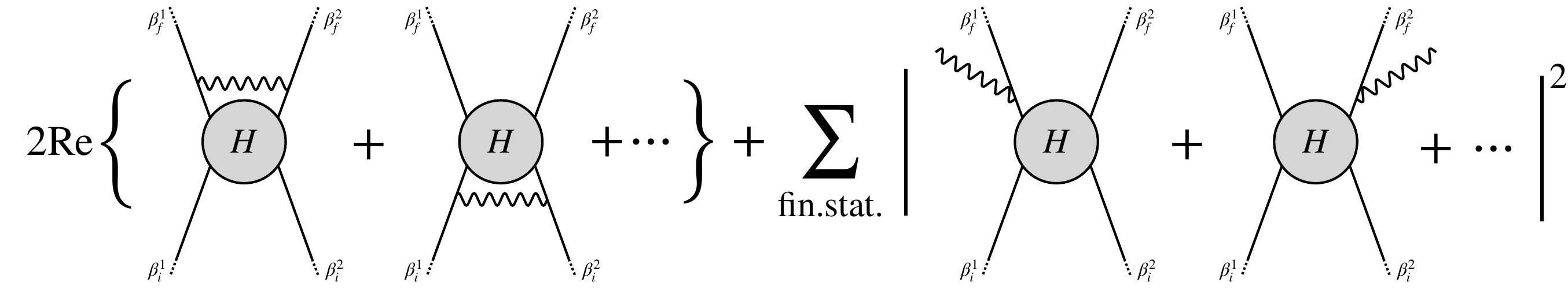}
    \caption{$\mathcal{O}(\alpha)$ diagrams with the exchange of a single virtual IR photon (left) and the emission of real IR photon (right) in  M\"oller scattering, in the Dyson S-matrix computation of the emission rate. 
    The ellipses represent diagrams with additional virtual or real IR photons attached in all possible ways to the \textit{in} and \textit{out} legs. }
    \label{fig:figure_8}
\end{figure}

At higher orders, virtual IR divergences from the exchange of many virtual IR photons will likewise be cancelled exactly by the real IR divergences  from the integrated phase space of the many emitted IR photons, reproducing our general result in  Eq.~\eqref{eq:Gamma_r_fi_s_result}, valid to all-loop orders in perturbation theory.

We will now consider exactly the same process from the perspective of the FK S-matrix $\bar{\mathcal{S}}_{fi}^{(r)}$ in  Eq.~\eqref{eq:s_fi_rell_nfgamma_nigamma_fk}. The electron worldlines at 
large distances are again well approximated by the cusped 4-trajectories of Fig.~\ref{fig:figure_7}. Using Eq.~\eqref{eq:cusp_worldlines}, and keeping finite 
$t_{f,i}^{1,2}$,  the net current of the system in the low momentum limit $k\to 0$ is given by
\ba 
\label{eq:j_mu_ir_k_fk_moller}
&\lim_{k\to 0} \tilde{J}_{\mu}^{(2,\ell)}(k)\simeq \lim_{k\to 0} \sum_{n=1}^2 g\int^{t_f^n}_{t_i^n} dt \dot{x}^n_\mu(t) e^{ik\wc x^n(t) -\varepsilon |t|}\nonumber\\
&=\lim_{k\to 0} \sum_{n=1}^2 \bigg\{\frac{g}{i} \frac{\beta_{f,\mu}^{n}}{k\wc \beta_{f}^n+i\varepsilon}e^{ik\wc x^n_f}
 -\frac{g}{i} \bigg\{\frac{\beta_{f,\mu}^{n}}{k\wc \beta_{f}^n+i\varepsilon}-\frac{\beta_{i,\mu}^{n}}{k\wc \beta_{i}^n-i\varepsilon}\bigg\}e^{ik\wc x^n_c}-\frac{g}{i}\frac{\beta_{i,\mu}^n}{k\wc \beta_i^n-i\varepsilon}e^{ik\wc x_i^n}\bigg\}\,.
\ea 
As emphasized previously, in Weinberg's derivation \cite{PhysRev.140.B516}, the first and last terms in the r.h.s. of Eq.~\eqref{eq:j_mu_ir_k_fk_moller} are dropped due to the rapid oscillations of their phases at asymptotic times, giving back the charged currents in Eq.~\eqref{eq:j_ir_mu_k_moller} corresponding to the Dyson S-matrix prescription. However for finite $x_{f,i}^n$ the phases accompanying the \textit{in} and \textit{out} asymptotic currents go to unity when $k\to 0$, thus cancelling the currents at the cusps. 

As discussed previously, one can split the net current in the IR limit into the two contributions,  
\ba
\label{eq:j_ir_mu_k_moller_fk}
&\lim_{k\to 0} \tilde{J}_\mu^{(2,\ell)}(k)=\lim_{k\to 0}\Big(\tilde{J}_\mu^{AS}(k)+\tilde{J}_\mu^{IR}(k)\Big)\,,\,\,\,\tilde{J}_\mu^{AS}(k) = +\frac{g}{i}\sum_{n'=1}^{4} \frac{\eta^{n'}\beta^{n'}_\mu}{k\wc \beta^{n'}+i\eta^{n'}\varepsilon}e^{ik\wc x^{n'}} \,.
\ea 
Using now Eqs.~\eqref{eq:j_ir_mu_k_moller} and \eqref{eq:j_ir_mu_k_moller_fk}, the soft factor in the FK S-matrix element in Eq.~\eqref{eq:s_fi_r_s_Nos_Nis_fk} corresponding to the amplitude of emission of the single real soft photon ($N_{o,s}=1$ and $N_{i,s}=0$) can be cast as
\ba 
\label{eq:s_fi_s_2_1_0_fk}
&\bar{\mathcal{S}}_{fi,s}^{(2)}(1,0)= g\frac{\epsilon^*_\mu(\v{k},\lambda)}{\sqrt{2\omega_{\v{k}}}(2\pi)^{3/2}}\sum_{n'=1}^4 \frac{\eta_{n'}\beta_{n'}^\mu}{k\wc \beta_{n'}-i\eta_{n'}\varepsilon}\Big(1-e^{-ik\wc x_{n'}}\Big)\nonumber\\
&\times\exp\bigg\{\sum_{n',m'=1}^4 \frac{1}{2}\int_{\Lambda'}^\Lambda\frac{d^4q}{(2\pi)^4}\,\, \frac{ig_{\mu\nu}}{q^2+i\varepsilon}\,\,\frac{\eta_{n'}\beta_{n'}}{q\wc \beta_{n'}-i\eta_{n'}\varepsilon}\,\, \frac{\eta_{m'}\beta_{m'}}{q\wc \beta_{m'}+i\eta_{m'}\varepsilon} \bigg\}\,.
\ea 
independently of the $\lambda$ cut-off, and is regular when $k\to 0$. Hence the effect of the asymptotic phases carried by the \textit{in} and \textit{out} electron currents in the scattering is to cancel both the IR divergences of the virtual soft photon exchanges, and the $\sim 1/k$ behavior of the amplitude created by the emission of the real photon. We refer the reader to Paper I for the details of the cancellations within the IR virtual exchanges in Eq.~\eqref{eq:s_fi_s_2_1_0_fk}. 
Both the real soft photon diagrams from Low's theorem for the Dyson S-matrix and the novel Faddeev-Kulish real asymptotic photon diagrams are shown here in Fig. \ref{fig:figure_9}. 
\begin{figure}[ht]
    \centering
    \includegraphics[scale=0.35]{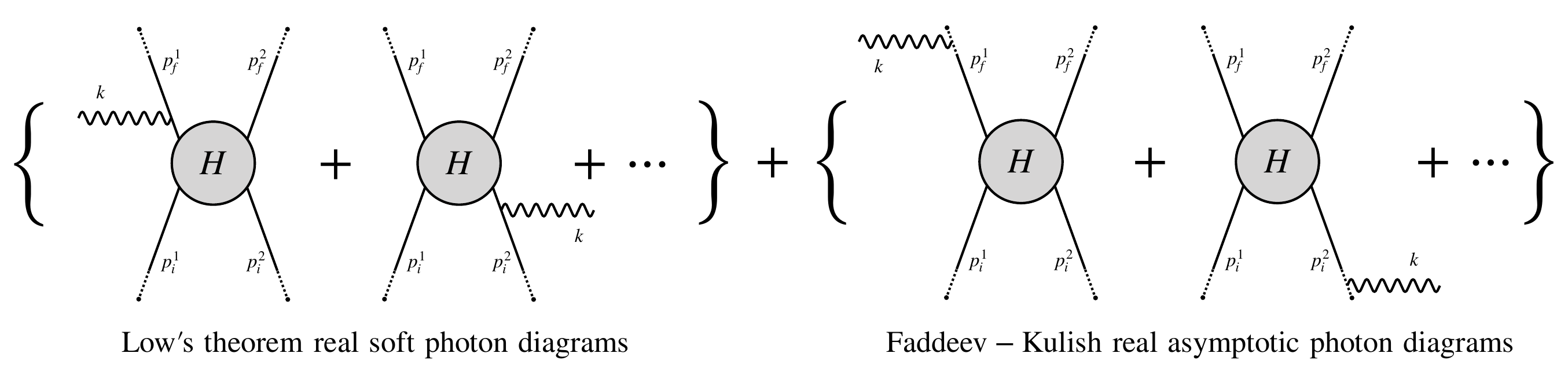}
    \caption{Real IR photon diagrams in M\"oller scattering representing Low's theorem within the  Dyson S-matrix approach 
    and asymptotic Faddeev-Kulish real photon diagrams, 
     with the ellipses indicating photons attached to any other of the \textit{in} and \textit{out} electron worldline legs. Each asymptotic photon diagram gives $\sim 1/k\wc p_{n'}$ and exactly cancels the $\sim 1/k\wc p_{n'}$ singularity of the corresponding real soft photon in Low's theorem in the $k\to 0$ limit where their relative phase vanishes.}
    \label{fig:figure_9}
\end{figure}

This amplitude level cancellation of the real and virtual IR divergences independently, should be contrasted with the conventional cross-section cancellation of Eq.~\eqref{eq:IR_cancellation_Dyson} shown schematically in Fig.~\ref{fig:figure_8}, involving the low energy phase space of the real photon emitted and the IR divergence created by the corresponding virtual photon loop. It is important to note, as emphasized in Paper I, that in the Faddeev-Kulish S-matrix $\bar{S}_{fi}^{(r)}$ the full cancellation of the virtual IR divergences requires one to consider fully the off-shell modes of the virtual exchanges. These are imaginary contributions and do not play any role in the conventional cross-section cancellation.

Eq.~\eqref{eq:s_fi_s_2_1_0_fk} shows transparently that the worldline formulation of any amplitude in QED as a first-quantized theory of charged currents with non-local interactions can be expressed in the low energy limit as a classical theory of worldline currents. With this understanding, the interpretation of the IR aspects of the interactions like the Faddeev and Kulish phases becomes straightforward. In the case of the cross-section for photon radiation in M\"oller scattering as per the FK S-matrix element prescription of the amplitude in Eq.~\eqref{eq:s_fi_s_2_1_0_fk}, the procedure follows exactly as that for the derivation for arbitrary number of charged particles $r$ in the transition detailed in Section \ref{sec:faddeev_kulish} by setting therein $r=2$.

\bibliography{main}
\bibliographystyle{utphys}

\end{document}